\documentclass[fleqn,usenatbib]{mnras}

\usepackage{mathptmx}

\usepackage[T1]{fontenc}

\usepackage[normalem]{ulem} 

\DeclareRobustCommand{\VAN}[3]{#2}
\let\VANthebibliography\thebibliography
\def\thebibliography{\DeclareRobustCommand{\VAN}[3]{##3}\VANthebibliography}

\usepackage{graphicx}	
\usepackage{amsmath}	
\usepackage{amssymb}	

\mathchardef\mhyphen="2D
\DeclareMathOperator{\sech}{sech}

\newcommand{\bfx}{\mathbf{x}}

\newcommand{\degree}{\ensuremath{^\circ}}

\newcommand{\bfv}{\mathbf{v}}
\newcommand{\pc}{\,{\rm pc}}
\newcommand{\kpc}{\,{\rm kpc}}
\newcommand{\Myr}{\,{\rm Myr}}

\newcommand{\kms}{\,{\rm km\, s^{-1}}}

\newcommand{\cs}{c_{\rm s}}

\newcommand{\hatn}{\hat{\textbf{n}}}
\newcommand{\hatx}{\hat{\textbf{x}}}
\newcommand{\haty}{\hat{\textbf{y}}}
\newcommand{\hatz}{\hat{\textbf{z}}}

\newcommand{\pa}{\partial}
\newcommand{\cmthree}{\,{\rm cm^{-3}}}
\newcommand{\Msun}{\,{\rm M_\odot }}
\newcommand{\Msunyr}{\,{\rm M_\odot \, yr^{-1}}}

\newcommand{\K}{\,{\rm K}}

\title[Gas dynamics in the CMZ]{Simulations of the Milky Way's central molecular zone - I. Gas dynamics}

\author[Tress et al.]{Robin G. Tress$^1$, Mattia C. Sormani$^{1}$, Simon C.O. Glover$^1$, Ralf S. Klessen$^{1,2}$,  \newauthor Cara D. Battersby$^3$, Paul C. Clark$^4$, H Perry Hatchfield$^3$ and Rowan J. Smith$^5$\\
$^1$Universit\"{a}t Heidelberg, Zentrum f\"{u}r Astronomie, Institut f\"{u}r theoretische Astrophysik, Albert-Ueberle-Str. 2, 69120 Heidelberg, Germany \\
$^2$Universit\"at Heidelberg, Interdiszipli\"ares Zentrum f\"ur Wissenschaftliches Rechnen, Im Neuenheimer Feld 205, 69120 Heidelberg, Germany \\
$^3$University of Connecticut, Department of Physics, 196 Auditorium Road, Unit 3046, Storrs, CT 06269, USA\\
$^4$School of Physics and Astronomy, Queen's Buildings, The Parade, Cardiff University, Cardiff, CF24 3AA, UK \\
$^5$Jodrell Bank Centre for Astrophysics, School of Physics and Astronomy, University of Manchester, Oxford Road, Manchester M13 9PL, UK \\
}

\begin{document}

\date{} 
\maketitle

\begin{abstract}
We use hydrodynamical simulations to study the Milky Way's central molecular zone (CMZ). The simulations include a non-equilibrium chemical network, the gas self-gravity, star formation and supernova feedback. We resolve the structure of the interstellar medium at sub-parsec resolution while also capturing the interaction between the CMZ and the bar-driven large-scale flow out to $R\sim 5\kpc$. Our main findings are as follows: (1) The distinction between inner ($R\lesssim120$~pc) and outer ($120\lesssim R\lesssim450$~pc) CMZ that is sometimes proposed in the literature is unnecessary. Instead, the CMZ is best described as single structure, namely a star-forming ring with outer radius $R\simeq 200$~pc which includes the 1.3$^\circ$ complex and which is directly interacting with the dust lanes that mediate the bar-driven inflow. (2) This accretion can induce a significant tilt of the CMZ out of the plane. A tilted CMZ might provide an alternative explanation to the $\infty$-shaped structure identified in Herschel data by Molinari et al. 2011. (3) The bar in our simulation efficiently drives an inflow from the Galactic disc ($R\simeq 3$~kpc) down to the CMZ ($R\simeq200$~pc) of the order of $1\rm\,M_\odot\,yr^{-1}$, consistent with observational determinations. (4) Supernova feedback can drive an inflow from the CMZ inwards towards the circumnuclear disc of the order of $\sim0.03\,\rm M_\odot\,yr^{-1}$. (5) We give a new interpretation for the 3D placement of the 20 and 50 km s$^{-1}$ clouds, according to which they are close ($R\lesssim30$~pc) to the Galactic centre, but are also connected to the larger-scale streams at $R\gtrsim100$~pc.
\end{abstract}


\section{Introduction} \label{sec:intro}

The central molecular zone (CMZ) is a concentration of molecular gas in the innermost few hundred parsecs ($R \lesssim 200 \pc)$ of our Galaxy. This accumulation of gas is believed to be created and fed from the outside by the Galactic bar, which drives an inflow of gas from large radii ($R \simeq 3$ kpc) inwards along the disc of the order of $\sim 1 \Msunyr$ \citep{SormaniBarnes2019}. The CMZ is the Milky Way (MW) counterpart of the star-forming nuclear rings commonly found at the centre of external barred galaxies such as NGC 1300 (see for example the atlas of nuclear rings of \citealt{Comeron+2010}).

The CMZ has a total gas mass of approximately $5 \times 10^7 \Msun$ \citep{Dahmen+1998}, which amounts to roughly 5\% of all the molecular gas in the Galaxy. The distribution of molecular gas in the CMZ is highly asymmetric with respect to the Galactic centre, with about $3/4$ of the gas seen in $^{13}$CO and CS residing at $l > 0$, and only $1/4$ at $l<0$ \citep{Bally+1988}. Our previous simulations have shown that such an asymmetry develops spontaneously when gas flows in a barred potential, even in the absence of stellar and/or other type of feedback: the highly non-axisymmetric potential of the bar drives large-scale turbulent flows, giving rise to fluctuations comparable to the observed asymmetry \citep{Sormani+2018a}.

The conditions in the CMZ are extreme. The CMZ has average densities \citep{Longmore+2017,Mills+2018}, temperatures \citep{Immer+2016,Ginsburg+2016,Krieger+2017,Oka+2019}, velocity dispersions \citep{Shetty+2012,Federrath+2016} and estimated magnetic field strengths \citep{Chuss+2003,Crocker+2010,Morris2015,Mangilli+2019} several orders of magnitudes higher than in the Galactic disc.

The inner Galaxy also hosts several high-energy processes. The most spectacular example is the presence of the Fermi Bubbles, two giant gamma-ray lobes extending $\sim 8 \kpc$ above and below the Galactic centre \citep{Su+2010}. These are part of a multiphase Galactic outflow that has been detected across most of the electromagnetic spectrum and that comprises most of the known interstellar gas phases including hot ionised ($T\sim10^6\K$, \citealt{Kataoka+2013,Ponti+2019,Nakashima+2019}), warm ionised ($T\sim 10^4\mhyphen10^5\K$, \citealt{Fox+2015,Bordoloi+2017}), cool atomic ($T\sim10^3\mhyphen10^4\K$, \citealt{McClureGriffiths+2013,DiTeodoro+2018}) and cool molecular ($T\sim100\K$, \citealt{DiTeodoro+2020}) gas.

Star formation and gas dynamics in the CMZ have been the subject of intense study during the past decade, which has produced vast advancements both on the observational \citep[e.g.][]{Molinari+2011,Jones+2012,Immer+2012,Immer+2016,Ginsburg+2016,Henshaw+2016,Longmore+2017,Krieger+2017,Kauffmann+2017a,Kauffmann+2017b,Mills+2018,Mangilli+2019,Oka+2019} and theoretical \citep[e.g.][]{SBM2015a,SBM2015c,Kruijssen+2015,KrumholzKruijssen2015,Krumholz+2017,Ridley+2017,Kruijssen+2019,Dale+2019,Sormani+2018a,Sormani+2019,Armillotta+2019,Armillotta+2020,Li+2020} sides. Despite these advancements, many important open questions remain. 

In the current study and in a companion paper (\citealt{Sormani+2020}, hereafter Paper II) we use high-resolution hydrodynamical simulations with sub-parsec resolution that include a non-equilibrium time-dependent chemical network, star formation and stellar feedback in order to address some of these questions. This paper focuses on describing the numerical methods and on the gas dynamics, while Paper II focuses on star formation.

Open questions that we address in the current work include: 
\begin{enumerate}
\item Is gas in the CMZ on $x_2$ orbits? (see Section \ref{sec:x2vsopen})
\item Is gas deposited by the bar-driven inflow onto an ``outer CMZ'' at $R\simeq 450 \pc$, or almost directly on the CMZ at $R \lesssim 200\pc$? Is there an ``outer CMZ'' which is physically distinct from an ``inner CMZ''? (see Section \ref{sec:innerouter})
\item What is the \emph{dynamical} origin of the $\infty$-shape discovered by \cite{Molinari+2011}? Is the CMZ disc tilted out of the Galactic plane? (see Section \ref{sec:tilt})
\item How is gas transported from the CMZ inwards to the central few parsecs? (see Section \ref{sec:feedbackinflow})
\item Are the 20 and 50 km s$^{-1}$ clouds very close ($R \leq 20 \pc$) to SgrA* or further out ($R = 50 \mhyphen 100 \pc$)? (see Section \ref{sec:2050})
\end{enumerate}

The paper is structured as follows. In Section \ref{sec:methods} we describe our numerical methods and the differences between the present and our previous simulations. In Section \ref{sec:global} we give a brief overview of the overall gas dynamics and morphology. In Section \ref{sec:thermal} we discuss the thermal and chemical properties of our simulated ISM. In Section \ref{sec:inflow} we discuss inflows and outflows. In Section \ref{sec:discussion} we discuss the implications of our results for the open questions raised above. We sum up in Section \ref{sec:conclusion}.

\section{Numerical methods} \label{sec:methods}

The simulations presented here are similar to those we previously discussed in \cite{Sormani+2018a,Sormani+2019}, with the following differences: (i) inclusion of the gas self-gravity; (ii) inclusion of a sub-grid prescription for star formation and stellar feedback. Besides that, we employ exactly the same externally imposed rotating barred potential, the same chemical/thermal treatment of the gas, and the same initial conditions as in \cite{Sormani+2019}. In Section \ref{sec:methods:overview} we give a brief overview that provides the minimum background necessary to read the remainder of this paper, while in Sections \ref{sec:methods:equations}-\ref{sec:IC} we describe our numerical methods in more detail.

\subsection{Overview} \label{sec:methods:overview}

We use the moving-mesh code {\sc arepo} \citep{Springel2010,Weinberger+2020}. A more detailed description of the strengths of this code in the context of modelling the multi-phase nature of the interstellar medium on galactic scales can be found for example in Sect 2.1 of \cite{Tress+2019}. The simulations are three-dimensional and unmagnetised, and include a live chemical network that keeps track of hydrogen and carbon chemistry (see Section \ref{sec:chemistry}). The simulations comprise interstellar gas in the whole inner disc ($R \leq 5 \kpc$) of the MW. This allows us to model the CMZ in the context of the larger-scale flow, which is important since the CMZ strongly interacts with its surrounding through the bar inflow \citep{Sormani+2018a}. The gas is assumed to flow in a multi-component external rotating barred potential $\Phi_{\rm ext}(\bfx,t)$ which is constructed to fit the properties of the MW. This potential is identical to that used in \cite{Sormani+2019} and is described in detail in the appendix of that paper.

The gas self-gravity is included. The process of star formation and stellar feedback processes are modelled as follows (see Sections \ref{sec:sinks} and \ref{sec:feedback} for more details):
\begin{enumerate}
\item When a high density region collapses and the resolution limit is reached, a sink particle (hereafter referred to simply as ``sink'') is created to replace the gas in this region. The sink does not represent an individual star, but a small cluster which contains both gas and stars.
\item Once a sink is created, a stellar population is assigned to it by drawing from an IMF according to the Poisson stochastic method described in \cite{Sormani+2017b}.
\item The sinks are allowed to accrete mass at later times. Their stellar population is updated every time mass is accreted.
\item For each massive star ($M \geq 8 \Msun$) assigned to the sink, we produce a supernova (SN) event with a time delay which depends on the stellar progenitor mass. Each SN event injects energy/momentum into the ISM and gives back to the environment part of the gas ``locked-up'' in the sink. SN feedback is the only type of feedback included in the simulation.
\item When all the SNe in a sink have exploded and all its gas content has been given back to the environment, the sink is converted into a collisionless N-body particle with a mass equal to the stellar mass of the sink. This N-body particle will continue to exist indefinitely in the simulation and will affect it through its gravitational potential.
\end{enumerate}

For simplicity, we ignore magnetic fields. Thus our simulations are unable to capture dynamical effects that may arise as a consequence of the strong and highly ordered magnetic field that is present in the CMZ \citep{Chuss+2003,Crocker+2010,Morris2015,Mangilli+2019}, which could for example induce radial mass and angular momentum transport \citep{BalbusHawley1998}, drive Galactic outflows \citep{PakmorSpringel2013}, shape the properties and structure of molecular clouds \citep{Pillai+2015,Girichidis+2018} and affect the star formation rate \citep{MacLowKlessen2004,KrumholzFederrath2019}.

When making projections onto the plane of the Sky, we assume an angle between the Sun-Galactic centre line and the bar major axis of $\phi = 20^\circ$, as in our previous papers \citep{Sormani+2018a,Sormani+2019}.

\subsection{Equations solved by the code}  \label{sec:methods:equations}

The code solves the equations of fluid dynamics:
\begin{align}
& \frac{\pa \rho}{\pa t} + \nabla \cdot (\rho \bfv) = 0, \\
& \frac{ \pa (\rho \bfv)}{\pa t} + \nabla \cdot \left( \rho \bfv \otimes \bfv + P \bold I \right) = - \rho \nabla \Phi, \\
& \frac{\pa(\rho e)}{\pa t} + \nabla \cdot \left[(\rho e + P) \bfv \right] = \dot{Q} + \rho \frac{\pa \Phi}{\pa t},  \label{eq:energy}
\end{align}
where $\rho$ is the gas density, $\bfv$ is the velocity, $P$ is the thermal pressure, $\bold I$ is the identity matrix, $e~=~e_{\rm therm} + \Phi + {\bfv^2}/{2} $ is the energy per unit mass, $e_{\rm therm}$ is the thermal energy per unit mass. We adopt the equation of state of an ideal gas, $P = (\gamma -1) \rho e_{\rm therm}$, where $\gamma=5/3$ is the adiabatic index. 

The term $\dot{Q}$ in Eq.~\eqref{eq:energy} contains the changes to the internal energy of the gas due to radiative, chemical and feedback processes ($\dot{Q}=0$ for an adiabatic gas). The following processes contribute to $\dot{Q}$: (i) the cooling function, which is part of the chemical network (Section \ref{sec:chemistry}) and which depends on the instantaneous chemical composition of the gas \citep{glo10,gc12}; (ii) heat released and absorbed by those chemical processes that occur in the interstellar medium and are tracked by our chemical network (Section \ref{sec:chemistry}), such as the formation of H$_2$ on dust grains; (iii) the averaged interstellar radiation field and cosmic ray ionisation rate, which represent external heating sources (Section \ref{sec:chemistry}); (iv) the SN feedback, which injects energy into the ISM if the Sedov-Taylor phase is resolved (see Section \ref{sec:feedback})  

The term $\Phi$ represents the sum of the externally imposed gravitational potential and of the self-gravity of the gas, sinks and N-body particles:
\begin{equation}
\Phi = \Phi_{\rm ext} + \Phi_{\rm sg}.
\end{equation}
The code calculates $\Phi_{\rm sg}$ by solving Poisson's equation at each timestep:
\begin{equation}
\nabla^2 \Phi_{\rm sg} = 4 \pi G \rho \, .
\end{equation}
{\sc Arepo} solves Poisson's equation by using a tree-based approach adapted from an improved version of {\sc gadget-2} \citep{Springel2005}. In this approach each gas cell is treated as a point mass located at the centre of the cell with an associated gravitational softening. The gas softening length $\epsilon_{\rm gas}$ is adaptive and depends on the cell size according to $\epsilon_{\rm gas} = 2 r_{\rm cell}$, where $r_{\rm cell}$ is the radius of a sphere with the same volume as the cell. {\sc Arepo} ensures that the gas cells remain quasi-spherical and so this radius is an accurate way of characterising the size of the cells. The lower limit of the softening length is set at $\epsilon_{\rm gas, min} = 0.1$~pc. The softening length of the sinks (which is the same as for N-body particles) is constant and is reported in Table \ref{tab:sinks}.

\subsection{Chemistry of the gas} \label{sec:chemistry}

We account for the chemical evolution of the gas using an updated version of the NL97 chemical network from \citet{gc12}, which itself was based on the work of \citet{gm07a,gm07b} and \citet{nl97}. With this network, we solve for the non-equilibrium abundances of H, H$_{2}$, H$^{+}$, C$^{+}$, O, CO and free electrons. This network is the same used in \citet{Sormani+2018a} and we refer to their Section~3.4 for a more extensive description of it.

One of the key parameters of the network is the strength of the spatially averaged interstellar radiation field (ISRF). This is set to the standard value $G_0$ measured in the solar neighbourhood \citep{draine78} diminished by a local attenuation factor which depends on the amount of gas present within 30~pc of each computational cell. This attenuation factor is introduced to account for the effects of dust extinction and H$_{2}$ self-shielding and is calculated using the {\sc Treecol} algorithm described in \cite{clark12}. The cosmic ray ionisation rate (CRIR) is fixed to $\zeta_{\rm H} = 3 \times 10^{-17} \: {\rm s^{-1}}$ \citep{gl78}. The values adopted for the strength of the ISRF and the size of the CRIR correspond to the `low' simulation of \cite{Sormani+2018a}. While absorption studies of H$_3^+$ \citep{LePetit+2016,Oka+2019} and the elevated molecular gas temperatures \citep{Clark+2013,Ginsburg+2016} indicate that both the ISRF and the CRIR are almost certainly higher than this in the CMZ, we have chosen these values in order to facilitate comparison with the previous simulation in \cite{Sormani+2019}, allowing us to isolate the effects of self-gravity and SN feedback on the star formation process, which as mentioned above are the only differences between \cite{Sormani+2019} and the simulation presented here. The main effects of a higher CRIR would be to decrease the amount of molecular gas and to make the gas warmer. In particular, since cosmic rays can penetrate much further than UV photons into high-column density regions, they can heat the interior of molecular clouds \citep{Clark+2013}. However, we have shown in \cite{Sormani+2018a} that the strength of the ISRF/CRIR makes little difference to the large-scale dynamics. Indeed, even if the ISRF field is a factor of a 1000 higher than in the solar neighbourhood, the thermal sound speed of the molecular gas never comes close to the values of $\cs=5\mhyphen10\kms$ which would be needed to significantly affect the dynamics of the gas \citep{SBM2015a}.  The effects of varying the strengths of the ISRF/CRIR in combination with the inclusion of self-gravity will be explored in future work. Finally, we impose a temperature floor $T_{\rm floor} =20$~K on the simulated ISM. Without this floor, the code occasionally produces anomalously low temperatures in cells close to the resolution limit undergoing strong adiabatic cooling, causing it to crash.

\subsection{Sink particles} \label{sec:sinks}

The sink particles implementation used here is identical to that in \cite{Tress+2019}. We therefore give only a brief description here and refer to their Section~2.3 for more details.

Accreting ``sink particles'' have been used mainly in small-scale simulations of \emph{individual} molecular clouds, in which the SF process can be spatially and temporally resolved reasonably well, in order to replace collapsing high-density region when the resolution limit is reached \cite[e.g.][]{Bate+1995,Federrath++2010}. In larger, galactic-scale simulations, in which the SF process is resolved less well, non-accreting stochastic ``star particles'' are instead more often employed \citep[e.g.][]{Katz1992, Katz+1996, Stinson+2006}. The former approach requires higher resolution but is more predictive, and allows one to assess the star formation efficiency and the mass distribution of fragments in a robust and quantitative way \citep{Federrath++2010}. The latter approach can be used to produce a healthy ISM matter cycle in lower-resolution simulations, but its predictive power is more limited, being often fine-tuned to reproduce the Schmidt-Kennicutt relation \citep{Schmidt1959, Kennicutt1998}, rather than ``predicting'' it.

Our simulations are in an intermediate regime. While we are able to resolve individual molecular clouds very well (see Section \ref{sec:resolution} and Figure \ref{fig:resolution}), we do not have the resolution to follow the formation of individual stars. In this intermediate regime, we choose to use sink particles, so that we can take advantage of their superior predictive power. However, in our simulations each sink represents a small stellar cluster rather than an individual star. We assume that only a fraction $\epsilon_{\rm SF} = 5\%$ of the mass accreted actually forms stars, while the rest is ``locked-up'' as gas in the sinks until SN feedback (see Section \ref{sec:feedback}) gives it back to the environment.  This is necessary for two important reasons. First, our resolution is not sufficient to fully resolve the turbulent cascade within individual molecular clouds or the generation of the associated density sub-structure. Because of this, the density of a Voronoi cell within a molecular cloud should properly be thought of as representing
the mean density of the gas associated with the region of space represented by the cell. Not all of the gas within that region of space will actually be at that density, and hence not all of the gas will be immediately available for star formation. Second, our simulations neglect a number of physical effects (e.g.\ magnetic field, protostellar outflows) that are known to make star formation less efficient \citep[see e.g.][]{Federrath+2015}. Because of this, the adoption of an efficiency of $\epsilon_{\rm SF} = 100$\% would actually be less realistic than our choice of a smaller value and would result in an unrealistically high star formation rate. We note that although the choice of $\epsilon_{\rm SF} = 5$\% is somewhat arbitrary, at our current resolution it does yield a depletion time for the CMZ in fairly good agreement with the observationally determined value (see Paper II). Moreover, our choice yields a star formation efficiency per free-fall time $\epsilon_{\rm ff} \simeq 1$\%, in good agreement with most observational determinations of this value \citep[see e.g.][]{Krumholz+2012,Evans+2014,Barnes+2017,Utomo+2018}.

Following \citet{Federrath++2010}, a sink particle is created if a region within an accretion radius $r_{\rm acc}$ and above a density threshold $\rho_{\rm c}$ simultaneously satisfies all the following criteria:
\begin{enumerate}
\item The gas flow is locally converging. To establish this, we require not only that the velocity divergence is negative ($\nabla \cdot v < 0$), but also that the divergence of the acceleration is negative ($\nabla \cdot a < 0$).  
\item The region is located at a local minimum of the potential $\Phi$.
\item The region is not situated within the accretion radius of another sink and also will not move within the accretion radius of a sink in a time smaller than the local free-fall time.
\item The region is gravitationally bound, i.e. $U > 2 (E_{\rm k} + E_{\rm th})$, where $U = G M^2 / r_{\rm acc}$ is the gravitational energy of the region within the accretion radius, $E_{\rm k} = 1/2 \sum_i m_i \Delta v_i^{2}$ is the total kinetic energy of all gas particles within the accretion radius with respect to the centre of collapse, $E_{\rm th} = \sum_i m_i e_{{\rm th}, i}$ is the total internal energy of the same region, the sum is extended over all cells within the region, and $m_i$ and $e_{{\rm th}, i}$ are the mass and the internal energy per unit mass of the single cells respectively.
\end{enumerate}
These criteria help to ensure that a region is only converted into a sink if it is truly self-gravitating and collapsing. Sink particles are collisionless, so they do not exert/feel pressure forces and interact with gas particles only through gravity. The softening length of the sink particles is reported in Table \ref{tab:sinks}.

Sink particles accrete mass during the simulation. If a gas cell is (i) denser than the threshold density $\rho_c$; (ii) within the accretion radius $r_{\rm acc}$ of the sink particle and (iii) gravitationally bound to it, then we move an amount of mass
\begin{equation}
\Delta m = \left(\rho_{\rm cell} - \rho_{\rm c}\right) V_{\rm cell}
\end{equation}
from the cell to the sink, where $\rho_{\rm cell}$ is the initial gas density in the cell and $V_{\rm cell}$ is its volume. Afterwards the new density of the cell is simply the threshold density $\rho_{\rm c}$. We also update appropriately any other quantities in the cell that depend on the mass, such as the total momentum or kinetic energy. In the case where a given gas cell is located within the accretion radii of multiple sink particles, we place the accreted mass from it onto the sink to which the gas is most strongly bound.

\begin{table}
	\centering
	\begin{tabular}{lcr} 
		parameter			& units			& value 		\\
		\hline
		$\rho_{\rm c}$ 		& g~cm$^{-3}$		 & $10^{-20}$ 	\\
		$r_{\rm acc}$ 		& pc       			 & $1.0$ 		\\
        		$r_{\rm soft}$ 		& pc    			 & $1.0$	 	\\
        		$\epsilon_{\rm SF}$  & 				 & $0.05$ 		\\
        		$r_{\rm sc}$    		& pc 				 & $5.0$ 		\\
		\hline
	\end{tabular}
        \caption{Parameters of the sink particles. $\rho_{\rm c}$ is the density threshold, $r_{\rm acc}$ is the accretion radius, $r_{\rm soft}$ is the softening length, $\epsilon_{\rm SF}$ is the SF efficiency, and $r_{\rm sc}$ is the scatter radius of SNe around the sink.}
\label{tab:sinks}
\end{table}

Ideally, we would like to have the threshold density $\rho_c$ as large as possible, but in practice this is set by what we can afford in terms of computational resources. In order to properly follow the hierarchical collapse and correctly follow the underlying fragmentation, we need to ensure that the local Jeans length is resolved by at least four resolution elements (\citealt{Truelove+1997}; see also \citealt{Federrath+2011} for further discussion). If the threshold density $\rho_{\rm c}$ is too high, the Jeans length can become prohibitively small. In this first set of simulation we choose set the threshold density at $\rho_{\rm c} = 10^{-20}$~g~cm$^{-3}$, and stop refining for densities above this threshold (see Section \ref{sec:resolution}). This is a good compromise between resolution in the collapsing regions, and computational performance. The chosen density threshold allows us to resolve star formation at the average densities of the CMZ.

The accretion radius is chosen such that at the given threshold density $\rho_{\rm c}$, several cells fall inside $r_{\rm acc}$ given the local size of the cells. The gravitational softening length of the collisionless sink particles is set to the same value as $r_{\rm acc}$, as this ensures that the gravitational potential is not altered much due to the infall of mass onto a sink, while at the same time limiting the size of the gravitational acceleration produced within $r_{\rm acc}$, which otherwise would have a detrimental effect on performance. The main parameters that characterise the sink particles used in our study are listed in Table~\ref{tab:sinks}. 

\begin{figure}
	\includegraphics[width=\columnwidth]{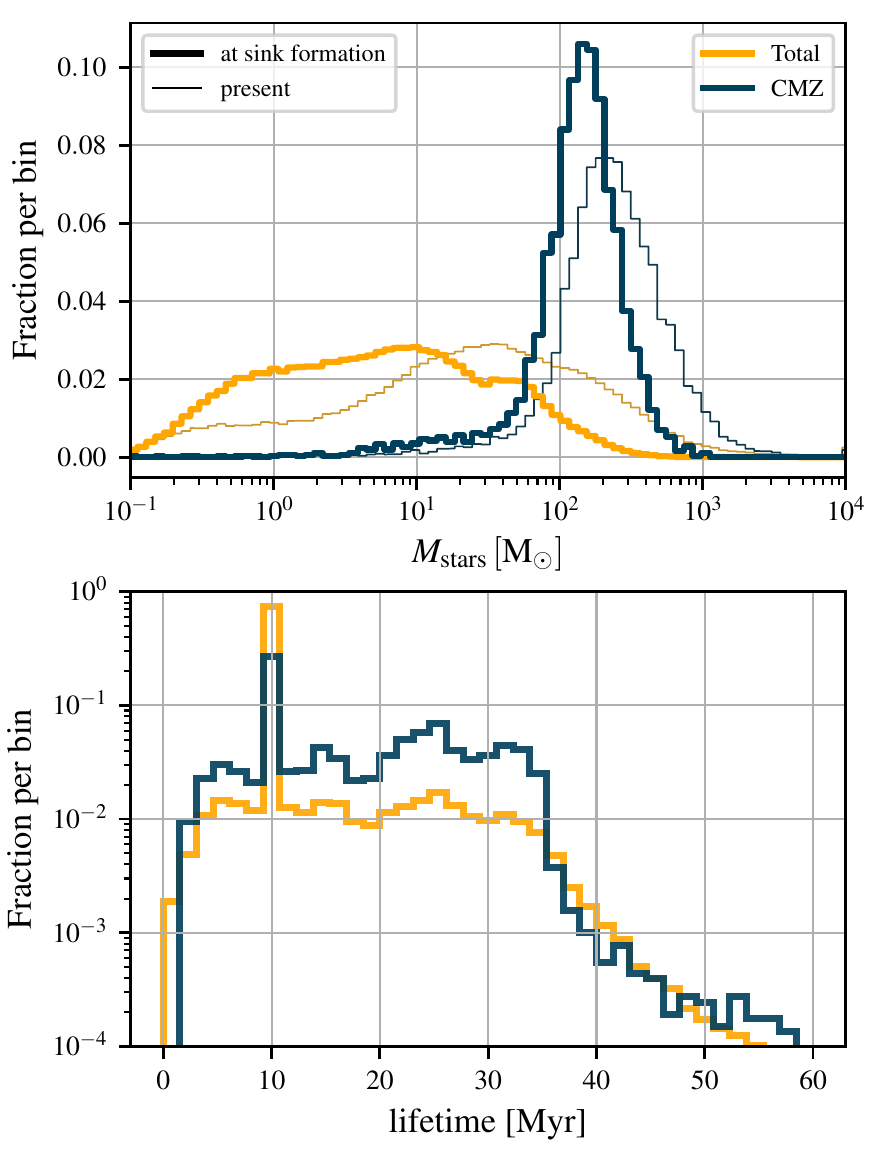}
    \caption{{\em Top panel}: mass distribution of the stellar mass of sink particles found at a representative snapshot in the simulations. We show the mass distribution at their formation ({\em thick lines}) and at the current time of the snapshot ({\em thin lines}) where the mass increased due to accretion. {\em Bottom panel}: distribution of the life-times of the sink particles in the simulation. The peak at $10$~Myr is due to sink particles that failed to generate massive stars and are then converted to a passive star particle (see Section \ref{sec:feedback}). Blue and yellow indicate sinks in the CMZ and disc respectively (see Section~\ref{sec:regions} and Figure~\ref{fig:regions} for definitions of CMZ and disc used in the analysis).}
    \label{fig:sinkmasses}
\end{figure}

The top panel \ref{fig:sinkmasses} illustrates the typical mass distribution of the sinks at the time of their formation (thick lines) and the instantaneous mass distribution at a representative time (thin lines). The figure shows that the sinks have on average higher masses in the CMZ compared to the Galactic disc (see Figure \ref{fig:regions} for the definition of CMZ and disc). The bottom panel in the figure shows a histogram of the sink lifetimes.

\begin{figure}
	\includegraphics[width=\columnwidth]{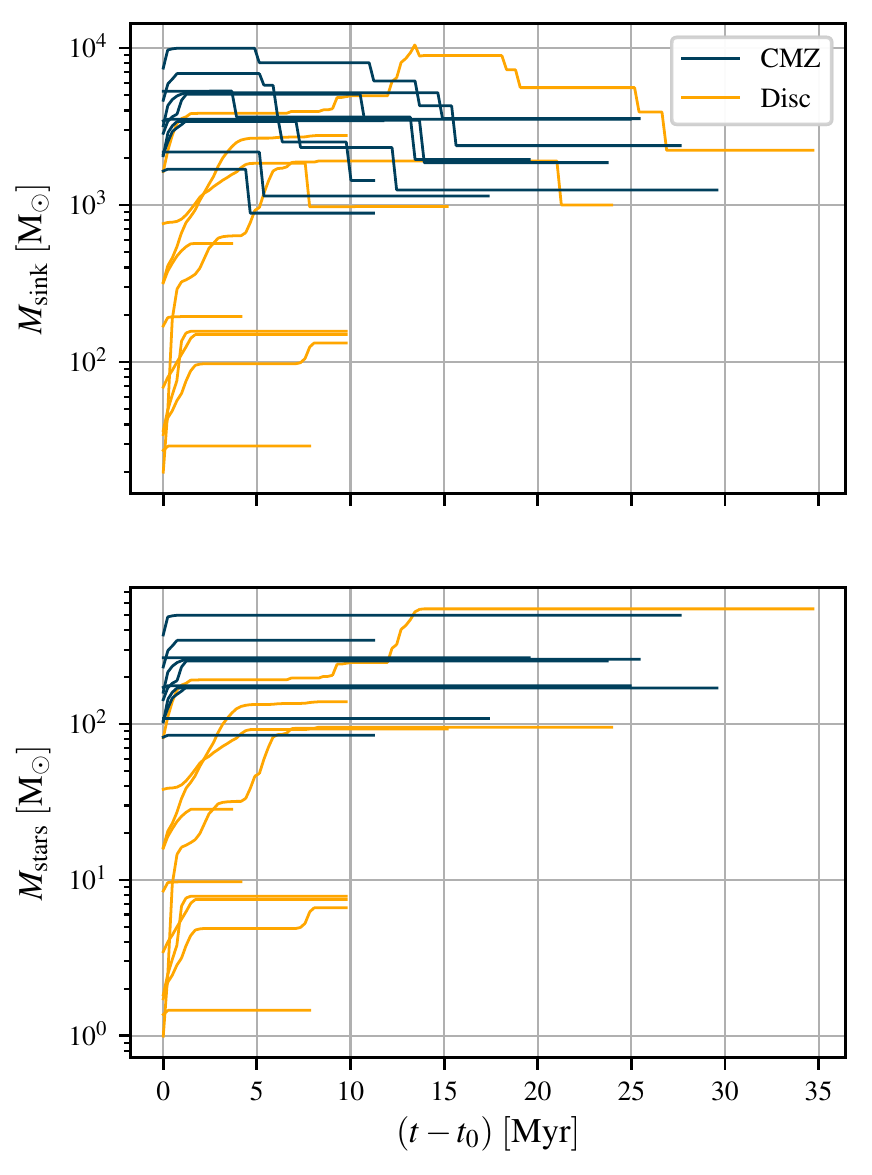}
    \caption{Accretion history for a sample of randomly selected sink particles in our simulation. \emph{Top panel}: total mass of the sink (gas + stars) as a function of time for randomly selected sinks (i) in the CMZ (blue lines) and (ii) in the disc (orange lines). $t_0$ denotes the sink formation time. Each increase in the mass corresponds to an accretion event, while each decrease corresponds to a SN explosion (which frees some of the gas locked-up in the sink). \emph{Bottom panel:} total \emph{stellar} mass as a function of time for the same sample of sinks shown in the top panel. The stellar mass does not decrease when a SN event occurs, contrary to the total mass, which decreases because gas is given back to the environment. This figure shows that (a) most of the accretion usually takes place during the first few Myr, when the sinks and the collapsing gas cloud from which they originated have not yet decoupled (see also discussion in Section~3.3 of Paper II), and (b) the sink formation mass is on average higher in the CMZ than in the disc (see also Figure \ref{fig:sinkmasses}). The typical lifetimes of the sinks can also be read off this figure: sinks without massive stars (and therefore without any associated SN) event die after $t=10 \Myr$, while sinks with massive stars die after the last SN event has taken place (see Section \ref{sec:feedback} for more details).}
    \label{fig:accretionhistory}
\end{figure}

The typical accretion histories of a representative sample of our sinks are shown in Figure \ref{fig:accretionhistory}. This figure shows that most of the accretion takes place during the first few Myr, when the sink is still immersed in the collapsing gas cloud from which it originated (see also discussion in \citealt{KlessenBurkert2000}). Accretion usually stops when the gas and sink decouple at later times (due to feedback and to gas and sinks following different equations of motion, see discussion in Section~3.3 of Paper II). 

\subsection{Stellar feedback} \label{sec:feedback}

The only type of feedback included in our simulation is type II supernova (SN) feedback. The implementation used here is identical to that described in Section~2.4 of \cite{Tress+2019},  to which we refer for a more detailed description.

Once a sink is created, it must be ``populated'' with stars. We attribute a discrete stellar population to the sink sampling from a \cite{Kroupa2001} initial mass function (IMF) using the method described in \citet{Sormani+2017b}. According to this method, we draw the IMF according to a Poisson distribution such that the average stellar mass attributed to a sink is $ \epsilon_{\rm SF} M_{\rm sink}$, where $M_{\rm sink}$ is the mass of the sink and $\epsilon_{\rm SF}$ its SF efficiency (see Table \ref{tab:sinks}). This approach ensures that, globally, the mass distribution of stars in the simulation follows the IMF. The same procedure is repeated every time a mass $\Delta M_{\rm sink}$ is accreted onto the sink, by drawing an IMF according to a Poisson distribution with average stellar mass $\Delta M_{\rm sink} \epsilon_{\rm SF}$. Our method ensures that the final stellar population of the sink (including the stars formed from mass accreted at later times) is independent of the particular accretion history of the sink.

For each star more massive than $8$ M$_\odot$ associated with the sink, we generate a SN event at the end of the lifetime of the star, which are calculated based on their mass from Table~25.6 of \citet{Maeder2009}. Since the sink represents an entire group of stars that can interact dynamically, we do not assume that the SN occurs exactly at the location of the sink. Instead, we randomly sample the SN location from a Gaussian distribution centred on the particle and with standard deviation $ r_{\rm sc} = 5$~pc. 

Since the assumed efficiency of SF within the sink is relatively small, most of the mass in the sink represents gas that should be eventually returned to the ISM. The ``gaseous'' mass locked-up in the sink is therefore gradually given back to the ISM with every SN event. Each event ejects a gas mass of $M_{\rm ej} = (M_{\rm sink} - M_{\rm stars}) / n_{\rm SN}$, where $M_{\rm sink}$ is the mass of the sink at the time that the supernova occurs, $M_{\rm stars}$ is the mass of stars contained within the sink at that time, and $n_{\rm SN}$ is the remaining number of SN events that the sink harbors. The mass is distributed uniformly within the energy injection region. The temperature of the injection cells is not altered at this stage.

Once the last massive star has reached the end of its lifetime, the sink has a final mass of $M_{\rm stars}$. At this point, we convert it into a collisionless $N$-body particle representing its evolved stellar population. It will then continue to exist until the end of the simulation as a collisionless particle which interacts with the gas only through gravity.  

It is not uncommon to have sinks that do not accrete enough mass to form a massive star. In this case we cannot return the gas mass trapped within the sink during an SN event. Instead, after a period of $10$~Myr, if the sink still did not manage to create a massive star, we convert it into a collisionless N-body particle and return the remaining mass ($95$\%) to the ISM by uniformly adding it to all gas particles in a surrounding sphere of $R=100$~pc. This is necessary because since gas and stars follow different equations of motion they will eventually decouple (stars are collisionless so that they do not feel pressure forces, while gas is collisional; see also Section 3.3 of Paper II). Typical lifetimes of the sinks can be read off Figure \ref{fig:sinkmasses}.

To model the supernova energy/momentum injection of each SN event, we calculate the radius of a supernova remnant at the end of its Sedov-Taylor phase based on an assumed SN energy of $10^{51}$~erg and the local mean density $\bar{n}$, which for solar metallicity is \citep{Blondin1998}
\begin{equation}    
R_{\rm ST} = 19.1 \left(\frac{\bar{n}}{1\mbox{ cm}^{-3}}\right)^{-7/17} \: {\rm pc},
\end{equation}
where in our case $\bar{n}$ is calculated including the contributions from both the ambient gas and also the mass loading of the SN event. We compare this with the radius of the injection region, $R_{\rm inj}$, defined as the size of the smallest sphere around the explosion site that contains 40 grid cells. If $R_{\rm ST} > R_{\rm inj}$, the Sedov-Taylor phase is resolved and we inject $E_{\rm SN} = 10^{51}$~erg into the injection region in the form of thermal energy and fully ionise the contained gas. If, on the other hand, the Sedov-Taylor phase of the SN remnant is unresolved, thermal energy would be radiated away too quickly, making it unable to generate a strong shock and deposit the correct amount of kinetic energy into the ISM. In this case, as is common practice in numerical simulations, we inject directly an amount of momentum given by \citep[see e.g.][]{Martizzi+2015, Gatto+2015, KimOstriker2015}
\begin{equation}
p_{\rm fin} = 2.6 \times 10^5 n^{-2/17} \; \mbox{ M}_\odot \mbox{ km s}^{-1},
\end{equation}
for a SN of energy $E_{\rm SN} = 10^{51}$~erg and solar metallicity. We do not change the temperature or the ionisation state of the region in this case as this would throw off-balance the energy budget in large unresolved regions. Since we lack early feedback, the initial SN of a star-forming region will be injected into a high density medium in which the Sedov-Taylor phase is often unresolved. Therefore most of the SNe in the simulation inject momentum, with only around $15$~\% injecting thermal energy.

Momentum injection alone cannot produce a hot phase in the ISM. By keeping the injection radius small we minimise the number of occasions on which we must inject momentum rather than thermal energy. On the other hand, we cannot take too small an injection radius since this would lead to anisotropic momentum injection due to the small number of cells amongst which to distribute the momentum. Even though SNe are observed to be anisotropic in some instances (e.g.\ \citealt{WangWheeler2008}), we have no reason to expect that the artificial anisotropy introduced by taking a very small accretion radius will resemble this real physical anisotropy. We therefore prefer to minimise the numerical noise that such a low cell count would introduce. We have found through experimentation that defining $R_{\rm inj}$ such that a total of $40$ grid cells are contained within a sphere of that radius seems to offer the best trade-off between minimising the number of momentum injection events and minimising the impact of grid noise and anisotropic expansion on the evolution of the individual remnants. We note that this mixed approach of injecting thermal energy in regions where $R_{\rm ST}$ is resolved and momentum in regions where this is not the case is not new. Similar methods have been successfully used by a number of other authors to study the impact of SN feedback on the ISM \citep[see e.g.][]{Kimm+2014,Hopkins+2014,Walch+2015,Simpson+2015,KimOstriker2017}.

Finally, we mention that SN is not the only type of feedback associated with SF. For example, stellar winds and radiation from young stars also play an important role in dispersing GMCs, particularly since they act much earlier than SN feedback \citep[e.g.][]{Dale+2014,Inutsuka+2015,Offner+2015,Rosen+2016,Gatto+2017,Rahner+2018, Rahner+2019,Kruijssen+2019b,Chevance+2020}. However, it remains computationally challenging to include all of these forms of feedback in simulations with the scale and resolution of those presented here. Therefore, in our initial study we restrict our attention to the effects of SN feedback and defer an investigation of other feedback processes to future work.

\subsection{Resolution} 
\label{sec:resolution}

\begin{figure}
	\includegraphics[width=\columnwidth]{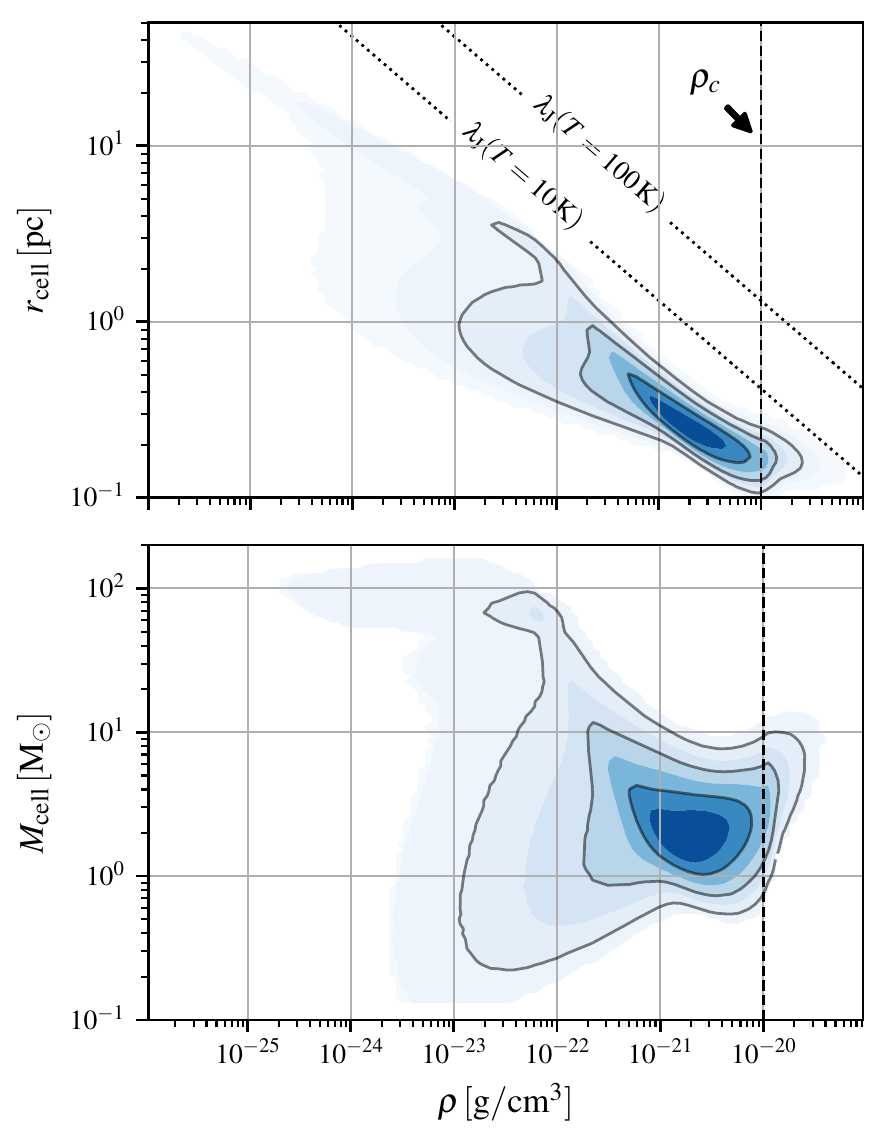}
    \caption{{\em Top}: spatial resolution as a function of density in our simulation. $r_{\rm cell}$ is the radius of a sphere with the same volume as the cell. {\em Bottom}: mass of the cells as a function of density. The black contours contain $(25, 50, 75)$\% of the total number of cells. The dotted lines show the Jeans length $\lambda_{\rm J}$ at the indicated temperature. The vertical dashed line denotes the sink density formation threshold $\rho_{\rm c}$  (See Table~\ref{tab:sinks}).}
    \label{fig:resolution}
\end{figure}

We use the system of mass refinement present in {\sc Arepo} so that the resolution depends on the local density and temperature according to the following criteria. We use a base target cell mass of  $100 \, \Msun$. This means that no cells in the simulation fall below this resolution (i.e., no cell in the simulations can have mass larger than this) within a tolerance factor $\sim 2$. Then, we ensure that the Jeans length is locally resolved by at least four resolution elements according to the criterion of \cite{Truelove+1997}. We stop refining for densities above the sink creation threshold $\rho_{\rm c}$ (see Table \ref{tab:sinks} and discussion in Section \ref{sec:sinks}). The resolution achieved in our simulation is displayed graphically in Figure \ref{fig:resolution}.

\subsection{Initial conditions} \label{sec:IC}

We initialise the density according to the following axisymmetric density distribution:
\begin{equation}
\rho(R,z) = \frac{\Sigma_0}{4 z_{\rm d}}  \exp\left(- \frac{R_{\rm m}}{R} - \frac{R}{R_{\rm d}}\right) \sech\left(\frac{z}{2 z_{\rm d}}\right)^2\, ,
\end{equation}
where $(R,\phi,z)$ denote standard cylindrical coordinates, $z_{\rm d} = 85 \pc$, $R_{\rm d} = 7 \kpc$, $R_{\rm m} = 1.5 \kpc$, $\Sigma_0 = 50 {\rm M_\odot} \pc^{-2}$, and we cut the disc so that $\rho=0$ for $R\geq 5\kpc$. This profile matches the observed radial distribution of gas in the Galaxy \citep{KalberlaDedes2008,HeyerDame2015}. The total initial gas mass in the simulation is $\simeq 1.5 \times 10^9 \Msun$. The computational box has a total size of $24 \times 24 \times 24 \kpc$ with periodic boundary conditions. The box is sufficiently large that the outer boundary has a negligible effect on the evolution of the simulated galaxy.

In order to avoid transients, we introduce the bar gradually, as is common practice in simulations of gas flow in barred potentials \citep[e.g.][]{Athan92b}. We start with gas in equilibrium on circular orbits in an axisymmetrised potential and then we turn on the non-axisymmetric part of the potential linearly during the first $146 \Myr$ (approximately one bar rotation) while keeping constant the total mass distribution which generates the underlying external potential. Therefore, only the simulation at $t \geq 146 \Myr$, when the bar is fully on, will be considered for the analysis in this paper.

\section{General properties of the gas flows} \label{sec:global}

\subsection{Subdivision in three regions: CMZ, DLR, disc} \label{sec:regions}

\begin{figure}
	\includegraphics[width=\columnwidth]{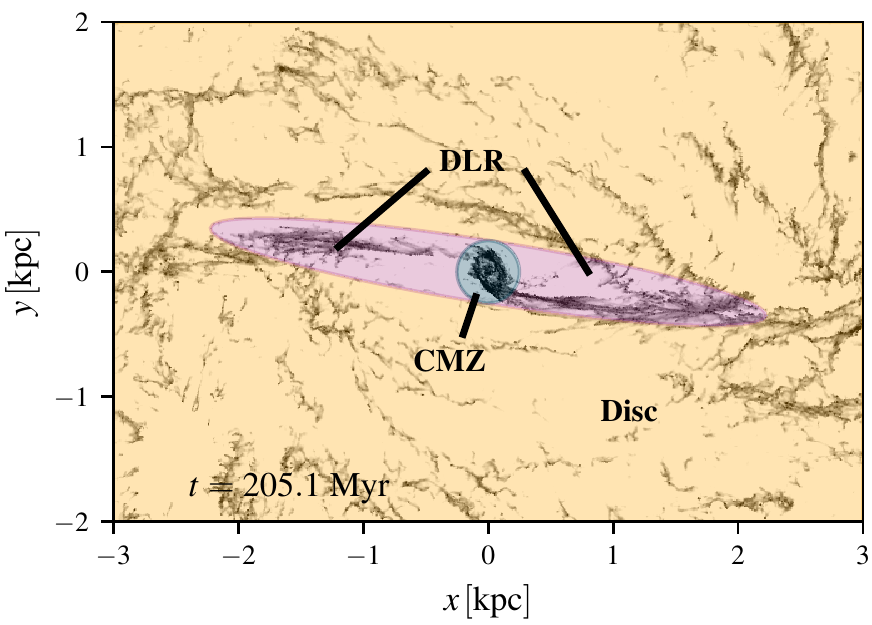}
    \caption{Definition of the three regions (CMZ, DLR, disc) into which we subdivide our simulated Galaxy for subsequent analysis. See Section \ref{sec:regions} for more details.}
    \label{fig:regions}
\end{figure}

In order to facilitate the analysis in the following sections, we subdivide our simulation into three spatial regions (see Figure \ref{fig:regions}): 
\begin{itemize}
\item The \emph{CMZ} is defined as the region within cylindrical radius $R \leq 250 \pc$. 
\item The \emph{dust lane region (DLR)} is the elongated transition region between the CMZ and the Galactic disc, where highly non-circular gas motions caused by the bar are present.\footnote{The name ``dust lane region'' comes from the fact that this is the region where in observations of external barred galaxies such as NGC 1300 or NGC 5383 one can see "the presence of two dust lanes leaving the nucleus one on each side of the bar and extending into the spiral arms" \citep{Sandage1961}. In the simulation, these correspond to the large-scale shocks which are clearly visible in Figure \ref{fig:regions} \citep[e.g.][]{Athan92b}.}
\item The \emph{disc} is defined as everything outside the DLR. In our simulation, the gaseous disc extends out to $R \simeq 5\kpc$ (see Section \ref{sec:IC}).
\end{itemize}

\subsection{Large-scale gas dynamics} \label{sec:largescale}

\begin{figure*}
	\includegraphics[width=\textwidth]{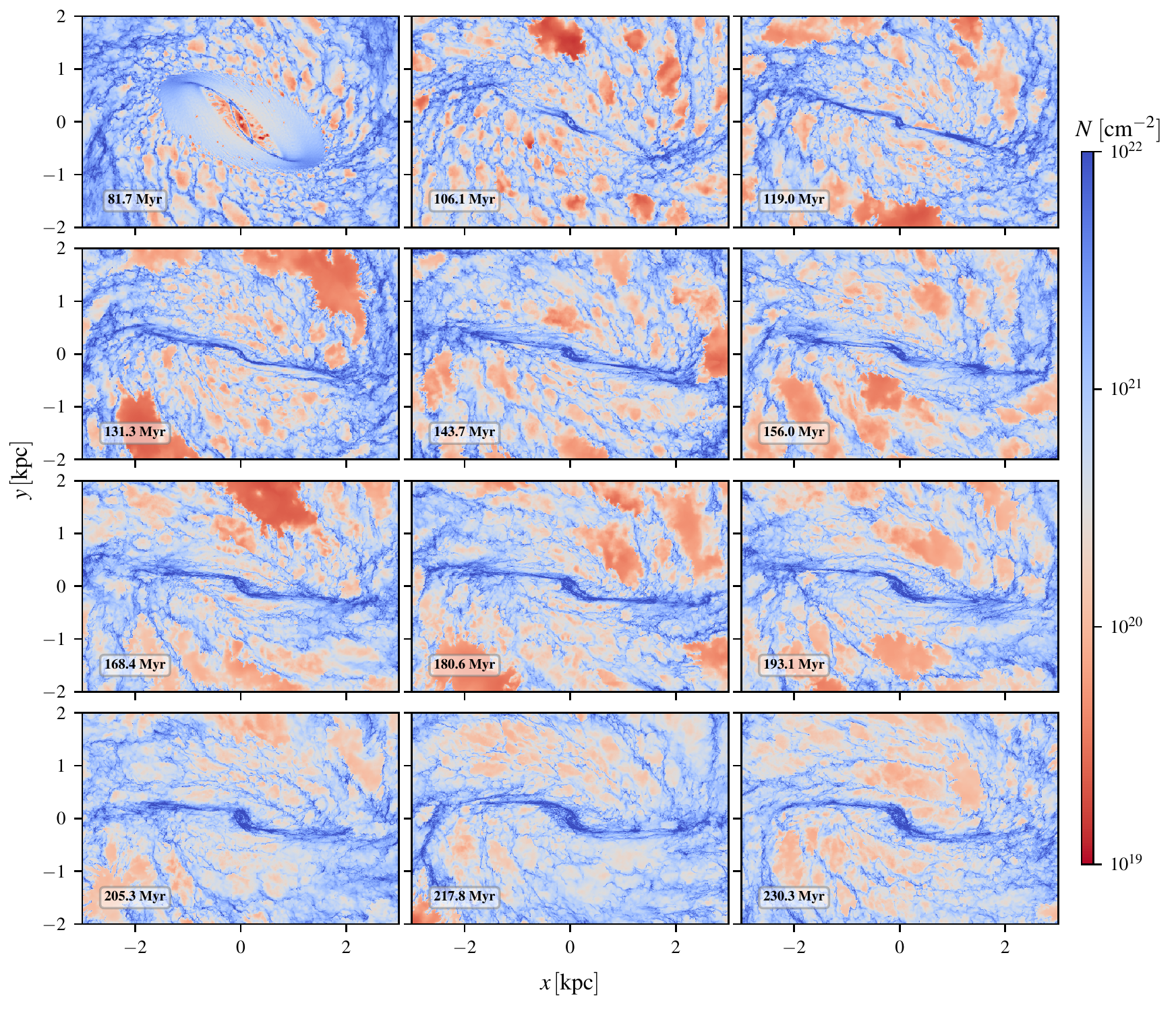}
    \caption{Total gas surface density at different times in the simulation. The simulated CMZ is the dense ring at the centre (see Figure \ref{fig:NTotvstime_CMZ}).  The bar potential is gradually turning on for $t<146 \Myr$ (see Section \ref{sec:IC}). Hence, only the times $t \geq 146 \Myr$, when the bar is fully on, should be considered when analysing the morphology. In all panels, the bar major axis is horizontal.}
    \label{fig:NTotvstime}
\end{figure*}

Figure \ref{fig:NTotvstime} shows the surface density of the gas at different times in the simulation and Figure \ref{fig:NTotvstime_CMZ} zooms onto the CMZ. The appendix provides further figures in different tracers. Figure \ref{fig:vfield} shows the instantaneous gas streamlines. These figures show that the large-scale gas flow approximately follows the $x_1/x_2$ orbit\footnote{This nomenclature has become standard in the literature after the work of \cite{ContopoulosGrosbol1989}.} dynamics, similar to our previous non-self gravitating simulations \citep{Sormani+2018a,Sormani+2019}. This has been described in detail for example in Section~4.1 of \cite{Sormani+2018a} and references therein, and will not be repeated here.

\subsection{Morphology of the CMZ} \label{sec:CMZmorphology}

\begin{figure*}
	\includegraphics[width=\textwidth]{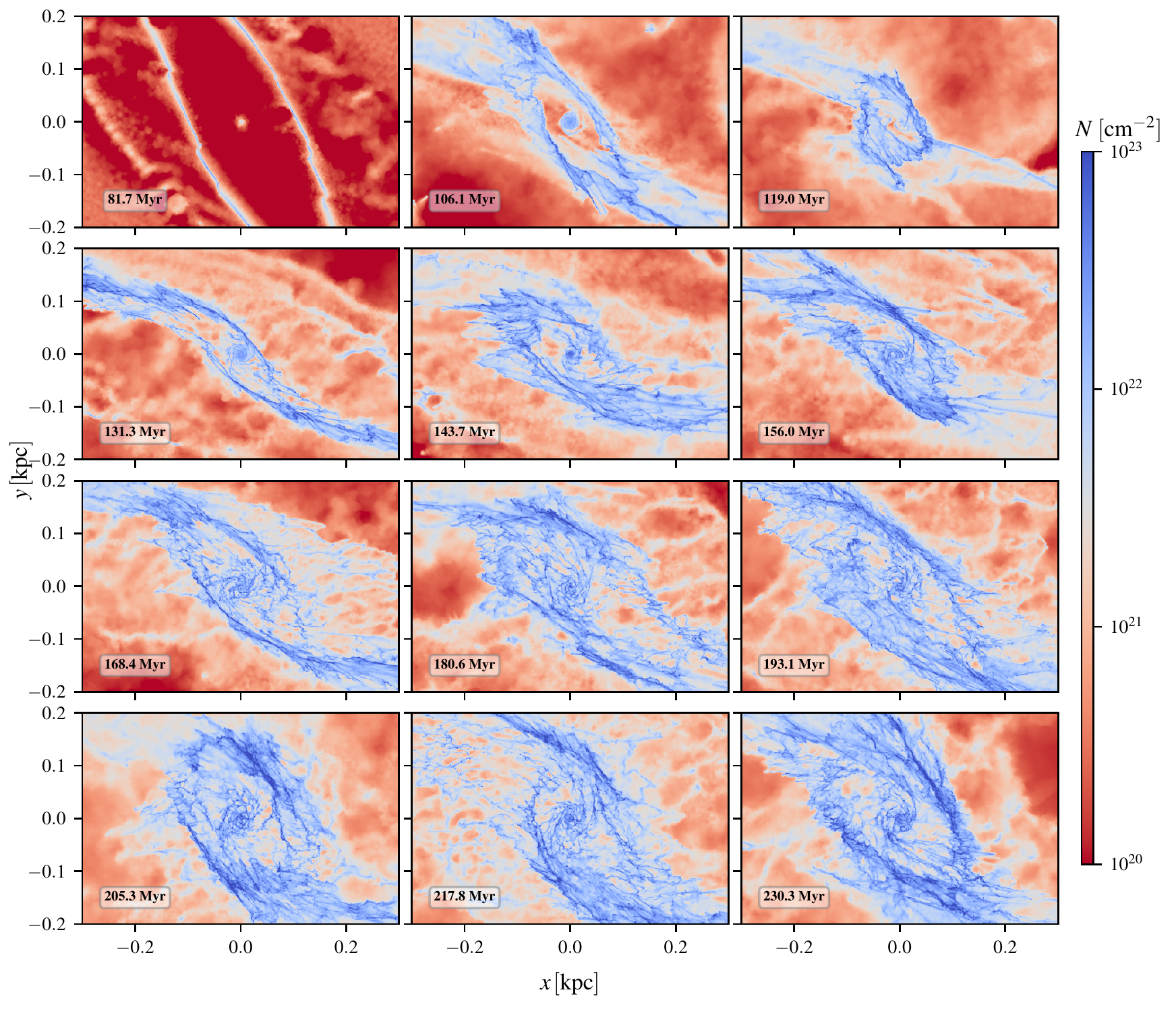}
    \caption{Zoom of Figure \ref{fig:NTotvstime} onto the CMZ}
    \label{fig:NTotvstime_CMZ}
\end{figure*}

\begin{figure*}
	\includegraphics[width=\textwidth]{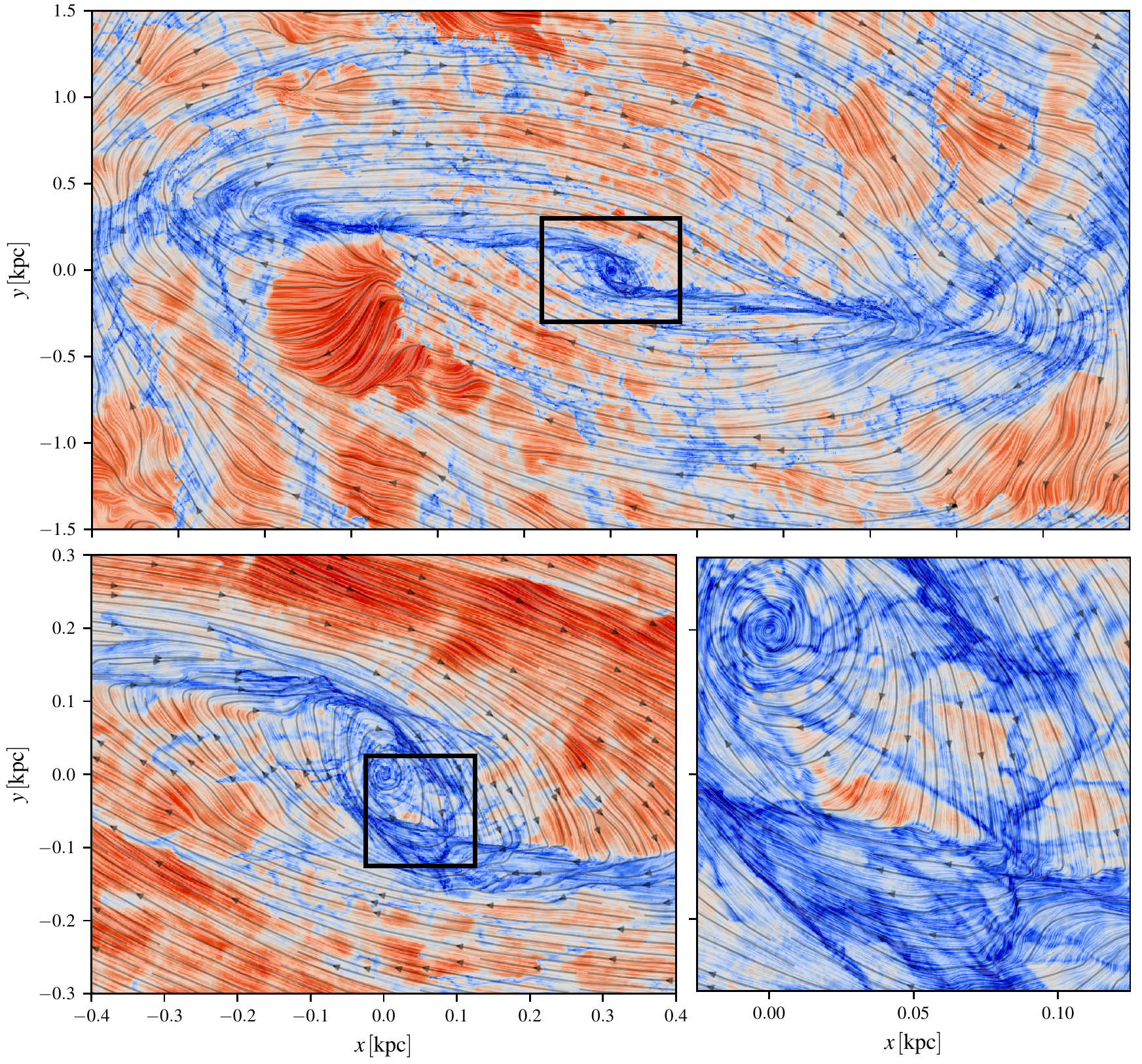}
    \caption{\emph{Top:} Instantaneous gas streamlines superimposed on the total gas density for the snapshot at $t=161 \Myr$. \emph{Bottom:} zoom onto the simulated CMZ. The velocities are shown in the frame corotating with the bar.}
    \label{fig:vfield}
\end{figure*}

\begin{figure}
	\includegraphics[width=\columnwidth]{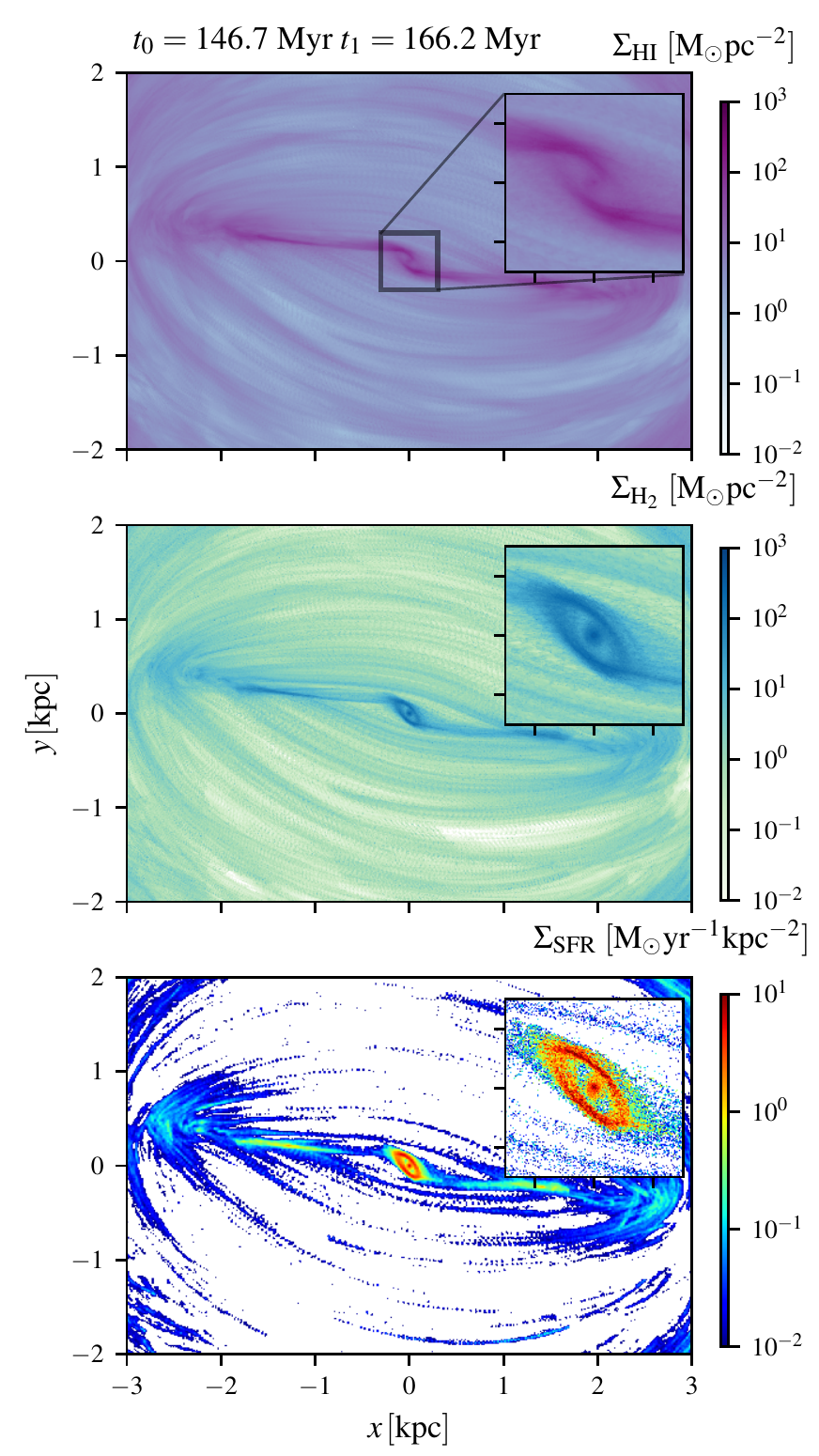}
    \caption{Time-averaged plot of \emph{top:}  H{\sc i} surface density; \emph{middle}: H$_2$ surface density; \emph{bottom}: star formation rate (SFR) surface density. The average is calculated over the time period $t = 146.7 \mhyphen 166.2 \Myr$. A comparison with Figures \ref{fig:NTotvstime} and \ref{fig:NTotvstime_CMZ} shows that while the time-averaged morphology is very regular, the instantaneous morphology displays a complex transient substructure. The SFR is calculated as a running average over the last 0.6 Myr and is analysed in more detail in Paper II.}
    \label{fig:average}
\end{figure}

The top and middle panels of Figure \ref{fig:average} show the time-averaged surface density of the gas. The time-averaged flow is very smooth and regular, and the CMZ shows up as a clearly defined ring. Comparing this figure with Figure \ref{fig:NTotvstime_CMZ} shows that while the time-averaged morphology is very regular and relatively simple, the instantaneous morphology of the CMZ is very rich in substructure that is transient in time. At any given time the CMZ can look very turbulent and irregular, to the point that the underlying regular structure is not obvious or even discernible. This illustrates that the most natural way to think of the current structure of the CMZ  is as the sum of a regular, time-averaged structure and a set of highly transient perturbations superimposed on it.

These considerations are another manifestation of what is shown in Figure~12 of \cite{Sormani+2018a}, where we display a typical molecular cloud orbiting in the CMZ at different times. We found that while the centre of mass of the cloud closely follows $x_2$ orbits on average, on top of this underlying ``guiding centre'' motion there are often significant turbulent fluctuations.\footnote{In \cite{Sormani+2018a} these fluctuations were entirely due to the bar inflow, while in the present work they are due to both supernova feedback and the bar inflow.} These fluctuations can make the instantaneous observed kinematics of the cloud look very different from what we may naively expect from a regular $x_2$ orbit. We continue the discussion on this point in Section~\ref{sec:x2vsopen}

Comparison of the top and middle panels of Figure \ref{fig:average} shows that the CMZ morphology also depends on which tracer one is considering. The CMZ morphology in H{\sc i} is more spiral-like, while in H$_2$ it is more ring-like. This is consistent with the fact that in external galaxies the morphologies of nuclear rings are complex, often showing both ring-like and spiral-like structures with a strong dependence on the tracer used \citep[e.g.][]{Izumi+2013}.

\begin{figure*}
	\includegraphics[width=\textwidth]{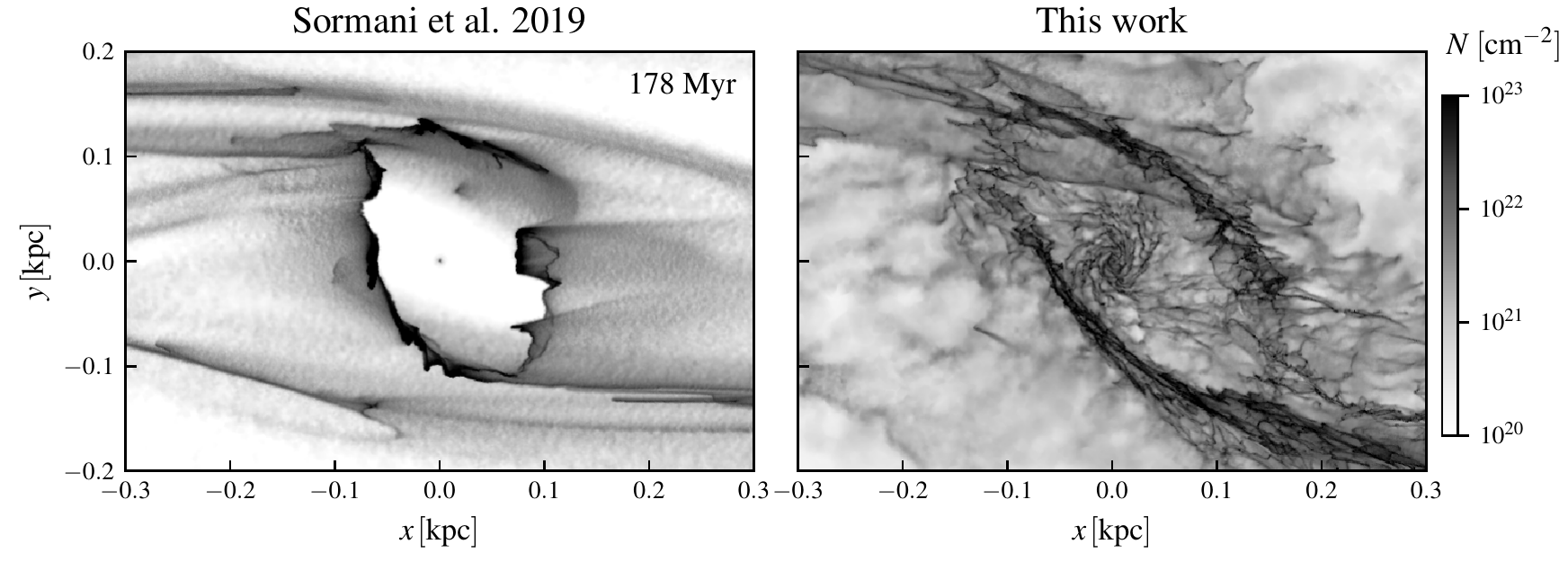}
    \caption{Comparison of the total gas surface density in the simulation from \citet{Sormani+2019} (which did not include the gas self gravity) and in the one presented in this work (which includes self-gravity, star formation and supernova feedback). The two simulations employ exactly the same external MW barred potential and are shown at the same time, $t=178\Myr$.}
    \label{fig:OldvsNew}
\end{figure*}

Figure \ref{fig:OldvsNew} compares the present simulations (that include gas self-gravity and stellar feedback) to our previous simulations (which did not). One of the most notable differences is that in the former gas can be found at radii inside the main CMZ ring, while in the latter there was essentially no gas inside the CMZ ring. This suggests that self-gravity and stellar feedback may play a major role in driving the gas from the CMZ inwards, a currently important open question since it is related to the fuelling of the super-massive black hole SgrA*. We investigate this in more detail in Section \ref{sec:feedbackinflow}.

\subsection{Resolving individual molecular clouds} \label{sec:clouds}

\begin{figure*}
	\includegraphics[width=1.0\textwidth]{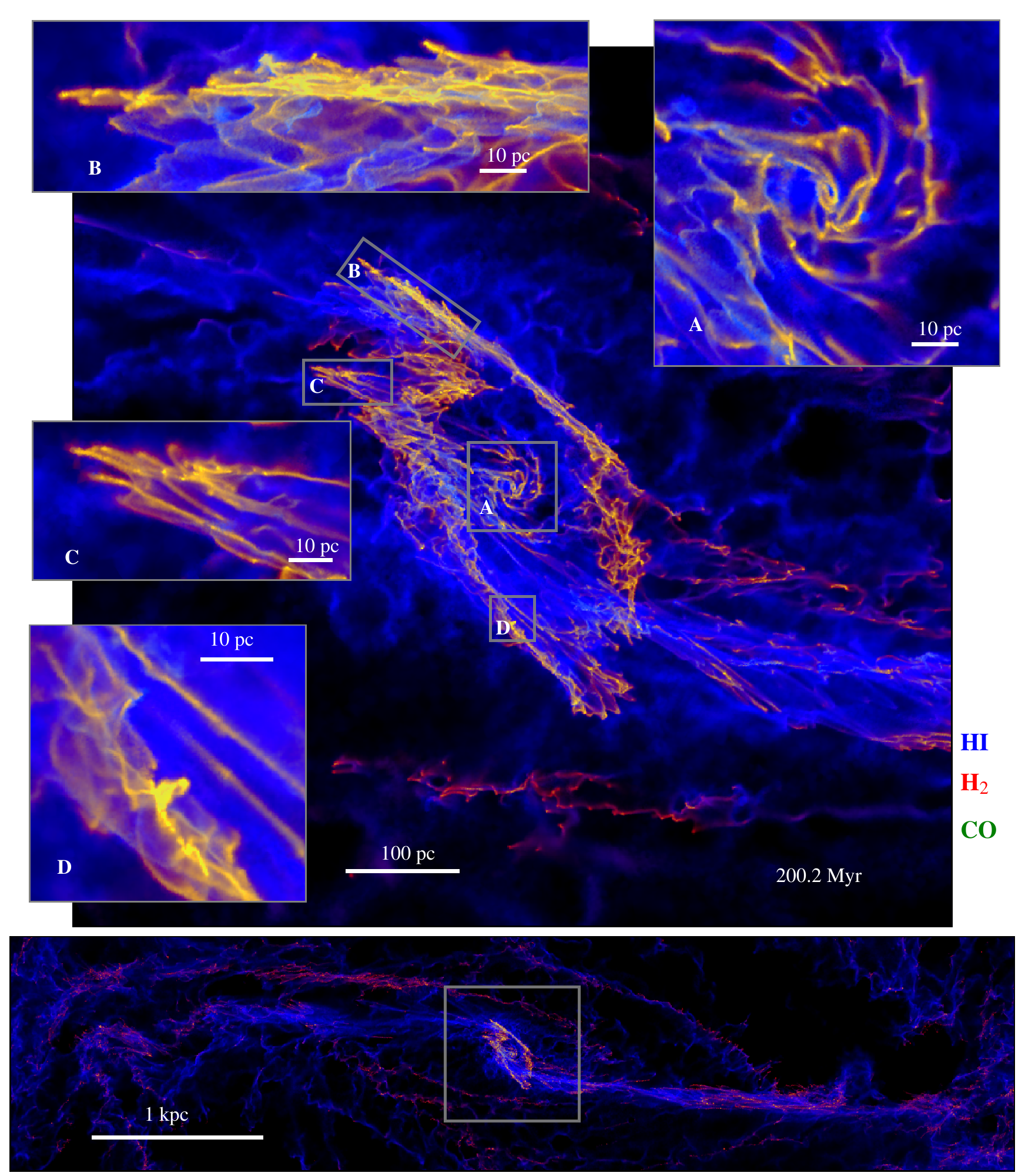}
    \caption{Three-colour composite showing the spatial relation of three chemical components: H{\sc i}, H$_2$ and CO. The zoom-in panels show various cloud complexes in the CMZ. The resolution of our simulation allows us to resolve individual molecular clouds, which are formed self-consistently from the large-scale flow.}
    \label{fig:RGB}
\end{figure*}

With this simulation we achieve resolutions that allow us to follow the formation of individual molecular clouds and start resolving their substructure. To illustrate this we show in Figure \ref{fig:RGB} a three-colour composite image of the CMZ with zoom-in panels onto individual cloud complexes. The dynamic nature of the complex environment in which the molecular clouds are embedded becomes clear from this figure. The clouds exhibit a complicated filamentary morphology, far from the idealised ``spherical cloud'' that is often used as a model. Clearly they cannot be idealised as individual and disconnected entities, but are part of a large interconnected network. Importantly, the clouds are part of and form self-consistently from the large-scale flow driven by the Galactic bar. Several processes unique to this environment shape their morphology, lifetime and evolution. Extreme shearing, stretching and compressive forces at the locations where clouds orbiting in the CMZ violently collide with the high velocity gas plunging in from the dust lanes produce dense and highly elongated structures around the apocentre (see Panels B and C of Figure \ref{fig:RGB}). These clouds can grow extremely compact and massive while moving downstream of the apocentre due to their self-gravity (for instance Panel D of Figure \ref{fig:RGB}), which will lead to intense star formation. Finally, supernova feedback leads to the formation of a filamentary gaseous disc in the innermost few tens of pc (Panel A of Figure \ref{fig:RGB}), which is fed by the main CMZ ring (see Section \ref{sec:feedbackinflow}). A more detailed analysis of the properties of the molecular cloud population will be presented in a future work.

\section{Thermal and chemical phases of the ISM} \label{sec:thermal}

\subsection{A three-phase medium} \label{sec:threephase}

The top panel in Figure \ref{fig:phasediagram1} shows that the ISM in the simulation is a three phase medium. The cold component ($T \leq 10^3 \K$) is predominantly molecular and is well traced by H$_2$ and CO. The warm component ($10^3 < T < 10^{4.5} \K$) is well traced by H. The hot component ($T \geq 10^{4.5} \K$) is produced by the SN feedback and is almost completely ionised. From the top panel in Figure \ref{fig:phasediagram1} one can read off the typical densities of the three phases. The transition from warm (atomic) to cold (molecular) gas occurs around $n\sim 10^2 \cmthree$, while the transition from warm to hot occurs around $n\sim 10^{-2} \cmthree$. From Figure \ref{fig:phasediagram2} one can see how the mass is distributed into the three phases. Globally, most of the mass is in the cold phase. The hot phase is the least massive phase, but as discussed below it has larger volume filling factors.

Figure \ref{fig:fillingfactors} shows the volume filling factors of the cold, warm and hot components in different regions. Most of the volume in the CMZ is occupied by the warm phase ($\sim 70\%$), followed by the hot phase ($\sim 25\%$), and by the cold phase ($\sim5\%$). These values are roughly consistent with the observational estimates of \cite{Oka+2019} stating that the volume filling factor of the dense molecular gas is $<10\%$ and of \cite{Ferriere+2007} who find that the filling factors of the cold molecular component is $\sim 3.3\%$. Figure \ref{fig:fillingfactors} also shows that there is significant scatter in the filling factors at different locations within the same region, so they can be expected to vary from sightline to sightline in observations. In the DLR, the warm ($\sim 65\%$) and hot ($\sim 35\%$) volume filling factors are similar to those of the CMZ, but there is a higher volume fraction of molecular gas ($\sim 10\%$) and it is possible to find $100\pc$-wide regions with values reaching $f_{\rm cold}\sim50\%$. In the Disc, the warm phase does not dominate the volume contrary to the CMZ ($\sim 35\%$), and most of the volume is occupied by the hot phase ($60\%$). The filling factor of the cold phase remains small ($\sim5\%$) although it is still possible to find $100\pc$-wide regions with high $f_{\rm cold}$, similarly to the DLR. In the Halo, the hot phase becomes dominant ($\sim 80\%$) while the cold phase is negligible ($\lesssim0.1\%$).

Figure \ref{fig:timeevolutionismphases} shows the time evolution of the various thermal and chemical components. From this plot, one can see that the amount of gas in each chemical/thermal phase is approximately constant in time in the second half of the simulation, indicating that a statistical steady state has been reached.

Figure \ref{fig:RGB} shows the spatial distribution and morphology of the three chemical phases relative to each other. Comparison between the different panels of Figure \ref{fig:slices} allows us to study the spatial and morphological connection between the thermal and chemical phases. Comparing the left column with the middle column one can clearly see the spatial correlation between the cold phase and H$_2$, while comparing the left and right columns one can see the correlation between the warm phase and H{\sc i}. The hot component is contained in large ionised cavities which are produced by the SN feedback.

Finally, we remark that the vertical dashed line in Figures \ref{fig:phasediagram1}-\ref{fig:phasediagram2} denotes the sink density formation threshold $\rho_{\rm c}$  (See Section~\ref{sec:sinks}). Above this density, gas in the simulation stars being converted into sink particles. This explains the decline in the PDF visible in the figures above this density.

\subsection{CMZ vs disc}

\begin{figure}
	\centering
	\includegraphics[width=\columnwidth]{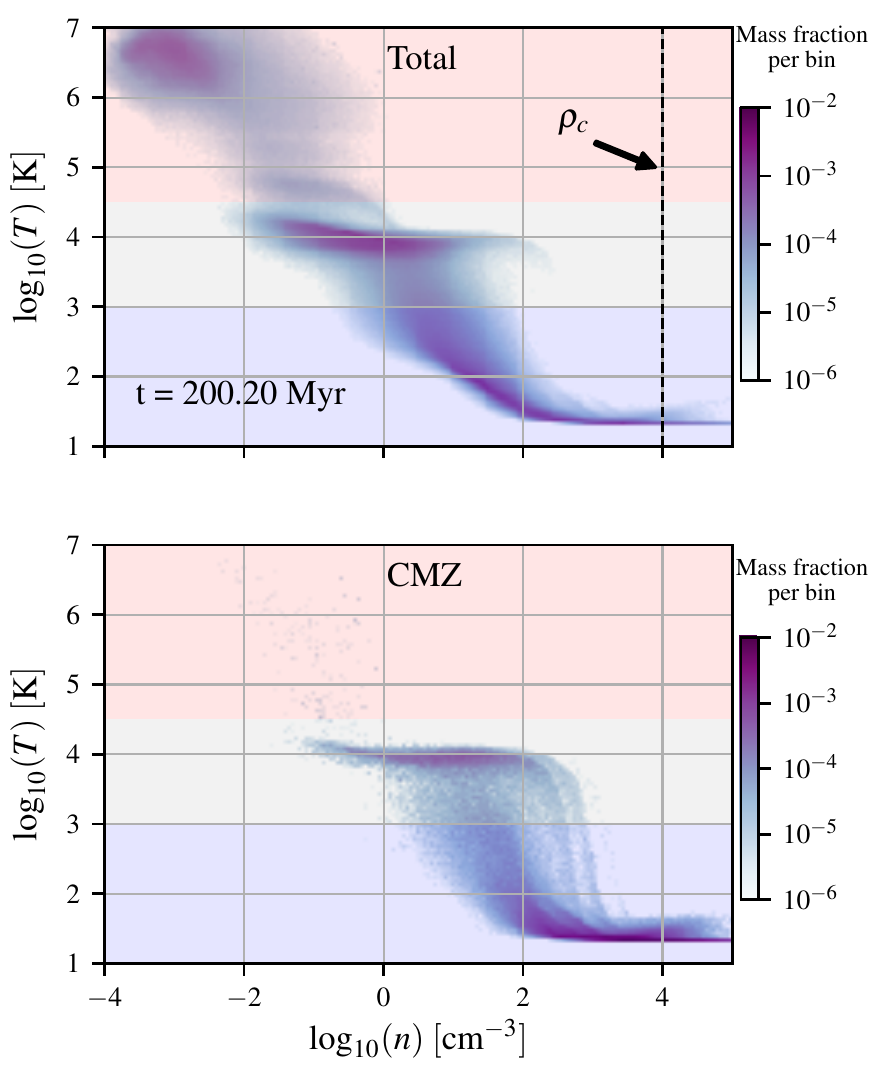}
    \caption{2D histograms showing the mass distribution in the $n\mhyphen T$ plane, where $T$ is the gas temperature and $n$ the gas number density. \emph{Top}: for all gas in the simulation. \emph{Bottom}: for the CMZ, defined as the region $R \leq 250\pc$ (see Figure \ref{fig:regions}). The vertical dashed line denotes the sink density formation threshold $\rho_{\rm c}$ (See Section \ref{sec:sinks} and Table \ref{tab:sinks}). Shaded coloured areas indicate the three thermal phases (cold, warm, hot).}
    \label{fig:phasediagram1}
\end{figure}

\begin{figure}
	\centering
	\includegraphics[width=\columnwidth]{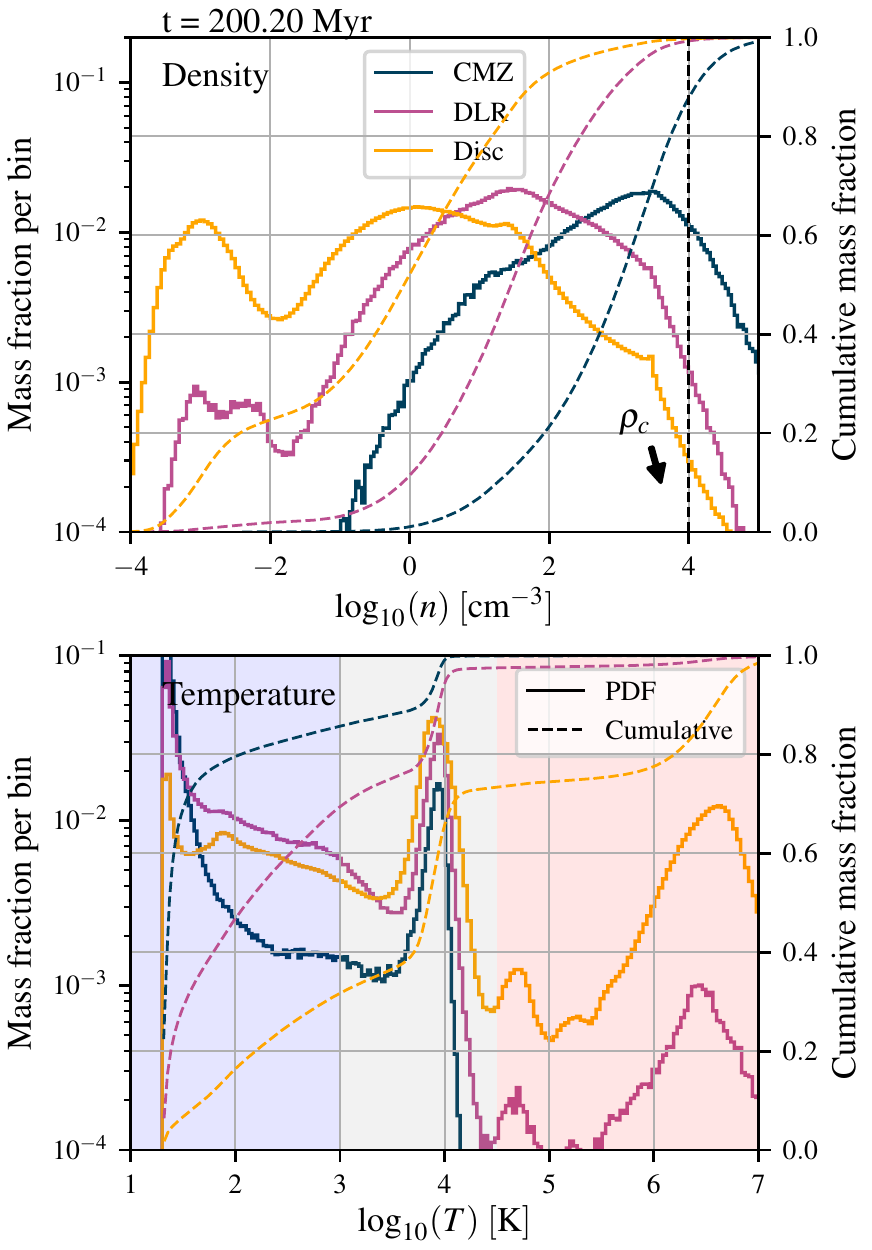}
    \caption{1D histograms showing the mass distribution as a function of total number density $n$ (\emph{top}) and temperature $T$ (\emph{bottom}). Full lines are probability distribution functions (PDFs), while dashed lines are cumulative distributions. The three different colours of the lines denote the three different regions defined in Figure \ref{fig:regions}. The vertical dashed line denotes the sink density formation threshold $\rho_{\rm c}$ (See Section \ref{sec:sinks} and Table \ref{tab:sinks}). Shaded coloured areas in the bottom panel indicate the three thermal phases (cold, warm, hot).}
    \label{fig:phasediagram2}
\end{figure}

\begin{figure*}
	\includegraphics[width=0.8\textwidth]{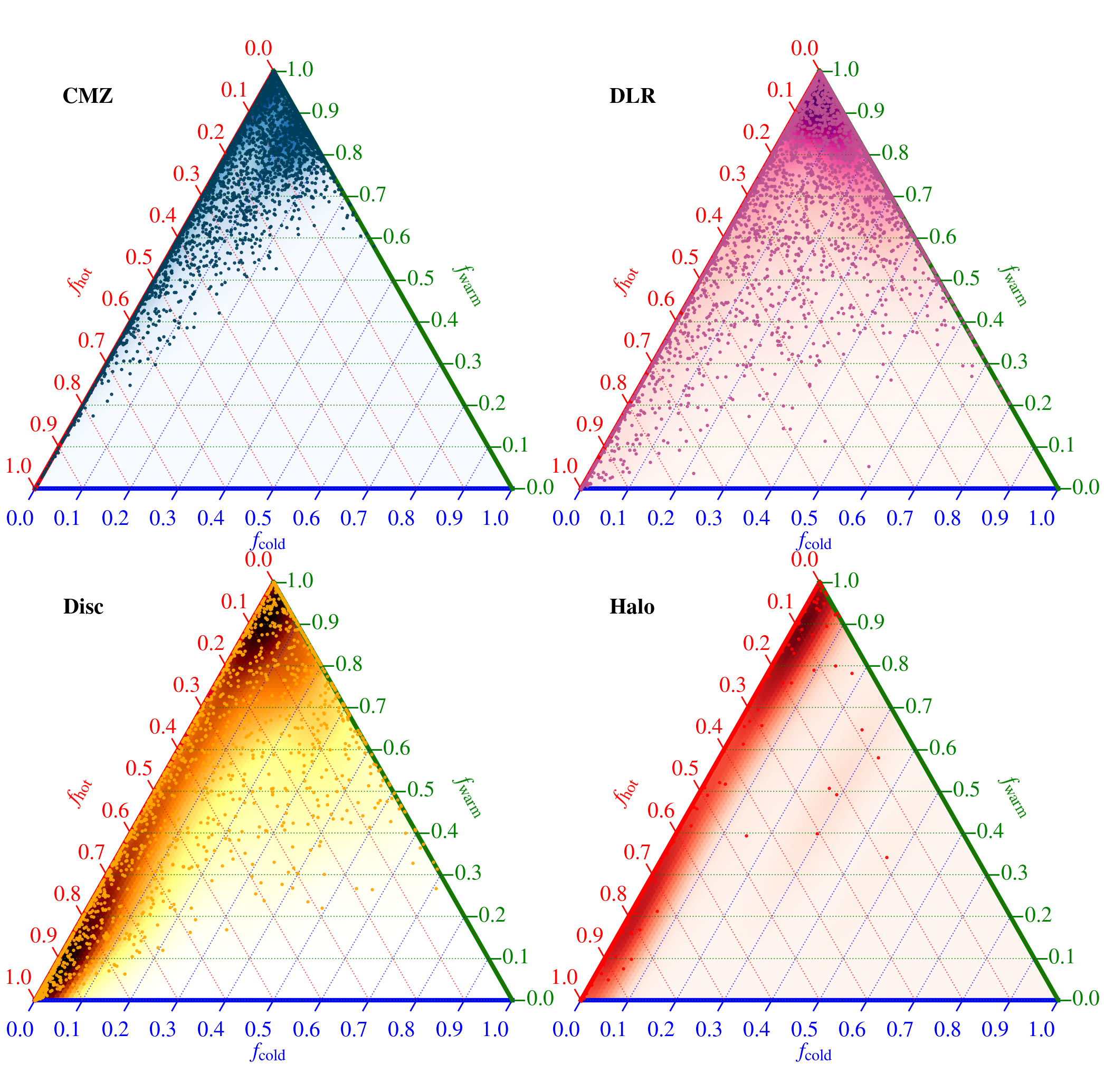}
    \caption{Triangle plot of the volume filling factors.  ``CMZ'', ``DLR'' and ``Disc'' refer to the three regions defined in Figure \ref{fig:regions}, while ``Halo'' is defined as the region $|z|>100\pc$. To construct the plot, we take 30 snapshots equally spaced in time between $t=175\Myr$ and $t=250\Myr$, and for each snapshot we draw 100 random points within each region (CMZ, DLR, Disc, Halo). We thus have a total of $30\times100$ points per region. We then measure the volume filling factor in a cube $100\pc$ on a side centred on these randomly selected points. Each dot in the plot represents one of these $100$-pc cubes.  The contour plot in background in each panel is obtained by a kernel density estimation of the dots. The three thermal phases (cold $T<10^3\K$;  warm $10^3<T<10^{4.5}\K$; hot $T>10^{4.5}\K$) are defined as in Section \ref{sec:threephase} and Figures \ref{fig:phasediagram1}-\ref{fig:timeevolutionismphases}, and satisfy the relation $f_{\rm cold} + f_{\rm warm} + f_{\rm hot} =1$. }
    \label{fig:fillingfactors}
\end{figure*}

\begin{figure}
	\centering
	\includegraphics[width=\columnwidth]{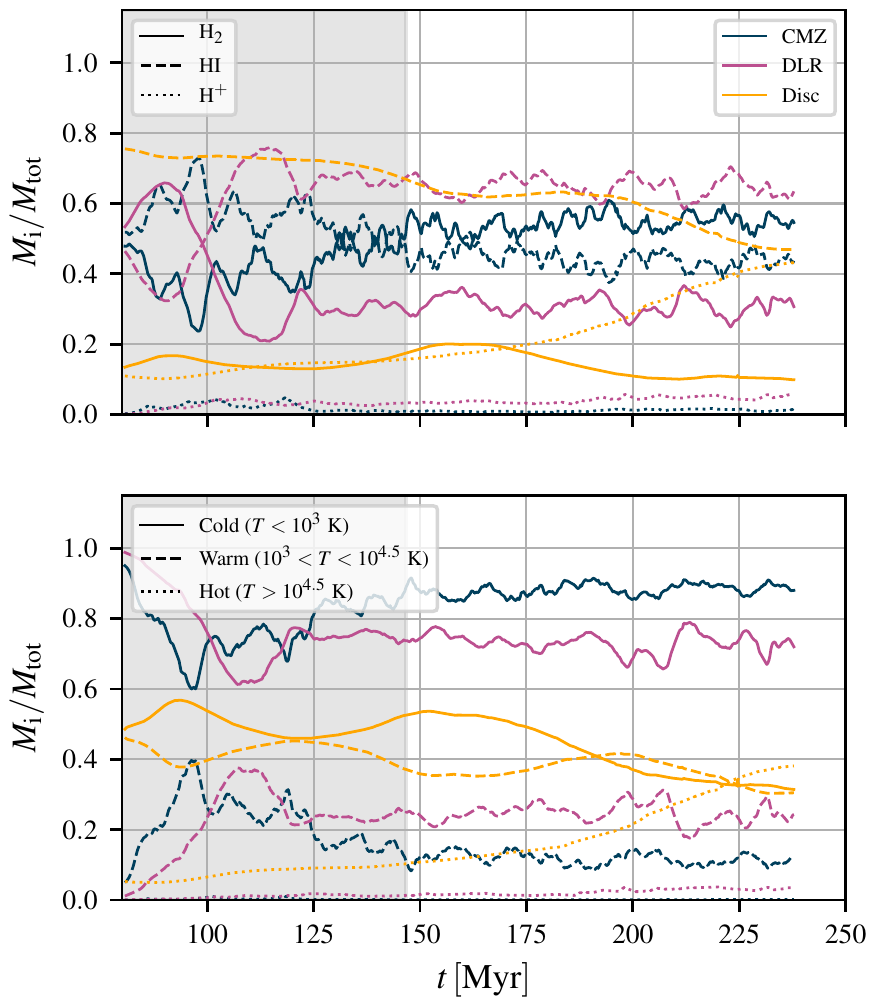}
    \caption{\emph{Top}: mass in each chemical (H$_2$, HI, H$^+$) component, normalised to the sum of the three components (the sum of three lines of the same colour is unity), as a function of time. \emph{Bottom:} mass fraction in each thermal (cold, warm, hot) component as a function of time. In both panels, different colours indicate the three regions defined in Figure \ref{fig:regions}.  }
    \label{fig:timeevolutionismphases}
\end{figure}

\begin{figure*}
	\includegraphics[width=\textwidth]{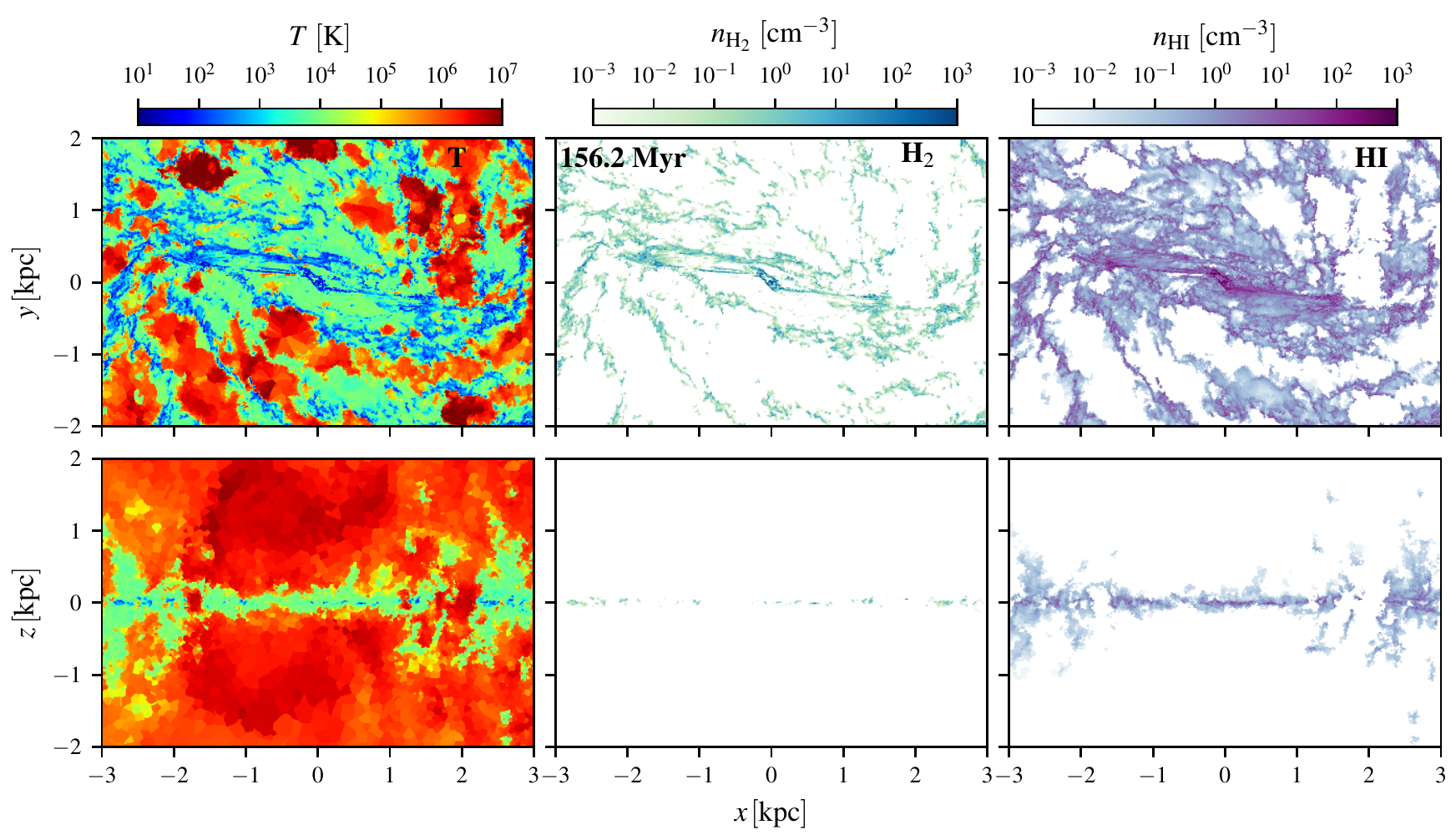}
	\includegraphics[width=\textwidth]{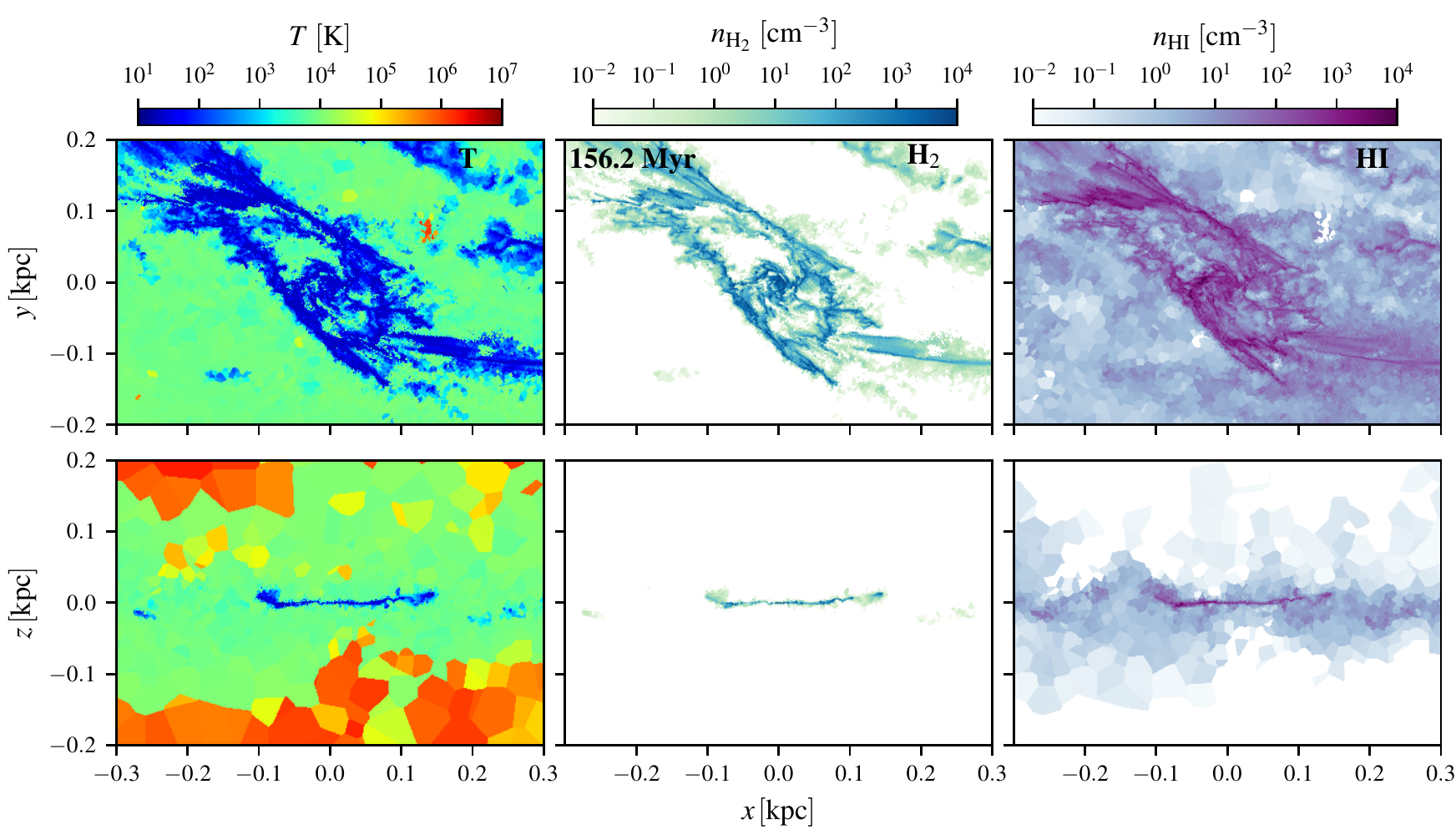}
    \caption{Temperature and density slices in the $z=0$ and $y=0$ planes. \emph{Top two rows}: large scales. \emph{Bottom two rows}: zoom on the CMZ. From this figure one can see that H$_2$ and CO trace the coldest and densest gas. Note the different scaling of the colourbars in the top two and bottom two rows.}
    \label{fig:slices}
\end{figure*}

Figures \ref{fig:phasediagram1}-\ref{fig:slices} allow us to compare the gas properties in the CMZ with those in the Galactic disc. Comparing the two panels in Figure \ref{fig:phasediagram1} we can see that the gas in the CMZ is much more predominantly cold/molecular than in the disc. The top panel in Figure \ref{fig:phasediagram2} shows that the gas in the CMZ has considerably higher average densities than the gas in the disc, qualitatively consistent with what is observed in the real CMZ \citep[e.g.][]{Mills+2018}. Figure \ref{fig:phasediagram2} also shows that our simulations underestimate the temperature of the cold gas compared to observations \cite[e.g.][]{Ginsburg+2016,Krieger+2017}, which is expected given the low ISRF/CRIR and the absence of radiation feedback from massive stars in our numerical scheme (see Sections \ref{sec:chemistry} and \ref{sec:feedback}). Figure \ref{fig:timeevolutionismphases} indicates that the fraction of cold gas in the CMZ makes up $\sim 90\%$ of its mass, compared to $\sim 50\%$ in the disc, and that the mass fraction of hydrogen which is in molecular form is $\sim 50\%$ in the CMZ vs $\sim 20\%$ in the disc. These trends are also qualitatively similar to those observed in the real Galaxy \citep[e.g.][]{MorrisSerabyn1996,Krieger+2017}.

\section{Inflows and Outflows} \label{sec:inflow}

The purpose of this section is to investigate inflows and outflows in our simulation, and to compare with our previous simulations in \cite{Sormani+2019}, which were identical except for the lack of gas self-gravity and SN feedback (see Section \ref{sec:methods} for more details).

\begin{figure*}
	\centering
	\includegraphics[width=\textwidth]{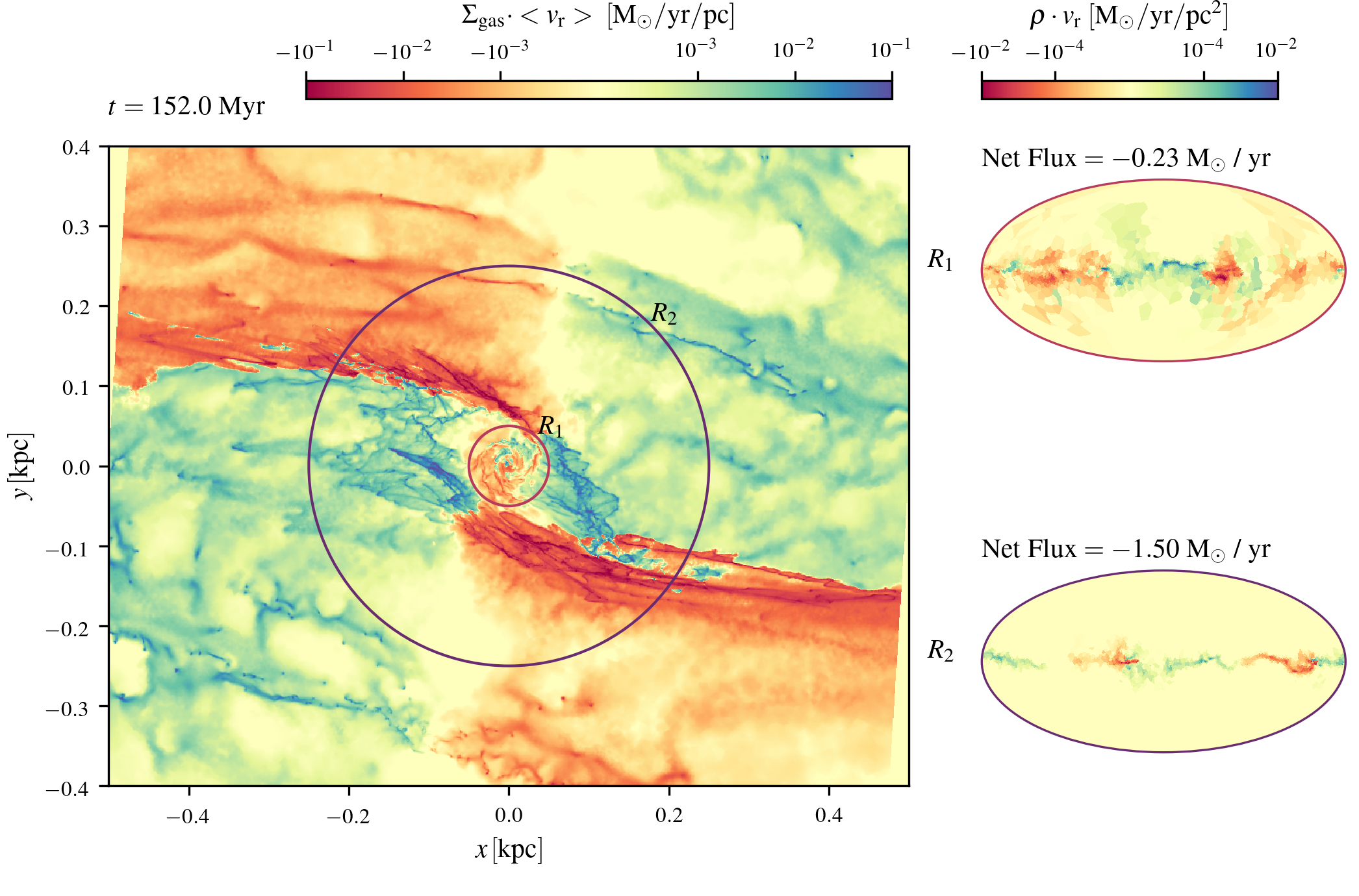}
    \caption{\emph{Left:} instantaneous radial mass flows. Red indicates spherically inward motion (inflow) while blue indicates spherically outwards motion (outflow).  Most of the inward-moving material (red) is concentrated in the two ``dust-lane features'' which mediate the bar-driven accretion. Embedded in the dust-lane features, some blue clouds can be spotted. These represent ``overshooting'' material which is moving outwards against the inward flow of the dust lane. These overshooting clouds are decelerating and eventually join the flow of the dust lanes (see text in Section \ref{sec:inflow} for more details). \emph{Right}: instantaneous mass fluxes through two spheres of radius $R_1= 50 \pc$ and $R_2 = 250 \pc$ in a Mollweide projection centred on the Galactic centre. Net flux indicates the instantaneous mass flow rate integrated over the surface of the sphere.}
    \label{fig:inflow_02}
\end{figure*}

Figure \ref{fig:inflow_02} shows the instantaneous radial mass flows. These are calculated as follows. Consider a spherical coordinate system $(R,\theta,\phi)$ centred on the Galactic centre. Then the radial mass flux per unit area and per unit time is:
\begin{equation}
F(R,\theta,\phi) = \hat{\mathbf{e}}_R \cdot \left( \rho \bfv \right) \, ,
\end{equation}
where $\hat{\mathbf{e}}_R$ is the unit vector in the $R$ direction, $\rho$ is the gas volume density and $\bfv$ is the gas velocity. $F>0$ indicates outflow, while $F<0$ indicates inflow.

The left panel of Figure \ref{fig:inflow_02} shows a face-on mass-weighted projection of the radial flows. At each instant, both radially inward (red) and outward (blue) flows are present. The radially inward motions (red) are concentrated along the two ``dust-lane features'' which mediate the bar-driven accretion, while most of the radially outward motions (blue) occur after the gas has passed the point of closest approach to the centre along its orbit and is on its way to the apocentre. 

It is interesting to note that embedded in the midst of the (red) dust lanes, some outflowing (blue) clouds can be spotted. These are clouds that are crashing against the dust lane after having moved in along the dust lanes on the opposite side and overshot passing by the CMZ. Because they are going ``against the current'', the hydrodynamic drag will cause them to decelerate and ultimately join the flow of the surrounding dust lane in which they are embedded.

The right panel in Figure \ref{fig:inflow_02} shows the instantaneous mass fluxes across the surfaces of two spheres of radii $R_1= 50 \pc$ and $R_2=250 \pc$. Because of the highly non-circular motions, there are both inflows and outflows, and some of the gas flows in and then back out. So it is not straightforward to tell whether the net flux is inwards or outwards. The instantaneous net fluxes for this particular snapshot, calculated by integrating over the surface of the spheres, are reported in Figure \ref{fig:inflow_02}. However, the net fluxes are highly variable in time, as can be seen from Figure \ref{fig:inflow_03}, which shows the net fluxes across the two spheres $R_1$ and $R_2$ as a function of time.

While it is instructive to look at the instantaneous fluxes, it is more meaningful to consider their time-averaged values. Figure \ref{fig:inflow_04} provides the time-averaged net fluxes across the two spheres defined in Figure \ref{fig:inflow_02}. In contrast to the instantaneous fluxes, which can be either positive or negative depending on the particular instant that we choose to look at them, the time-averaged net contributions are always negative. The net inflow through $R_1$ is much smaller than the net inflow through $R_2$. We now analyse in more detail the implications of this finding.

\subsection{Bar inflow from the disc to the CMZ} \label{sec:barinflow}

It is well known that the Galactic bar is very efficient at driving the gas from large radii ($R \simeq 4$ kpc) inwards along the dust lanes down to the CMZ ring at $R \simeq 200 \pc$. \cite{SormaniBarnes2019} use a simple geometrical model applied to CO data to provide an observational estimate of this bar-driven inflow. They report a mass inflow rate of $2.7^{+1.5}_{-1.7}\, \Msunyr$. As they discuss in their Section 4.1.4, this value should be multiplied by an overshooting factor $f \leq 1$ to take into account the fact that only part of the gas falling along the dust lanes accretes onto the CMZ, while partly misses it and `overshoots', eventually hitting the dust lane on the opposite side. The value quoted in \cite{SormaniBarnes2019} assumed $f=1$ since at the time this paper was written the value of $f$ was unknown. Hatchfield et al. (in preparation) has measured the value of $f$ by adding tracer particles to the simulation of \cite{Sormani+2019}, and found $f \simeq 50\%$. This implies that the observationally determined value should be corrected to $1.35^{+0.75}_{-0.85}\, \Msunyr$.

The bar-driven inflow in our simulation is quantified by the net flux across $R_2$ (see Figure \ref{fig:inflow_02}). The time averaged net flux is $\dot{M} = 0.987 \Msunyr$ (Figure \ref{fig:inflow_04}). This is consistent with the observationally determined inflow rate reported above. In addition to the average value, the time variability of the inflow in our simulation is also roughly consistent with the one inferred from observations (compare Figure \ref{fig:inflow_03} with Figure 3 in \citealt{SormaniBarnes2019}). Our inflow value is also consistent with the numerical result of \cite{Armillotta+2019}, which is $\dot{M}=1\mhyphen 3 \Msunyr$.

The blue lines in Figure \ref{fig:inflow_01} show the gaseous mass of the CMZ as a function of time, which grows according to the inflow rate determined above. The figure also shows that most of the mass in the CMZ is locked-up in the sinks, which is expected given that the average density of the CMZ is significantly higher than that of the disc (see Figure \ref{fig:phasediagram2}), so that a larger fraction of the gas lies at densities above the sink formation threshold $\rho_{\rm c}$ (see Table \ref{tab:sinks}). The total gas mass of the CMZ in the simulation at $t>146 \Myr$ (after the bar potential is fully on, see Section \ref{sec:IC}) is comparable to the observationally determined mass of $\sim 5 \times 10^7 \Msun$. 

Where does the gas go? Mass conservation requires that (see also Section 3.3 of \citealt{Crocker2012}):
\begin{equation}
\dot{M}_{\rm inflow}=\dot{M}_{\rm outflow} + {\rm SFR} + \dot{M}_{\rm CMZ}\,.
\end{equation}
In our simulation, the outflow rate is small (Section \ref{sec:outflow}), and the total mass of the CMZ will continue to grow until the SFR approximately matches the bar-driven inflow rate or becomes high enough to power a more significant outflow rate (see discussion in the last paragraph in Section 3.1 of Paper II). However, in the real Galaxy it might be possible that the Galactic outflow efficiently removes the gas at a rate which is up to $\sim 90\%$ of the inflow rate through high-energy processes that are not included in our simulation (see the discussion in Section 4.2 of \citealt{SormaniBarnes2019}).

The bar-driven inflow rate determined here is very similar ($\simeq 1 \Msunyr$) to the one that is obtained using our previous non-self gravitating simulations in \cite{Sormani+2019}. Recall that these are identical to those presented here except that they do not include gas self-gravity and stellar feedback (see Section \ref{sec:methods}). This result suggests that the inflow from the disc to the CMZ is driven by the non-axisymmetric potential of the Galactic bar, and not by the gas self-gravity and/or the SN feedback.

\subsection{Feedback driven inflow from the CMZ inwards} \label{sec:feedbackinflow}

As discussed in the previous section, the bar is very efficient in driving the gas from large radii ($R \simeq 4$ kpc) inwards along the dust lanes down to the CMZ ring at $R \simeq 200 \pc$. However, a long-standing and important open question is how the gas continues its journey from the CMZ inwards and how (and to what extent) these large-scale flows reach the supermassive black-hole (SgrA*) at the centre of our Galaxy \citep[e.g.][]{Phinney1994}. Indeed, our previous simulations show that the bar is ineffective in driving the gas from the CMZ inwards: in the absence of any form of stellar feedback/magnetic fields, the gas just piles up in the CMZ ring and stalls there, without going further in \citep{Sormani+2018a,Sormani+2019}. Understanding how the gas flows from the CMZ inwards is key to understand the formation of the circum-nuclear disc (CND, e.g. \citealt{Mills+2013,Trani+2018}) and to learn more about the fuelling of SgrA* \citep[e.g.][]{Ghez+2008,Genzel+2010,Gillessen+2017}.

One possibility to drive mass transport from the CMZ inwards is that stellar feedback associated with the intense CMZ SF activity could stochastically launch parcels of gas towards the centre \citep[e.g.][]{Davies+2007}. Given the right initial condition, a cloud might plunge almost undisturbed towards the centre, since the mean free path of molecular clouds is comparable to the radius of the CMZ \citep{SormaniLi2020}. The gas self-gravity can also play a role in driving an inflow by creating gravitational torques that are additional to those of the Galactic bar. The simulations presented here are ideal to test these hypotheses, because we can compare them with the previous non-self gravitating simulations in \cite{Sormani+2019}. Since the only difference between the current simulations and those in \cite{Sormani+2019} (in which, as we have checked, there was essentially no net inflow from the CMZ inwards) is the presence of self-gravity and SN feedback (see Section \ref{sec:methods}), all the inflow present in the newer simulations can be attributed to these two ingredients.

As mentioned in Section \ref{sec:CMZmorphology}, the difference between the old and new simulations is already evident from Figure \ref{fig:OldvsNew}. In the old simulation there is essentially no gas inside the CMZ ring, while in the new simulation a rich morphology is found inside the CMZ ring.

\begin{figure}
	\centering
	\includegraphics[width=\columnwidth]{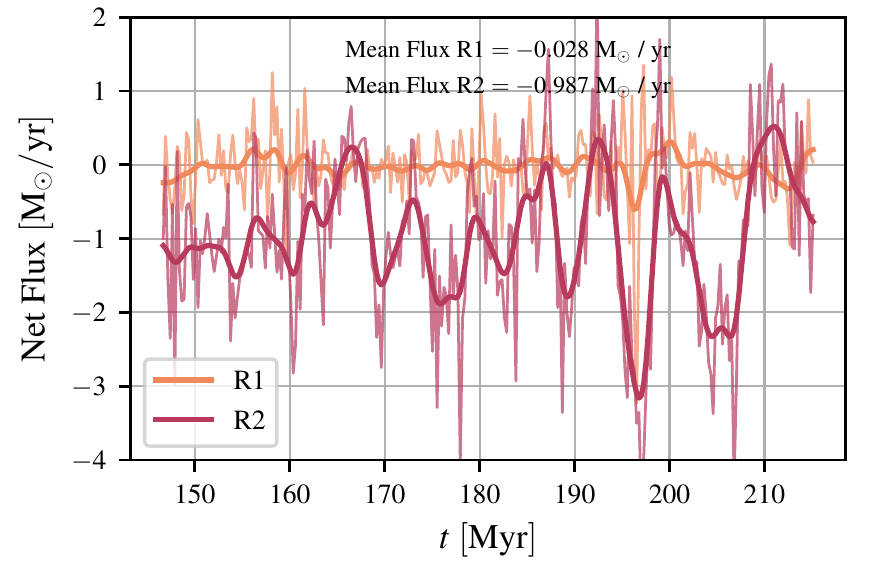}
    \caption{Mass flow rates integrated over the surface of the two spheres with radius $R_1=50\pc$ and $R_2=250\pc$ (see Figure \ref{fig:inflow_02}) as a function of time. Thin lines are the instantaneous flows, while thick lines are the same lines smoothed with a gaussian filter of width $0.75\Myr$. Mean flux indicates the mean over the period shown in the figure.}
    \label{fig:inflow_03}
\end{figure}

The inflow from the CMZ inwards can be quantified by the averaged net flux across $R_1$ in Figure \ref{fig:inflow_02}. This is highly variable in time (Figure \ref{fig:inflow_03}), but there is a time averaged net inward flux of $\dot{M} = 0.03 \Msunyr$ (see Figure \ref{fig:inflow_04}). While this is a factor of 30 smaller than the bar-driven inflow onto the CMZ  (Section \ref{sec:barinflow}), it is quite significant. Observations show that at radii of $1.5 \lesssim R \lesssim 10\pc$, there is a concentration of molecular gas known as the circum-nuclear disc (CND) \citep[e.g.][]{Mills+2013,Tsuboi+2018}. The CND has an estimated mass of $M=10^4\,\mhyphen\,10^5 \Msun$ \citep{RequenaTorres+2012}, and is the closest large reservoir of molecular gas to SgrA*, whose gravitational potential becomes dominant at $R\lesssim1\pc$ \citep{Ghez+2008,Genzel+2010,Gillessen+2017}, and it is therefore critical in understanding its fuelling. Our determined inflow rate of $\dot{M} = 0.03 \Msunyr$ is significant because at this rate the CND would take only $0.3 \mhyphen 3 \Myr$ to build up, a relatively short time (see Section 3.2 of Paper II for a discussion of star formation associated with this inflow).

SN feedback is the dominant process in driving the nuclear inflow. Torques from the gas-self gravity appear to be negligible since we have checked that they are consistently one order of magnitude smaller than those from the Galactic bar at all radii. Note that although the Galactic bar alone is ineffective in driving the inflow in the absence of SN feedback, it still constitutes the ultimate ``sink'' of angular momentum even when the feedback is present. SN feedback cannot change the total amount of angular momentum, but can only redistribute it in such a way that it becomes available for removal by the bar. For example, a SN explosion in a CMZ cloud might send parts of it towards the centre, contributing to the inflow, and parts of it at larger radii, where the bar can then remove its angular momentum and eventually send it back to the CMZ.

We conclude that SN feedback is able to contribute $\dot{M} \simeq 0.03 \Msunyr$ to the inflow from the CMZ towards the CND. The question of how the gas migrates from the CND to SgrA* cannot be addressed with the current simulations because the resolution is not sufficient to resolve the gas flows at $R\leq 1\pc$.

\subsection{Outflow} \label{sec:outflow}

\begin{figure}
	\centering
	\includegraphics[width=\columnwidth]{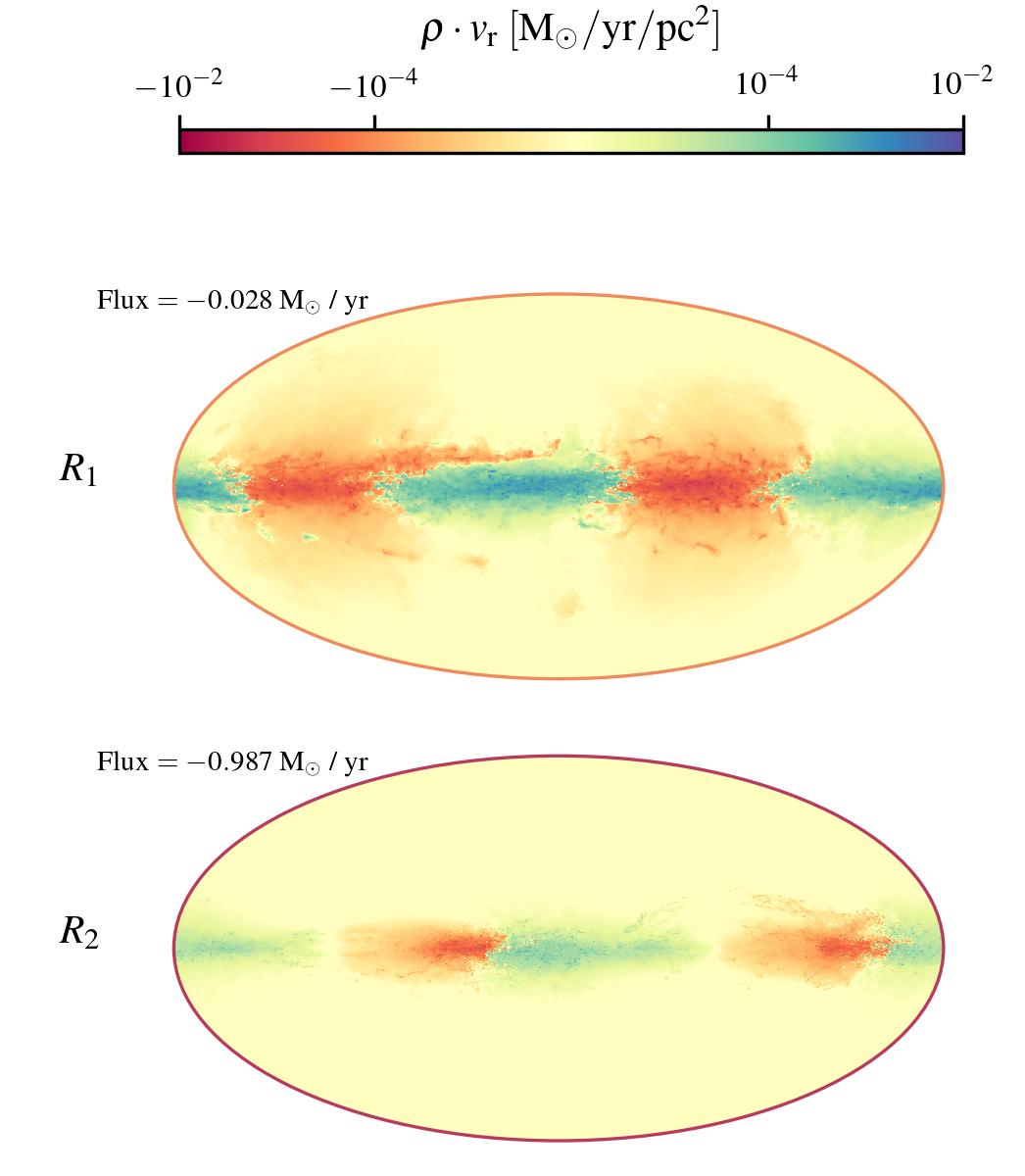}
    \caption{Time-averaged mass flow rates across the two spheres of radius $R_1=50\pc$ and $R_2=250\pc$ (see Figure \ref{fig:inflow_02}). The time average is extended over the period $147 \leq t \leq 215 \Myr$. Flux indicates the total time-averaged flux integrated over the surface of the sphere (the same numbers are obtained by averaging the lines in Figure \ref{fig:inflow_03}).}
    \label{fig:inflow_04}
\end{figure}

\begin{figure}
	\centering
	\includegraphics[width=\columnwidth]{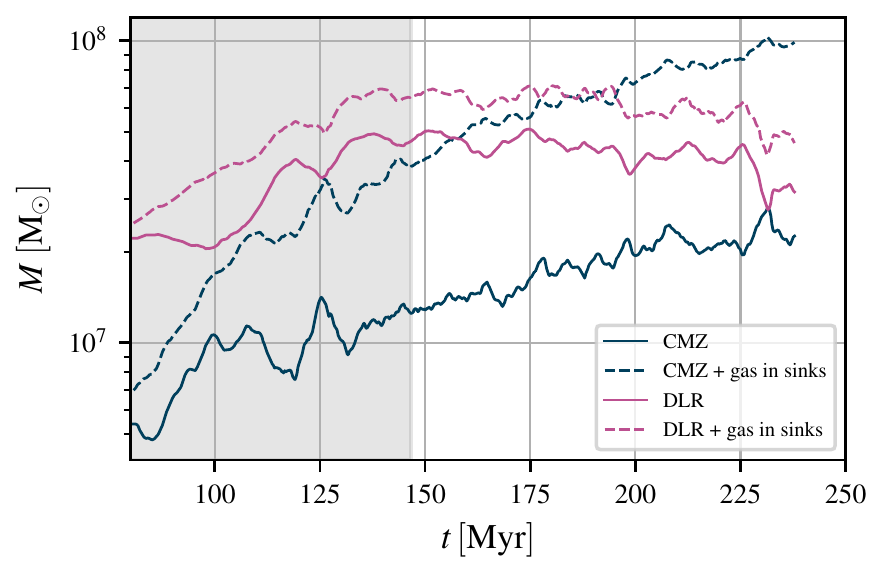}
    \caption{Total gas mass in the CMZ and in the DLR (defined in Figure \ref{fig:regions}) as a function of time.}
    \label{fig:inflow_01}
\end{figure}

\begin{figure}
	\centering
	\includegraphics[width=\columnwidth]{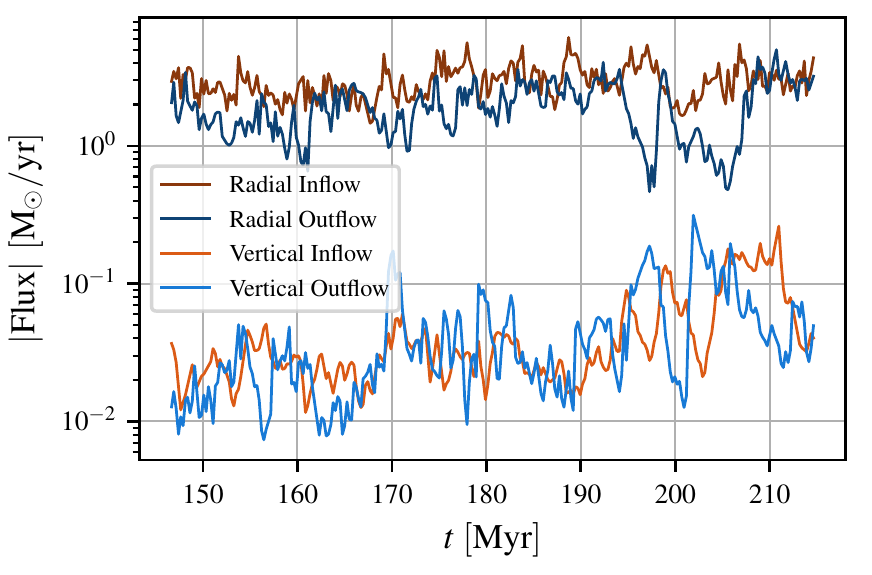}
    \caption{Inflowing and outflowing gas integrated across the lateral surface and the top/bottom surface of a cylinder of radius $R=250\pc$ and $z=\pm 100 \pc$. The net flux is obtained by subtracting the two.}
    \label{fig:outflow}
\end{figure}

An interesting question is to what extent the supernova feedback is able to drive an outflow in the vertical direction. However, this cannot be studied from Figure \ref{fig:inflow_04} since the mass flux in this figure is dominated by motions in the plane $z=0$. Hence, in order to investigate the presence of vertical outflows, we move from the spherical geometry of Figure \ref{fig:inflow_04} to a cylindrical geometry. Figure \ref{fig:outflow} shows the total inward and outward fluxes integrated across the lateral and the top/bottom surfaces of a cylinder with radius $R=250\pc$ and $z=\pm 100\pc$. The corresponding fluxes are obtained by integrating over all points with velocity vectors pointing inside and outside the cylinder respectively. From this figure it can be seen that the stellar feedback can drive a moderate outflow of about $\sim 0.1 \Msunyr$, most of which falls back at a later time in a fountain flow, since its velocity is smaller than the escape velocity. This finding is similar to the predictions of the models of \cite{Crocker2012} and  \cite{Crocker+2015} in the context of of modelling the high-energy phenomenology of the inner Galaxy. The outflow is on average stronger at later ($t>190\Myr$) than at earlier ($147 < t <190\Myr$) times, indicative of a correlation with the SFR, which also increases at later times (see Figure 2 in Paper II). All these findings are similar to those obtained in the simulation of \cite{Armillotta+2019}; compare our Figure \ref{fig:outflow} with their Figure 10.

\section{Discussion} \label{sec:discussion}

\subsection{$x_2$ orbits vs open orbit: a false dichotomy?} \label{sec:x2vsopen}

We have seen in Sections \ref{sec:largescale} and \ref{sec:CMZmorphology} that clouds in CMZ roughly follow $x_2$ orbits, as first proposed by \cite{Binney+1991}. However, \cite{Kruijssen+2015} (hereafter KDL) report an inconsistency (see KDL Section 5.2.5) between their best-fitting open orbit and the $x_2$ orbits. What is the relation between the present work and the KDL orbit? Are the two pictures really inconsistent? The purpose of this section is to show that the KDL open orbit scenario is not necessarily in contradiction with the \cite{Binney+1991} picture. Instead, the KDL open orbit should be considered as one of the several possible ways to refine the \cite{Binney+1991} model by taking into account epicyclic excursions around $x_2$ orbits. Indeed, from a theoretical point of view, any open orbit like the best-fitting open orbit of KDL is most naturally understood as a libration around an underlying $x_2$ orbit. In order to understand the theoretical arguments that lead to this conclusion, we need to digress a little bit to discuss the theory of orbits in barred potential.

\begin{figure}
	\includegraphics[width=\columnwidth]{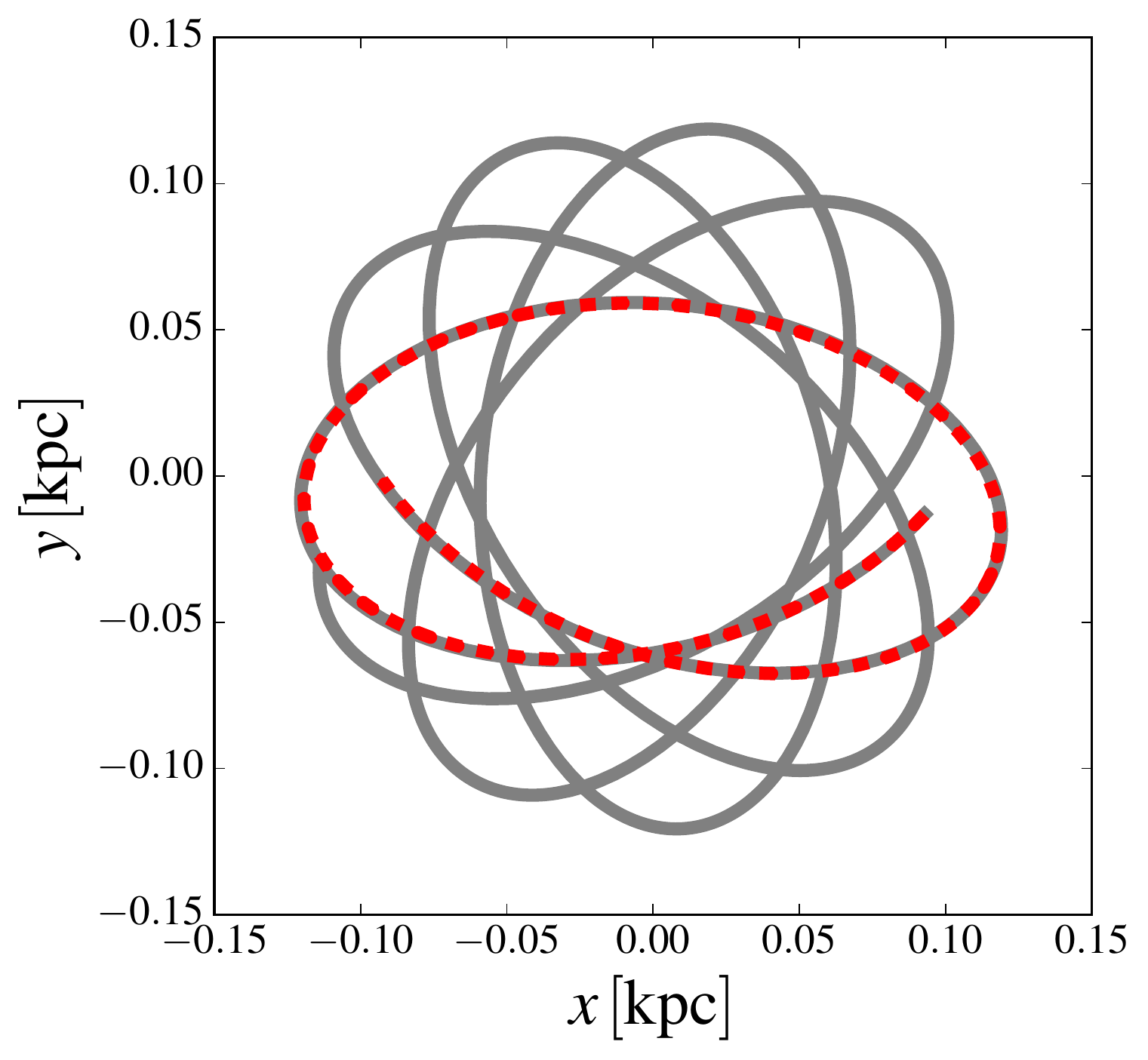}
    \caption{The best-fitting orbit of \citet{Kruijssen+2015} (KDL) is a portion of a rosette. \emph{Dashed red:} the best-fitting orbit of KDL, which is integrated for $5\Myr$. \emph{Gray solid}: continuation of the KDL orbit if it is integrated for further $15 \Myr$.}
    \label{fig:KDL}
\end{figure}

In an \emph{axisymmetric} potential, such as the one used by KDL, the only stable planar \emph{closed} orbits are circular orbits. The typical (non-closed) orbit is a rosette \citep[see for example Figure 3.1 in][]{BT2008}. Any rosette orbit is naturally described as the sum of the motion of a guiding centre which follows the circular orbit plus excursions around this circular orbit \citep{BT2008}. Indeed, the KDL orbit, if integrated forward or backwards in time for longer than shown in Figure~6 of KDL, would be a rosette which oscillates between the maximum and minimum radii of 59 pc and 121 pc respectively (see Figure \ref{fig:KDL} and KDL Table 1).

In the central regions of a \emph{non-axisymmetric} rotating barred potential, such as the one of the MW, the typical orbit still looks like a rosette, but circular orbits are replaced by a family of mildly elongated stable \emph{closed} orbits called $x_2$ orbits. What do closed orbits look like in a sequence of potentials which start from an axisymmetric potential and then gradually become more barred? The answer to this question can be seen for example in Figure 6 of \cite{SBM2015b}: the circular orbits gradually morph into the $x_2$ orbit. How can we understand \emph{open} orbits (i.e., rosettes) in the central regions of a barred potential? Just as any rosette orbit in an axisymmetric potential can be understood as a libration around a circular orbit, any rosette (non-closed) orbit in a bar potential can be understood as a libration around an underlying ``parent'' $x_2$ orbit which acts as a guiding centre. The parent closed orbit can be identified through the use of surfaces of section (see e.g. Chapter 3 of \citealt{BT2008}).  This is the most natural generalisation to non-axisymmetric potentials of axisymmetric framework described in the previous paragraph  \citep{BT2008}.

Strictly speaking, since KDL have used an axisymmetric approximation to the real gravitational potential, their orbit should be considered as a libration around an underlying \emph{circular} orbit. However, since the real gravitational potential of the Galaxy \emph{is} barred, any open orbit which resembles the KDL orbit in the real galaxy has a parent $x_2$ (not circular) orbit, and should be considered as an excursion around it. Therefore, \emph{physically}, the KDL scenario says that the gas follows an open orbit with a specific type of excursion around an underlying $x_2$ orbit. This is \emph{not} in contradiction with the \cite{Binney+1991} paper, because at the time the authors were only concerned with the \emph{average} motion of the gas, i.e. with the guiding centre motions, which as we just argued is physically an $x_2$ orbit even in the KDL scenario. \cite{Binney+1991} were aware that there must be deviations from this average motion, but they were not attempting to describe it. 

The excursions described by KDL model are not the only type of excursions possible, and there are other ways to generalise the \cite{Binney+1991} picture by adding excursions around $x_2$ orbits. For example, in the isothermal simulations of \cite{Ridley+2017} excursions around $x_2$ orbits driven by gas pressure collectively give rise to nuclear spiral density waves. As another example, in the simulations of \cite{Sormani+2018a}, stochastic excursions are driven by larger-scale turbulent flows. In the calculations presented in this paper, large excursions are driven by the large-scale flows as well as by the stellar feedback. Note that in all these simulations, the centres of mass of the clouds follow well $x_2$ orbits on average (see e.g. Figure 12 of \citealt{Sormani+2018a}). This confirms that \emph{the most natural theoretical framework} to think of orbital motions of gas in the CMZ is as the sum of the motion of a guiding centre which follows a closed $x_2$ orbit plus excursions around this guiding centre \citep{Sormani+2018a}. The question of understanding the orbital dynamics of the CMZ is therefore reduced to understanding what kind of excursions around the underlying $x_2$ orbits of the Galactic potential are present in the real CMZ, and what causes them.

KDL do not discuss the physical origin of the excursions in their model or how the gas ends up on their stream. Their is a phenomenological model obtained by fitting a ballistic orbit to the observed $(l,b,v_{\rm los})$ distribution of dense gas. In our simulations, we do not find gas streams with properties similar to the KDL orbit, because the hydrodynamical interactions with the bar inflow and/or stellar feedback typically disturb the coherence and ballistic nature of the stream in less than one orbital time (typically twice per orbit, when collisions with dust lane infall happen). Moreover, reproducing the KDL orbit would require that a sequence of clouds are launched with exactly the same initial conditions from the same point in space during a time interval of 5 Myr, which is unlikely to happen given the stochastic nature of the perturbations. \cite{Armillotta+2020} also report that they are unable to reproduce a stream with the characteristics of the KDL orbit in their simulations. At present, it remains to be shown that excursions compatible with the KDL orbit can occur in a self-consistent simulation of gas flow in a barred potential.

\subsection{Is there an inner and outer CMZ?} \label{sec:innerouter}

In the model of \cite{KrumholzKruijssen2015} material is deposited by the bar inflow at $R \simeq 450\pc$, at the outer edge of what they call the ``outer CMZ'', defined as the region $120 \lesssim R \lesssim 450 \pc$, and is then transported by acoustic instabilities to the ``inner CMZ'', defined as the region $0 \lesssim R \lesssim 120 \pc$. In their model the distinction between the inner and outer CMZ has a physical meaning as the radius which separates the acoustically unstable region from the gravitationally unstable region. They argued that the distinction is necessary since the bar-driven inflow mediated by the dust lanes is only able to transport the gas down to $R\simeq 450\pc$, not down to the the inner CMZ. The same distinction between inner and outer CMZ is used in \cite{Kruijssen+2015} and \cite{Krumholz+2017}. 

\begin{figure}
	\includegraphics[width=\columnwidth]{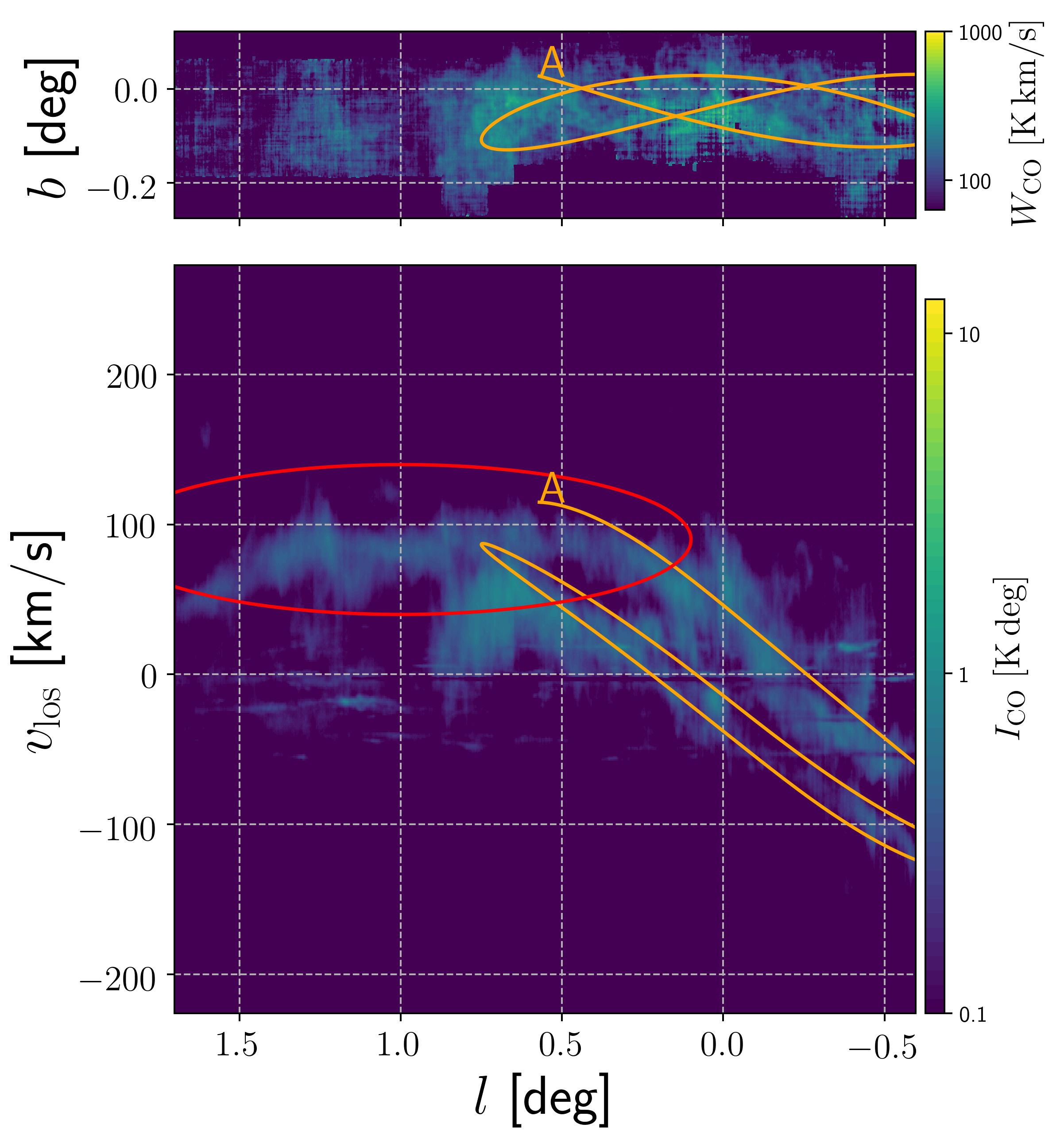}
    \caption{\emph{Top:} $^{13}$CO 2-1 longitude-latitude map of the CMZ, integrated over velocity, from the APEX CMZ SHFI-1 survey (\citealt{Ginsburg+2016}, \url{https://doi.org/10.7910/DVN/27601}). \emph{Bottom:} longitude-velocity map of the same data, integrated over latitude. The orange line shows the \citet{Kruijssen+2015} (KDL) best-fitting orbit. Point A shows the ending point of the orbit as provided in KDL. The red ellipses highlights a feature that in $(l,b,v)$ space appears to be continuously connected with stream 1 of KDL (see discussion in Section \ref{sec:innerouter}).}
    \label{fig:Apex13CO}
\end{figure}

\begin{figure}
	\includegraphics[width=\columnwidth]{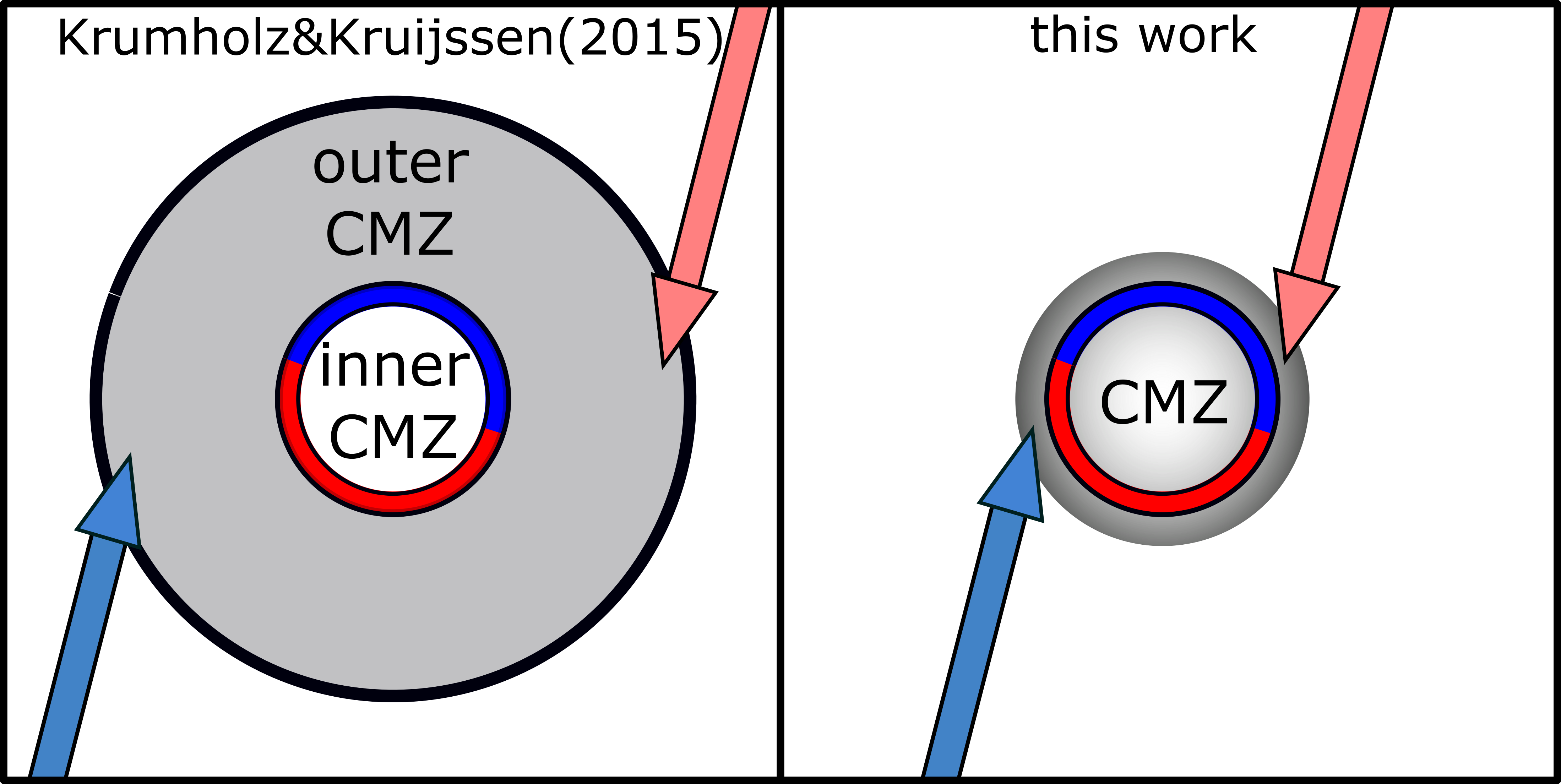}
    \caption{Schematic distinction between the scenarios in \citet{KrumholzKruijssen2015} and the one presented here. In the former, the bar inflow mediated by the dust lanes deposits gas at the outer edge of the outer CMZ at $R\simeq 450\pc$, while acoustic instabilities drive the gas down to the inner CMZ at $R\simeq 100\pc$. In the latter, there is just a single structure called CMZ, defined as the region $R \lesssim 200\pc$, and there is no distinction between outer and inner CMZ. The dust lanes deposit the gas at $R\simeq 200\pc$, and acoustic instabilities play no role. See discussion in Section \ref{sec:innerouter}.}
    \label{fig:innerouter}
\end{figure}

In the model presented here, the distinction between outer and inner CMZ is not necessary. Instead, the CMZ is a star-forming ring at $R\lesssim 200\pc$ which is the natural continuation of the gas flow in a barred potential and is directly interacting with the dust lanes that mediate the bar inflow, as schematically depicted in Figure \ref{fig:innerouter}. This view is supported by the following facts:
\begin{enumerate}
\item Contrary to the scenario described in \cite{KrumholzKruijssen2015}, in the simulations presented here the gas is deposited by the bar dust lanes at $R\simeq200\pc$, where it interacts directly with the ring (see for example Figure \ref{fig:RGB}). There is no axisymmetric disc outside this radius. Acoustic instabilities play no role in our model. Indeed, \citet{SormaniLi2020} show that acoustic instabilities are a spurious result and cannot drive turbulence in the ISM.\footnote{The ``acoustic instability'' was proposed by \cite{Montenegro+99}, who studied the propagation of small perturbations in a differentially rotating fluid disc. These authors derived a dispersion relation (see their Equation 10) in the WKB approximation, similar to the classic \cite{LinShu1964} dispersion relation, but keeping one extra order when expanding in the small quantity $1/|k R|$ (assumed small in the WKB approximation), where $k$ is the radial wavenumber and $R$ is the radius. In the axisymmetric case ($m=0$) the \cite{Montenegro+99} dispersion relation reduces to that of \cite{LinShu1964} and yields the standard \cite{Toomre1964} criterion. In the non-axisymmetric case ($m\neq0$) their dispersion relation contains some extra terms and predicts new unstable modes that are absent in \cite{LinShu1964} and which were dubbed ``acoustic'' because they are driven by pressure rather than self-gravity. However, \cite{SormaniLi2020} used hydrodynamical simulations to show that modes that are predicted to be acoustically unstable by \cite{Montenegro+99} are actually stable. The physical reason is exactly the same reason as to why non-axisymmetric modes that are formally predicted to be unstable by the standard \cite{LinShu1964} dispersion relation turn out to be stable. As shown by \cite{GoldreichLyndenbell1965} and \cite{JulianToomre1966} (see also footnote 6 at pg. 494 of \citealt{BT2008}), such non-axisymmetric disturbances in a fluid disc wind up and propagate off to infinity, leaving behind a smooth disc. This is in hindsight not surprising, since the \cite{Montenegro+99} is just a refinement of the \cite{LinShu1964} dispersion relation. Hence the acoustic instability is a spurious result and cannot drive turbulence in the interstellar medium.}

\item In the observations, the dust lanes can be seen to reach down to at least $l=1.7^\circ$ on the positive side (see point C in Figure 1 of \citealt{SormaniBarnes2019}) and to $l=-1.5^\circ$ on the negative longitude side (see point D in the same figure), which, assuming a Galactic centre distance of $8.2\kpc$ \citep{BlandhawthornGerhard2016,Gravity2019}, correspond to projected radial distances of $240\pc$ and $210\pc$ respectively, in good agreement with our simulations. Moreover, the dust lanes can also be seen to interact directly with the dense gas in the CMZ at $l\leq1.5^\circ$, in particular with the $l=1.3^\circ$ complex through one of the most prominent ``extended velocity features'' (EVF, see Figure 1 and Section 5.2 in \citealt{Sormani+2019}).
\item In our simulation, copious star formation occurs as soon as the dust lanes touch the ring (see Paper II). If the dust lanes entered the CMZ at $R\simeq450\pc$, we should see widespread star formation at those radii, which is not seen in observations \citep{Longmore+2013a}.
\item The orbital model of KDL only fits gas in the ``inner'' CMZ and omits the widespread gas emission at higher longitudes than SgrB2 (i.e. it omits the 1.3$^\circ$ complex). In their picture, the $1.3^\circ$ complex belongs to the outer CMZ and is physically disconnected from the inner CMZ which is fitted by their orbit. However, molecular gas observations indicate that the inner CMZ may be physically connected to the $1.3^\circ$ complex. Figure \ref{fig:Apex13CO} shows $^{13}$CO observations of the CMZ. The red ellipse highlights a feature in the $1.3^\circ$ complex that appears to be continuously connected in $(l,b,v_{\rm los})$ space to the inner CMZ, and in particular to ``Stream 1'' in the notation of Figure 4 of KDL. This suggests that Stream 1 and the $1.3^\circ$ complex are part of the same coherent structure, blurring the distinction between outer and inner CMZ. This connection cannot be explained in the KDL model. The orange line in Figure \ref{fig:Apex13CO} shows the KDL orbit, and point A indicates its final point as provided by KDL. If integration is continued in time beyond this point, the orbit will turn back towards smaller longitudes, and will never reach the feature highlighted by the red ellipse.
\item Other features at higher longitudes, such as Bania Clump 2 $\simeq 3.2\degree$, which might be seen as indicating the presence of an ``outer CMZ'', can be more naturally interpreted as material that is crashing onto the dust lanes \citep{Sormani+2019}. If this interpretation is correct, it makes the presence of an outer axisymmetric CMZ disc unnecessary, since no gas in the observations can be clearly attributed to it.
\end{enumerate}

We conclude that both theory and observations support the view that the distinction between inner and outer CMZ is unnecessary. Instead, we argue that it is more useful to think of the CMZ as a ring-like structure which is interacting directly with the inflow along the dust lanes. This ring-like structure is surrounded for a few tens of parsecs by clumpy scattered material (see e.g. left panel in Figure~\ref{fig:OldvsNew}), which is created partly by the $\sim 50\%$ dust lane infall that overshoots the main CMZ ring, and partly from gas that is pushed out the CMZ ring via SN feedback. This scattered material will later fall back on the main CMZ ring, or get picked up by the opposite dust lane. In our interpretation, the $1.3^\circ$ degree complex is part of the CMZ and represents an interaction region that is located at the intersection between the ring-like structure and the dust lanes, rather than being a structure which is physically separate from the CMZ.

\subsection{The 20 and 50 km s$^{-1}$ clouds} \label{sec:2050}

The 20 and 50 km s$^{-1}$ clouds are two prominent ($M \gtrsim 10^5 \Msun$) molecular clouds located in projection close to SgrA* ($\simeq 12$ pc and $\simeq 7$ pc respectively, see for example Figure~3 of \citealt{Molinari+2011}). They take their names from their line of sight velocities. 

In the orbital model of \cite{Kruijssen+2015}, these clouds are placed at $R = 50\mhyphen100\pc$ from the Galactic centre. This conclusion is based on the fact that the bulk emission of the two clouds is perfectly overlapping in $(l,b,v_{\rm los})$ space with one of the extended larger-scale streams that these authors fit with their orbit (compare black triangles and orange in Figure~9 of \citealt{Henshaw+2016}).

On the other hand, many authors agree in placing the 20 and 50 km s$^{-1}$ clouds close to SgrA* ($R \lesssim 20 \pc$), based on the following lines of evidence:
\begin{enumerate}
\item They appear to be interacting with the SgrA East supernova remnant \citep{Genzel+1990,McGary+2001,HerrnsteinHo2005,Sjouwerman+2008,Lee+2008}
\item They appear to be morphologically and kinematically connected to gaseous structures in the immediate vicinity SgrA*, such as the circum-nuclear disc \citep{Okumura+1991,CoilHo2000,Liu+2012,Takekawa+2017,Hsieh+2017,Tsuboi+2018,Ballone+2019}.
\item They have a comparable or larger column density than that of the Brick cloud \citep{Longmore+2012,Marsh+2016,Lu+2017,Barnes+2017,Henshaw+2019}, however they are not seen as dark extinction features at 8 micron with Spitzer (GLIMPSE: \citealt{Benjamin+2003}), whereas the Brick is a highly prominent infrared dark cloud (IRDC e.g. \citealt{Longmore+2012}). It is thought that the Brick is such a prominent IRDC since it is a cold, dense cloud in front of the very bright Galactic centre background. Given this logic, if the 20 and 50 km s$^{-1}$ clouds were also 50-100 pc in front of the Galactic centre, as well as being very dense, they should also be seen as prominent extinction features. A straightforward solution to this problem would arise if these clouds were closer to SgrA*, with more bright infrared Galactic centre emission in the foreground.
\end{enumerate}

\begin{figure}
	\includegraphics[width=\columnwidth]{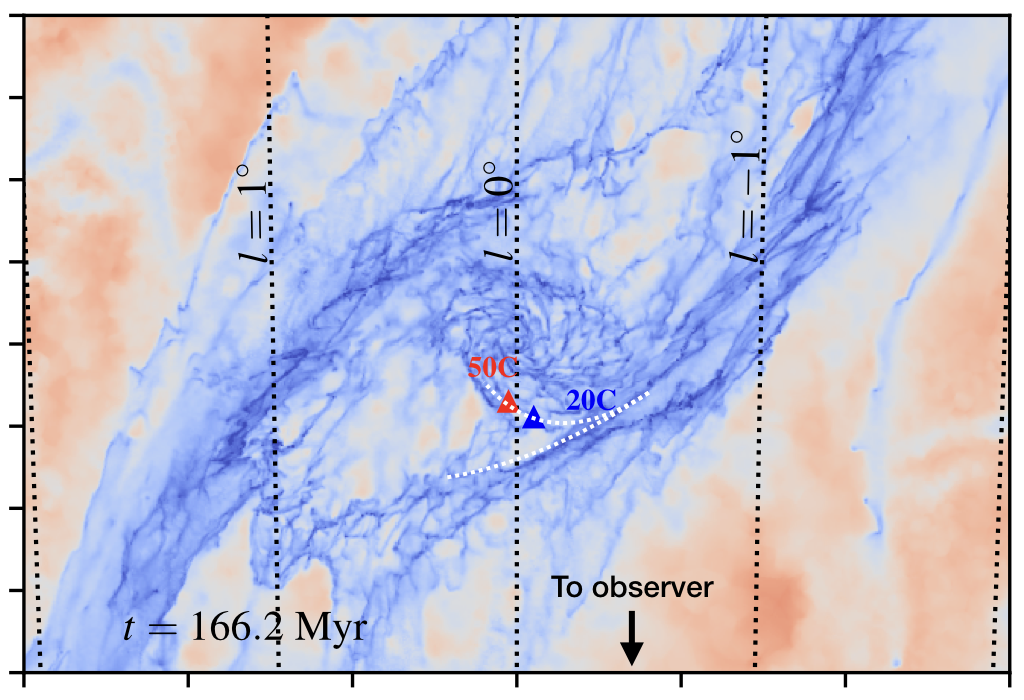}
    \caption{Placement of the 20 and 50 km s$^{-1}$ cloud according to our interpretation (see Section \ref{sec:2050}). The white dashed line indicates a bifurcation in the 100-pc stream.}
    \label{fig:2050}
\end{figure}

How can we reconcile the fact that the clouds seem to be connected to \emph{both} to gas on the larger scale ($\sim 100\pc$) stream and to gas close to SgrA*? Our simulation suggests a way to reconcile these apparently contradictory observational facts. Figure \ref{fig:2050} shows a snapshot of our simulation with a possible placement of the 20 and 50 km s$^{-1}$ clouds. The two clouds are placed where the line-of-sight velocity in the simulation matches their observed one at their observed longitude values with respect to SgrA*.\footnote{We have assumed that SgrA* coincides with the centre in our simulations and we have corrected for the fact that SgrA* is not exactly at $l=b=0$.} In the figure it can be seen that while the clouds are orbiting close to the centre ($R \lesssim 20\pc$), at the same time they are also continuously connected to the large scale streams that form the CMZ ring at $R \gtrsim 100\pc$ in our simulation.

The fact that the 20 and 50 km s$^{-1}$ clouds in our simulation are connected to both the CMZ ring-like stream at $R \simeq 100\pc$ and to the gas in the immediate surrounding of SgrA* as well is possible because the CMZ stream bifurcates into two branches (see white dashed line in Figure \ref{fig:2050}). One of the two branches connects to the 20 and 50 km s$^{-1}$ clouds, while the other continues its orbit along the main ring. A similar behaviour is found in the observations: when the 100-pc stream passes in front of the SgrA* clouds, at the location of the 20 and 50 km s$^{-1}$ clouds, it appears to bifurcate into three independent velocity structures (at least one of which is presumably from gas on the far side). This bifurcation is not explained in the \cite{Kruijssen+2015} model, who choose to fit only the two brightest velocity components and ignore the third one (see item iv in their Section 2.2). The bifurcation is instead naturally explained in our model. 

Finally, we note that our interpretation is also supported by observations of external galaxies which often show gaseous ``feathers'' that emerge from inside the nuclear rings (e.g. NGC 4314, see Figure~1 in \citealt{Benedict+2002}), consistent with being bifurcations similar to the one discussed above.

Our interpretation implies that the 3D geometry of the CMZ is even more complicated than previously suggested \citep[e.g.][]{Sofue95,Molinari+2011,Kruijssen+2015,Henshaw+16a,Ridley+2017}, and that we should take into account that gas in the CMZ ring is continuously connected through secondary branches to gas near SgrA*.

\subsection{The tilt and the $\infty$-shaped ring} \label{sec:tilt}

In the orbital models of \cite{Molinari+2011} and \cite{Kruijssen+2015}, the observed $\infty$-shape clearly visible in Figure~3 of \cite{Molinari+2011} is explained as the result of a vertically oscillating orbit. An alternative explanation is provided by \cite{Ridley+2017}, who suggested that the $\infty$-shape is created by a tilted gas distribution (see their Figure~12). According to this interpretation, the gas rotates in a ring/disc whose rotation axis forms an angle with respect to the rotation axis of the Galaxy at large. The tilt angle required to explain observations in their model is $\theta \simeq 5 \degree$. In this section we discuss whether a tilt is present in our simulations and the implications for observations.

The bottom row in Figure \ref{fig:slices} shows a slice of the CMZ. It can be seen from this figure that the CMZ does not lie exactly in the plane $z=0$, but is slightly tilted. We now analyse this in a more quantitative way. Following \cite{Ridley+2017}, we can specify the orientation of the normal to the disc $\hat{n}$ with respect to a coordinate system $xyz$ centred on the Galactic centre,
\begin{equation}
\hatn = \sin \theta \cos \phi \, \hatx + \sin \theta \sin \phi \, \haty + \cos \theta \, \hatz \, ,
\end{equation}
where $\hatz$ points towards the North Galactic Pole.

The three-dimensional orientation of the CMZ can be measured by diagonalising the moment of inertia $I$ of the molecular gas distribution,
\begin{equation}
I_{i j} = \sum_{\rm c} \rho_{\rm H2} x_{\rm c}^i x_{\rm c}^j \, ,
\end{equation}
where the sum is extended over all cells within the CMZ ($R \leq 250 \pc$). The direction $\hat{\bf n}$ normal to the CMZ disc will be the eigenvector of $I_{i j}$ associated with the largest eigenvalue.

\begin{figure}
	\includegraphics[width=\columnwidth]{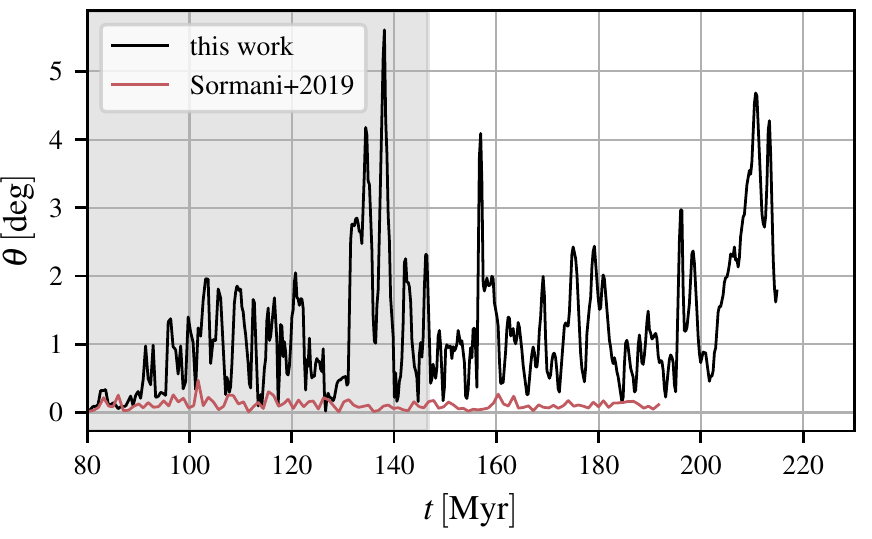}
    \caption{The tilt of the CMZ ring measured in the simulation presented in this paper (thick black line) and in the previous non-self gravitating simulations of \citet{Sormani+2019} (thin red line). See discussion in Section \ref{sec:tilt}.}
    \label{fig:tilt}
\end{figure}

Figure \ref{fig:tilt} shows the tilt angle $\theta$ as a function of time, measured using the procedure outlined above. The black line indicates the simulation presented in this work, while the red line compares with the previous non-self gravitating simulations of \cite{Sormani+2019}. The CMZ in the newer simulation shows a significant tilt which is typically of the order of $\theta \simeq 2^\circ$ and can occasionally reach values of $\theta \simeq 4 \mhyphen 5^\circ$. The latter values are marginally compatible with the value required by the model of \cite{Ridley+2017}. A similar tilt was also found in the hydrodynamical simulations of \cite{Shin+2017}.

What causes the CMZ tilt in the present simulations? Self-gravity and stellar feedback must play a role since these are the only differences from the previous simulations of \cite{Sormani+2019}, which did not show a tilt. However, if the SNe were directly responsible for creating the tilt, we would expect various parts of the orbit to be independently vertically displaced, because the supernovae randomly explode along the CMZ ring and only disturb the gas locally. In contrast, we often find that the tilt is coherent over the whole CMZ ring. 

While we cannot rule out that SN feedback directly causes the tilt, closer inspection of the movies of the simulations suggests that the SNe may be causing the tilt in an indirect way. SN feedback creates turbulence and causes the gas layer in the disc outside the bar to be thicker than in previous non-self gravitating simulations. Therefore, the gas that enters the dust lanes typically does not do so at $z=0$, but rather at some height above or below the Galactic plane. As the gas flows down along the dust lanes towards the CMZ, it acquires some vertical momentum. When this gas accretes onto the CMZ it transfers some of the momentum to it, causing the CMZ ring to wobble. Since the scale of the dust lanes is much larger ($2\mhyphen 3 \kpc$) than the scale of the CMZ ring ($ \simeq 100\pc$), this mechanism causes oscillations that have more coherence than those that can be expected from direct supernovae driving. This suggest that the tilt can be caused by accretion-driven torques. 

\begin{figure}
	\includegraphics[width=\columnwidth]{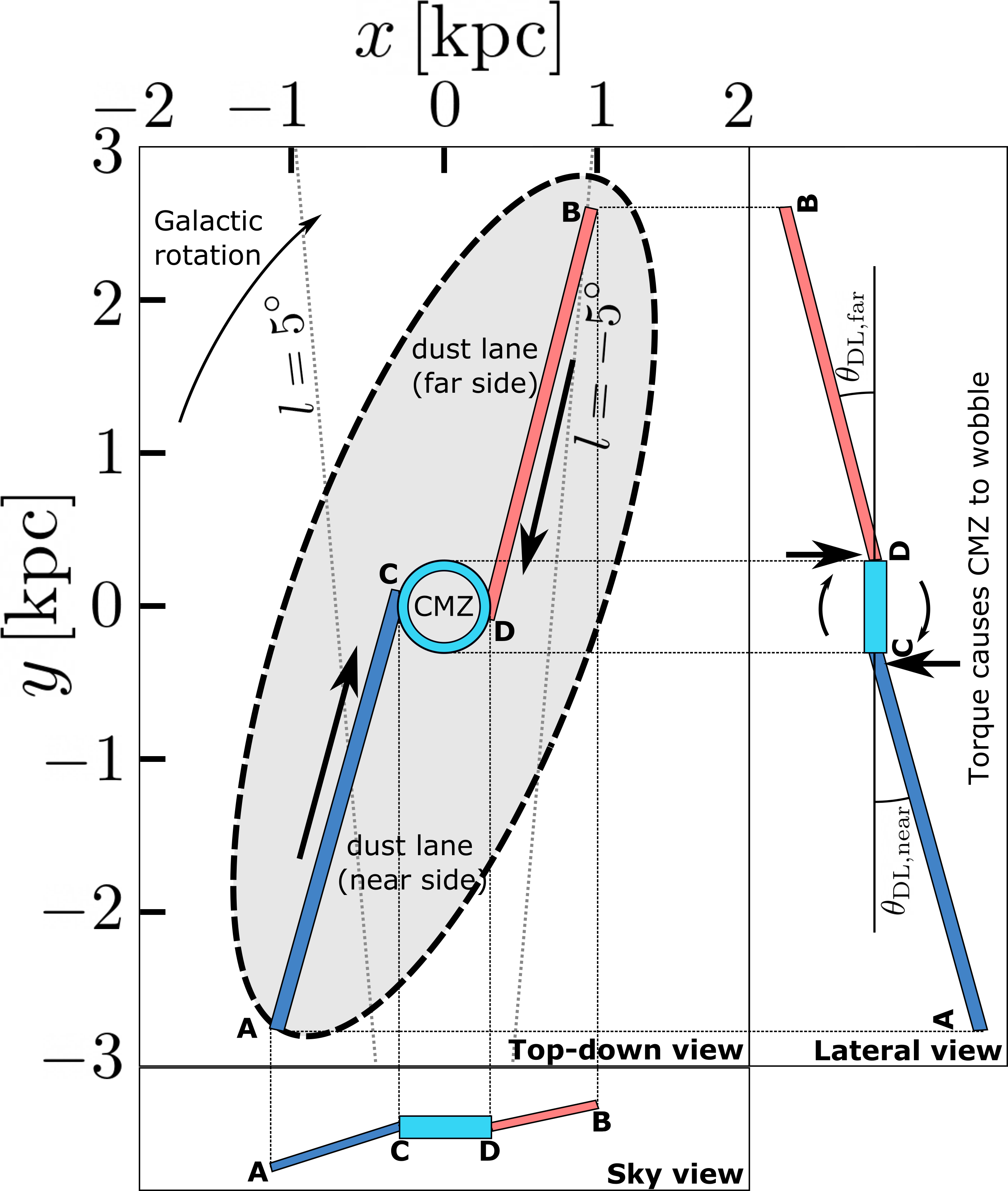}
    \caption{Schematic representation of how the dust lanes exert a torque on the CMZ along an axis that lies within the plane of the Galaxy, causing the CMZ to wobble. See discussion in Section \ref{sec:tilt}.}
    \label{fig:torque}
\end{figure}

What do the observations say? Interestingly, the dust lanes \emph{are} very clearly tilted. This is obvious from Figure~1 of \cite{SormaniBarnes2019} (see also \citealt{Marshall+2008}). Referring to that figure, the near side dust lane shows a clear gradient towards negative latitudes in going from point C to point A, while the far side dust lane shows a gradient towards increasing latitude from point D to point B. We can use the geometrical model provided in \cite{SormaniBarnes2019} and schematically reproduced in Figure \ref{fig:torque} to estimate the tilt angle. Both dust lanes have a length of $l \simeq 3\kpc$. In going from where the near side dust lane touches the CMZ (point C) to its endpoint (point A) there is a $\Delta b \simeq 1^\circ$, and its endpoint is at a distance of $d \simeq 5\kpc$ from us. This implies a tilt of $\theta_{\rm DL, near} \simeq  d  \Delta b / l \simeq 2^\circ$ (see Figure \ref{fig:torque} for the definition of this angle). For the far side dust lane, $\Delta b \simeq 1^\circ$ and $d \simeq 10 \kpc$, which implies $\theta_{\rm DL, far} \simeq  d  \Delta b / l \simeq 3^\circ$. These results are consistent with the dust lanes lying on a common tilted plane with average tilt $\theta_{\rm DL} \simeq 2.5^\circ$. 

We can estimate the time required for the torque\footnote{Note that this is a torque around an axis that lies within the plane of the Galaxy, not around the $z$ axis. See ``lateral view'' in Figure \ref{fig:torque}.} caused by the dust lane inflow to tilt the CMZ from $\theta_i=0$ to $\theta_f = 5 \degree$. The vertical momentum transferred from the dust lanes to the CMZ per unit time is $\dot{p}_{\rm z} = \dot{M} v \sin(\theta_{\rm DL})$ where  $\dot{M} \simeq 1 \Msunyr$ is the observationally determined bar-driven mass inflow rate (see discussion in Section \ref{sec:barinflow}) and $v \simeq 250\kms$ is the speed of the gas in the dust lane as it enters the CMZ. The lever arm, which we can take to be the radius of the CMZ, is $R \simeq 100\pc$. This gives a torque of $\tau = \dot{M}  v R \sin(\theta_{\rm DL}) \simeq 1.1 \Msunyr \kms \kpc$. The mass of the CMZ is $M = 5 \times 10^7 \Msun$ and its moment of inertia along an axis lying within the plane, assuming the CMZ mass can be approximated as distributed uniformly in a disc for the purposes of this estimation, is $I = M R^2/4 \simeq 1.25 \times 10^5 \Msun \kpc^2 $. The time required to generate the tilt is then $t = \sqrt{2 \theta_f I / \tau } \simeq 4 \Myr$. This is a short time, comparable to the dynamical orbital time of the CMZ ring ($\simeq 5\Myr$). The tilt angle cannot grow more than a few degrees because as the gas moves away from the plane $z=0$, the vertical restoring force of the Galactic potential brings it back to the plane (this restoring force is also what makes the CMZ wobble in a way similar to that of a coin spun on a table under the effect of the Earth's gravitational field). We conclude that the observed bar inflow should induce a significant and coherent tilt of the CMZ.

Although in our simulations the tilt in the dust lanes is driven by the SN feedback in the disc region outside the bar, we reckon this is unlikely to produce a tilt of both dust lanes along a common plane, because of the stochastic nature of the SN feedback. It is more likely that the dust lane tilt in the MW is produced by some other mechanism, perhaps the same mechanism that produces the warp of the outer H{\sc I} layer \citep{OstrikerBinney1989}. This unidentified mechanism would act in addition to the SN feedback that is at work in our simulation. Regardless of what causes the origin of the dust lane tilt, our simulations show that the dust lane inflow is a viable mechanism for producing a tilt in the CMZ.

Summarising the discussion, we conclude that:
\begin{itemize}
\item Our simulations show that accretion through the the dust lanes can induce a significant tilt in the CMZ.
\item Observations show that both the MW dust lanes are currently tilted by an angle of $\theta = 2 \mhyphen 3^\circ$ along a roughly common plane. An order of magnitude calculation shows that this will induce a significant tilt of the CMZ within $\sim 4 \Myr$ (roughly one orbital time).
\item It is likely that the CMZ is currently tilted because of the bar-driven inflow. A CMZ tilted out of the Galactic plane may explain the $\infty$-shaped structure seen in the observations.
\end{itemize}

\section{Summary and conclusions} \label{sec:conclusion}

We have presented high-resolution hydrodynamical simulations of the Milky Way innermost $5 \kpc$, and have used these to investigate gas dynamics in the central molecular zone (CMZ). The simulations include a realistic external barred potential, a time-dependent chemical network that keeps track of hydrogen and carbon chemistry (see Section \ref{sec:chemistry}), a physically motivated model for the formation of new stars using sink particles (see Section \ref{sec:sinks}), and supernova feedback (see Section \ref{sec:feedback}). The simulations allow us to follow the large-scale flow, which conforms to the usual $x_1/x_2$ orbit dynamics (see Section \ref{sec:largescale}), while at the same time reaching sub-parsec resolution in the dense regions (see Figure~\ref{fig:resolution}), thus allowing us to resolve individual molecular clouds formed self-consistently from the large-scale flow.

Our main conclusions are as follows:
\begin{itemize}
\item The time-averaged morphology of the CMZ is very smooth and regular (Figure \ref{fig:average}), while the instantaneous morphology of the CMZ is very rich in substructure that is transient in time (see Figure \ref{fig:NTotvstime_CMZ}). The instantaneous CMZ morphology can depart significantly from the average morphology, to the point that the underlying regular structure is not obvious or even discernible (see Section \ref{sec:CMZmorphology}). The CMZ morphology also depends on the tracer used: in H{\sc i} is more spiral-like, while in H$_2$ it is more ring-like (see Section \ref{sec:CMZmorphology}). The CMZ dynamics is best understood as gas following $x_2$ orbits plus librations (see Section \ref{sec:x2vsopen}).
\item Molecular clouds exhibit a highly complex filamentary morphology and do not resemble idealised ``spherical clouds''. They enter the CMZ already in this filamentary form, which suggests that the often-employed idealised idea of a roughly spherical cloud orbiting in the CMZ \citep[e.g.][]{Dale+2019} may be an oversimplification which should be applied with care when studying the dynamical evolution of clouds in the Galactic centre environment (see Figure \ref{fig:RGB} and Section \ref{sec:clouds}). This is consistent with the highly filamentary and almost fractal structure observed in the CMZ streams and in real molecular clouds \cite[e.g.][]{Jones+2012,Rathborne+2015,Henshaw+2019}.
\item The bar efficiently drives an inflow from the Galactic disc ($R \simeq 3\kpc$) down to the CMZ ($R \simeq 200\pc$) of the order of $1 \Msunyr$, consistent with observational determinations and previous theoretical findings (see Section \ref{sec:barinflow})
\item Stellar feedback can drive an inflow from the CMZ inwards towards the CND of $\sim 0.03 \Msunyr$ (see Section \ref{sec:feedbackinflow}). This number is significant for the formation of the circum-nuclear disc (CND) since at this rate it would take only $0.3\mhyphen 3 \Myr$ to reach its current mass. It is also important for understanding the fuelling of the supermassive black hole SgrA* since the CND is its closest large reservoir of molecular gas.
\item Both theory and observations support the view that there is no ``outer CMZ'' quasi-axisymmetric disc at radii $120 \lesssim R \lesssim 450 \pc$ in between the CMZ 100-pc ring and the point where the dust lanes deposit the inflow. Instead, the dust lanes deposit the inflow almost directly into the ring, which in our interpretation extends to the $1.3^\circ$ degree complex (see Section \ref{sec:innerouter}).
\item We give a new interpretation for the 3D placement of the 20 and 50 km s$^{-1}$ clouds, according to which they are close $(R \lesssim 30\pc)$ to SgrA*, while at the same time being connected to the main gas streams orbiting in the CMZ ring-like structure at $R \gtrsim 100\pc$ (see Section \ref{sec:2050}).
\item Accretion through the dust lanes can induce a significant tilt of the CMZ out of the Galactic plane. Observations show that the MW dust lanes are tilted by an angle of $\theta = 2 \mhyphen 3^\circ$, and an order of magnitude calculation shows that this will induce a significant CMZ tilt within $\simeq 4 \Myr$. A CMZ tilted out of the Galactic plane may provide an alternative explanation to the $\infty$-shaped figure discovered by \cite{Molinari+2011} (See Section \ref{sec:tilt}).
\end{itemize}

\section*{Acknowledgements}
RGT and MCS both acknowledge contributing equally to this paper. We thank the anonymous referee for constructive comments that improved the paper. MCS thanks Ashley Barnes, Adam Ginsburg, Jonathan Henshaw, Zhi Li, Diederik Kruijssen and Elisabeth Mills for useful comments and discussions. RGT, MCS, SCOG, and RSK acknowledge financial support from the German Research Foundation (DFG) via the collaborative research center (SFB 881, Project-ID 138713538) ``The Milky Way System'' (subprojects A1, B1, B2, and B8). CDB and HPH gratefully acknowledges support for this work from the National Science Foundation under Grant No. (1816715). PCC acknowledges support from the Science and Technology Facilities Council (under grant ST/K00926/1) and StarFormMapper, a project that has received funding from the European Union's Horizon 2020 Research and Innovation Programme, under grant agreement no. 687528. HPH thanks the LSSTC Data Science Fellowship Program, which is funded by LSSTC, NSF Cybertraining Grant No. 1829740, the Brinson Foundation, and the Moore Foundation; his participation in the program has benefited this work. RJS gratefully acknowledges an STFC Ernest Rutherford fellowship (grant ST/N00485X/1) and HPC from the Durham DiRAC supercomputing facility (grants ST/P002293/1, ST/R002371/1, ST/S002502/1, and ST/R000832/1). RSK furthermore thanks for financial support from the European Research Council via the ERC Synergy Grant ``ECOGAL -- Understanding our Galactic ecosystem: from the disk of the Milky Way to the formation sites of stars and planets'' (grant 855130). The authors acknowledge support by the state of Baden-W\"urttemberg through bwHPC and the German Research Foundation (DFG) through grant INST 35/1134-1 FUGG. The authors gratefully acknowledge the data storage service SDS@hd supported by the Ministry of Science, Research and the Arts Baden-W\"urttemberg (MWK) and the German Research Foundation (DFG) through grant INST 35/1314-1 FUGG.

\section*{Data availability} The data underlying this article will be shared on reasonable request to the corresponding author. Movies of the simulations can be found at the following link: \url{https://www.youtube.com/channel/UCwnzfO-xLxzRDz9XsexfPoQ}.

\def\aap{A\&A}\def\aj{AJ}\def\apj{ApJ}\def\mnras{MNRAS}\def\araa{ARA\&A}\def\aapr{Astronomy \&
 Astrophysics Review}\def\apjs{ApJS}\def\apjl{ApJ}\def\pasj{PASJ}\def\nat{Nature}\def\prd{Phys. Rev. D}
\def\ssr{Space Sci. Rev.}\def\pasp{PASP}\def\aaps{A\&AS}\def\pasa{Publications of the Astronomical Society of Australia}\def\rmp{Rev. Mod. Phys.}\def\apss{Astrophysics and Space Science}\def\baas{Bulletin of the American Astronomical Society}
\bibliographystyle{mnras}
\bibliography{bibliography}

\begin{thebibliography}{}
\makeatletter
\relax
\def\mn@urlcharsother{\let\do\@makeother \do\$\do\&\do\#\do\^\do\_\do\%\do\~}
\def\mn@doi{\begingroup\mn@urlcharsother \@ifnextchar [ {\mn@doi@}
  {\mn@doi@[]}}
\def\mn@doi@[#1]#2{\def\@tempa{#1}\ifx\@tempa\@empty \href
  {http://dx.doi.org/#2} {doi:#2}\else \href {http://dx.doi.org/#2} {#1}\fi
  \endgroup}
\def\mn@eprint#1#2{\mn@eprint@#1:#2::\@nil}
\def\mn@eprint@arXiv#1{\href {http://arxiv.org/abs/#1} {{\tt arXiv:#1}}}
\def\mn@eprint@dblp#1{\href {http://dblp.uni-trier.de/rec/bibtex/#1.xml}
  {dblp:#1}}
\def\mn@eprint@#1:#2:#3:#4\@nil{\def\@tempa {#1}\def\@tempb {#2}\def\@tempc
  {#3}\ifx \@tempc \@empty \let \@tempc \@tempb \let \@tempb \@tempa \fi \ifx
  \@tempb \@empty \def\@tempb {arXiv}\fi \@ifundefined
  {mn@eprint@\@tempb}{\@tempb:\@tempc}{\expandafter \expandafter \csname
  mn@eprint@\@tempb\endcsname \expandafter{\@tempc}}}

\bibitem[\protect\citeauthoryear{{Armillotta}, {Krumholz}, {Di Teodoro}  \&
  {McClure-Griffiths}}{{Armillotta} et~al.}{2019}]{Armillotta+2019}
{Armillotta} L.,  {Krumholz} M.~R.,  {Di Teodoro} E.~M.,   {McClure-Griffiths}
  N.~M.,  2019, \mn@doi [\mnras] {10.1093/mnras/stz2880}, \href
  {https://ui.adsabs.harvard.edu/abs/2019MNRAS.490.4401A} {490, 4401}

\bibitem[\protect\citeauthoryear{{Armillotta}, {Krumholz}  \& {Di
  Teodoro}}{{Armillotta} et~al.}{2020}]{Armillotta+2020}
{Armillotta} L.,  {Krumholz} M.~R.,   {Di Teodoro} E.~M.,  2020, \mn@doi
  [\mnras] {10.1093/mnras/staa469}, \href
  {https://ui.adsabs.harvard.edu/abs/2020MNRAS.493.5273A} {493, 5273}

\bibitem[\protect\citeauthoryear{{Athanassoula}}{{Athanassoula}}{1992}]{Athan92b}
{Athanassoula} E.,  1992, \mnras, \href
  {http://ukads.nottingham.ac.uk/abs/1992MNRAS.259..345A} {259, 345}

\bibitem[\protect\citeauthoryear{{Balbus} \& {Hawley}}{{Balbus} \&
  {Hawley}}{1998}]{BalbusHawley1998}
{Balbus} S.~A.,  {Hawley} J.~F.,  1998, \mn@doi [Reviews of Modern Physics]
  {10.1103/RevModPhys.70.1}, \href
  {https://ui.adsabs.harvard.edu/abs/1998RvMP...70....1B} {70, 1}

\bibitem[\protect\citeauthoryear{{Ballone}, {Mapelli}  \& {Trani}}{{Ballone}
  et~al.}{2019}]{Ballone+2019}
{Ballone} A.,  {Mapelli} M.,   {Trani} A.~A.,  2019, \mn@doi [\mnras]
  {10.1093/mnras/stz2147}, \href
  {https://ui.adsabs.harvard.edu/abs/2019MNRAS.488.5802B} {488, 5802}

\bibitem[\protect\citeauthoryear{{Bally}, {Stark}, {Wilson}  \&
  {Henkel}}{{Bally} et~al.}{1988}]{Bally+1988}
{Bally} J.,  {Stark} A.~A.,  {Wilson} R.~W.,   {Henkel} C.,  1988, \mn@doi
  [\apj] {10.1086/165891}, \href
  {http://adsabs.harvard.edu/abs/1988ApJ...324..223B} {324, 223}

\bibitem[\protect\citeauthoryear{{Barnes}, {Longmore}, {Battersby}, {Bally},
  {Kruijssen}, {Henshaw}  \& {Walker}}{{Barnes} et~al.}{2017}]{Barnes+2017}
{Barnes} A.~T.,  {Longmore} S.~N.,  {Battersby} C.,  {Bally} J.,  {Kruijssen}
  J.~M.~D.,  {Henshaw} J.~D.,   {Walker} D.~L.,  2017, \mn@doi [\mnras]
  {10.1093/mnras/stx941}, \href
  {https://ui.adsabs.harvard.edu/abs/2017MNRAS.469.2263B} {469, 2263}

\bibitem[\protect\citeauthoryear{{Bate}, {Bonnell}  \& {Price}}{{Bate}
  et~al.}{1995}]{Bate+1995}
{Bate} M.~R.,  {Bonnell} I.~A.,   {Price} N.~M.,  1995, \mn@doi [\mnras]
  {10.1093/mnras/277.2.362}, \href
  {https://ui.adsabs.harvard.edu/abs/1995MNRAS.277..362B} {277, 362}

\bibitem[\protect\citeauthoryear{{Benedict}, {Howell}, {J{\o}rgensen}, {Kenney}
   \& {Smith}}{{Benedict} et~al.}{2002}]{Benedict+2002}
{Benedict} G.~F.,  {Howell} D.~A.,  {J{\o}rgensen} I.,  {Kenney} J. D.~P.,
  {Smith} B.~J.,  2002, \mn@doi [\aj] {10.1086/338895}, \href
  {https://ui.adsabs.harvard.edu/abs/2002AJ....123.1411B} {123, 1411}

\bibitem[\protect\citeauthoryear{{Benjamin} et~al.,}{{Benjamin}
  et~al.}{2003}]{Benjamin+2003}
{Benjamin} R.~A.,  et~al., 2003, \mn@doi [\pasp] {10.1086/376696}, \href
  {https://ui.adsabs.harvard.edu/abs/2003PASP..115..953B} {115, 953}

\bibitem[\protect\citeauthoryear{{Binney} \& {Tremaine}}{{Binney} \&
  {Tremaine}}{2008}]{BT2008}
{Binney} J.,  {Tremaine} S.,  2008, {Galactic Dynamics: Second Edition}.
Princeton University Press

\bibitem[\protect\citeauthoryear{{Binney}, {Gerhard}, {Stark}, {Bally}  \&
  {Uchida}}{{Binney} et~al.}{1991}]{Binney+1991}
{Binney} J.,  {Gerhard} O.~E.,  {Stark} A.~A.,  {Bally} J.,   {Uchida} K.~I.,
  1991, \mnras, \href {http://adsabs.harvard.edu/abs/1991MNRAS.252..210B} {252,
  210}

\bibitem[\protect\citeauthoryear{{Bland-Hawthorn} \&
  {Gerhard}}{{Bland-Hawthorn} \& {Gerhard}}{2016}]{BlandhawthornGerhard2016}
{Bland-Hawthorn} J.,  {Gerhard} O.,  2016, \mn@doi [\araa]
  {10.1146/annurev-astro-081915-023441}, \href
  {http://adsabs.harvard.edu/abs/2016ARA%26A..54..529B} {54, 529}

\bibitem[\protect\citeauthoryear{{Blondin}, {Wright}, {Borkowski}  \&
  {Reynolds}}{{Blondin} et~al.}{1998}]{Blondin1998}
{Blondin} J.~M.,  {Wright} E.~B.,  {Borkowski} K.~J.,   {Reynolds} S.~P.,
  1998, \mn@doi [\apj] {10.1086/305708}, \href
  {http://adsabs.harvard.edu/abs/1998ApJ...500..342B} {500, 342}

\bibitem[\protect\citeauthoryear{{Bordoloi} et~al.,}{{Bordoloi}
  et~al.}{2017}]{Bordoloi+2017}
{Bordoloi} R.,  et~al., 2017, \mn@doi [\apj] {10.3847/1538-4357/834/2/191},
  \href {https://ui.adsabs.harvard.edu/abs/2017ApJ...834..191B} {834, 191}

\bibitem[\protect\citeauthoryear{{Chevance} et~al.,}{{Chevance}
  et~al.}{2020}]{Chevance+2020}
{Chevance} M.,  et~al., 2020, \mn@doi [\mnras] {10.1093/mnras/stz3525}, \href
  {https://ui.adsabs.harvard.edu/abs/2020MNRAS.493.2872C} {493, 2872}

\bibitem[\protect\citeauthoryear{{Chuss}, {Davidson}, {Dotson}, {Dowell},
  {Hildebrand}, {Novak}  \& {Vaillancourt}}{{Chuss} et~al.}{2003}]{Chuss+2003}
{Chuss} D.~T.,  {Davidson} J.~A.,  {Dotson} J.~L.,  {Dowell} C.~D.,
  {Hildebrand} R.~H.,  {Novak} G.,   {Vaillancourt} J.~E.,  2003, \mn@doi
  [\apj] {10.1086/379538}, \href
  {https://ui.adsabs.harvard.edu/abs/2003ApJ...599.1116C} {599, 1116}

\bibitem[\protect\citeauthoryear{{Clark}, {Glover}  \& {Klessen}}{{Clark}
  et~al.}{2012}]{clark12}
{Clark} P.~C.,  {Glover} S.~C.~O.,   {Klessen} R.~S.,  2012, \mn@doi [\mnras]
  {10.1111/j.1365-2966.2011.20087.x}, \href
  {http://adsabs.harvard.edu/abs/2012MNRAS.420..745C} {420, 745}

\bibitem[\protect\citeauthoryear{{Clark}, {Glover}, {Ragan}, {Shetty}  \&
  {Klessen}}{{Clark} et~al.}{2013}]{Clark+2013}
{Clark} P.~C.,  {Glover} S.~C.~O.,  {Ragan} S.~E.,  {Shetty} R.,   {Klessen}
  R.~S.,  2013, \mn@doi [\apjl] {10.1088/2041-8205/768/2/L34}, \href
  {http://adsabs.harvard.edu/abs/2013ApJ...768L..34C} {768, L34}

\bibitem[\protect\citeauthoryear{{Coil} \& {Ho}}{{Coil} \&
  {Ho}}{2000}]{CoilHo2000}
{Coil} A.~L.,  {Ho} P. T.~P.,  2000, \mn@doi [\apj] {10.1086/308650}, \href
  {https://ui.adsabs.harvard.edu/abs/2000ApJ...533..245C} {533, 245}

\bibitem[\protect\citeauthoryear{{Comer{\'o}n}, {Knapen}, {Beckman},
  {Laurikainen}, {Salo}, {Mart{\'\i}nez-Valpuesta}  \& {Buta}}{{Comer{\'o}n}
  et~al.}{2010}]{Comeron+2010}
{Comer{\'o}n} S.,  {Knapen} J.~H.,  {Beckman} J.~E.,  {Laurikainen} E.,  {Salo}
  H.,  {Mart{\'\i}nez-Valpuesta} I.,   {Buta} R.~J.,  2010, \mn@doi [\mnras]
  {10.1111/j.1365-2966.2009.16057.x}, \href
  {https://ui.adsabs.harvard.edu/abs/2010MNRAS.402.2462C} {402, 2462}

\bibitem[\protect\citeauthoryear{{Contopoulos} \& {Grosbol}}{{Contopoulos} \&
  {Grosbol}}{1989}]{ContopoulosGrosbol1989}
{Contopoulos} G.,  {Grosbol} P.,  1989, \mn@doi [\aapr] {10.1007/BF00873080},
  \href {http://adsabs.harvard.edu/abs/1989A%26ARv...1..261C} {1, 261}

\bibitem[\protect\citeauthoryear{{Crocker}}{{Crocker}}{2012}]{Crocker2012}
{Crocker} R.~M.,  2012, \mn@doi [\mnras] {10.1111/j.1365-2966.2012.21149.x},
  \href {https://ui.adsabs.harvard.edu/abs/2012MNRAS.423.3512C} {423, 3512}

\bibitem[\protect\citeauthoryear{{Crocker}, {Jones}, {Melia}, {Ott}  \&
  {Protheroe}}{{Crocker} et~al.}{2010}]{Crocker+2010}
{Crocker} R.~M.,  {Jones} D.~I.,  {Melia} F.,  {Ott} J.,   {Protheroe} R.~J.,
  2010, \mn@doi [\nat] {10.1038/nature08635}, \href
  {https://ui.adsabs.harvard.edu/abs/2010Natur.463...65C} {463, 65}

\bibitem[\protect\citeauthoryear{{Crocker}, {Bicknell}, {Taylor}  \&
  {Carretti}}{{Crocker} et~al.}{2015}]{Crocker+2015}
{Crocker} R.~M.,  {Bicknell} G.~V.,  {Taylor} A.~M.,   {Carretti} E.,  2015,
  \mn@doi [\apj] {10.1088/0004-637X/808/2/107}, \href
  {https://ui.adsabs.harvard.edu/abs/2015ApJ...808..107C} {808, 107}

\bibitem[\protect\citeauthoryear{{Dahmen}, {Huttemeister}, {Wilson}  \&
  {Mauersberger}}{{Dahmen} et~al.}{1998}]{Dahmen+1998}
{Dahmen} G.,  {Huttemeister} S.,  {Wilson} T.~L.,   {Mauersberger} R.,  1998,
  \aap, \href {https://ui.adsabs.harvard.edu/abs/1998A&A...331..959D} {331,
  959}

\bibitem[\protect\citeauthoryear{{Dale}, {Ngoumou}, {Ercolano}  \&
  {Bonnell}}{{Dale} et~al.}{2014}]{Dale+2014}
{Dale} J.~E.,  {Ngoumou} J.,  {Ercolano} B.,   {Bonnell} I.~A.,  2014, \mn@doi
  [\mnras] {10.1093/mnras/stu816}, \href
  {http://adsabs.harvard.edu/abs/2014MNRAS.442..694D} {442, 694}

\bibitem[\protect\citeauthoryear{{Dale}, {Kruijssen}  \& {Longmore}}{{Dale}
  et~al.}{2019}]{Dale+2019}
{Dale} J.~E.,  {Kruijssen} J.~M.~D.,   {Longmore} S.~N.,  2019, \mn@doi
  [\mnras] {10.1093/mnras/stz888}, \href
  {https://ui.adsabs.harvard.edu/abs/2019MNRAS.486.3307D} {486, 3307}

\bibitem[\protect\citeauthoryear{{Davies}, {M{\"u}ller S{\'a}nchez}, {Genzel},
  {Tacconi}, {Hicks}, {Friedrich}  \& {Sternberg}}{{Davies}
  et~al.}{2007}]{Davies+2007}
{Davies} R.~I.,  {M{\"u}ller S{\'a}nchez} F.,  {Genzel} R.,  {Tacconi} L.~J.,
  {Hicks} E.~K.~S.,  {Friedrich} S.,   {Sternberg} A.,  2007, \mn@doi [\apj]
  {10.1086/523032}, \href
  {https://ui.adsabs.harvard.edu/abs/2007ApJ...671.1388D} {671, 1388}

\bibitem[\protect\citeauthoryear{{Di Teodoro}, {McClure-Griffiths}, {Lockman},
  {Denbo}, {Endsley}, {Ford}  \& {Harrington}}{{Di Teodoro}
  et~al.}{2018}]{DiTeodoro+2018}
{Di Teodoro} E.~M.,  {McClure-Griffiths} N.~M.,  {Lockman} F.~J.,  {Denbo}
  S.~R.,  {Endsley} R.,  {Ford} H.~A.,   {Harrington} K.,  2018, \mn@doi [\apj]
  {10.3847/1538-4357/aaad6a}, \href
  {https://ui.adsabs.harvard.edu/abs/2018ApJ...855...33D} {855, 33}

\bibitem[\protect\citeauthoryear{{Di Teodoro}, {McClure-Griffiths}, {Lockman}
  \& {Armillotta}}{{Di Teodoro} et~al.}{2020}]{DiTeodoro+2020}
{Di Teodoro} E.~M.,  {McClure-Griffiths} N.~M.,  {Lockman} F.~J.,
  {Armillotta} L.,  2020, arXiv e-prints, \href
  {https://ui.adsabs.harvard.edu/abs/2020arXiv200809121D} {p. arXiv:2008.09121}

\bibitem[\protect\citeauthoryear{{Draine}}{{Draine}}{1978}]{draine78}
{Draine} B.~T.,  1978, \mn@doi [\apjs] {10.1086/190513}, \href
  {http://adsabs.harvard.edu/abs/1978ApJS...36..595D} {36, 595}

\bibitem[\protect\citeauthoryear{{Evans}, {Heiderman}  \&
  {Vutisalchavakul}}{{Evans} et~al.}{2014}]{Evans+2014}
{Evans} Neal~J. I.,  {Heiderman} A.,   {Vutisalchavakul} N.,  2014, \mn@doi
  [\apj] {10.1088/0004-637X/782/2/114}, \href
  {https://ui.adsabs.harvard.edu/abs/2014ApJ...782..114E} {782, 114}

\bibitem[\protect\citeauthoryear{{Federrath}}{{Federrath}}{2015}]{Federrath+2015}
{Federrath} C.,  2015, \mn@doi [\mnras] {10.1093/mnras/stv941}, \href
  {https://ui.adsabs.harvard.edu/abs/2015MNRAS.450.4035F} {450, 4035}

\bibitem[\protect\citeauthoryear{{Federrath}, {Banerjee}, {Clark}  \&
  {Klessen}}{{Federrath} et~al.}{2010}]{Federrath++2010}
{Federrath} C.,  {Banerjee} R.,  {Clark} P.~C.,   {Klessen} R.~S.,  2010,
  \mn@doi [\apj] {10.1088/0004-637X/713/1/269}, \href
  {http://adsabs.harvard.edu/abs/2010ApJ...713..269F} {713, 269}

\bibitem[\protect\citeauthoryear{{Federrath}, {Sur}, {Schleicher}, {Banerjee}
  \& {Klessen}}{{Federrath} et~al.}{2011}]{Federrath+2011}
{Federrath} C.,  {Sur} S.,  {Schleicher} D. R.~G.,  {Banerjee} R.,   {Klessen}
  R.~S.,  2011, \mn@doi [\apj] {10.1088/0004-637X/731/1/62}, \href
  {https://ui.adsabs.harvard.edu/abs/2011ApJ...731...62F} {731, 62}

\bibitem[\protect\citeauthoryear{{Federrath} et~al.,}{{Federrath}
  et~al.}{2016}]{Federrath+2016}
{Federrath} C.,  et~al., 2016, \mn@doi [\apj] {10.3847/0004-637X/832/2/143},
  \href {https://ui.adsabs.harvard.edu/abs/2016ApJ...832..143F} {832, 143}

\bibitem[\protect\citeauthoryear{{Ferri{\`e}re}, {Gillard}  \&
  {Jean}}{{Ferri{\`e}re} et~al.}{2007}]{Ferriere+2007}
{Ferri{\`e}re} K.,  {Gillard} W.,   {Jean} P.,  2007, \mn@doi [\aap]
  {10.1051/0004-6361:20066992}, \href
  {https://ui.adsabs.harvard.edu/abs/2007A&A...467..611F} {467, 611}

\bibitem[\protect\citeauthoryear{{Fox} et~al.,}{{Fox} et~al.}{2015}]{Fox+2015}
{Fox} A.~J.,  et~al., 2015, \mn@doi [\apjl] {10.1088/2041-8205/799/1/L7}, \href
  {https://ui.adsabs.harvard.edu/abs/2015ApJ...799L...7F} {799, L7}

\bibitem[\protect\citeauthoryear{{Gatto} et~al.,}{{Gatto}
  et~al.}{2015}]{Gatto+2015}
{Gatto} A.,  et~al., 2015, \mn@doi [\mnras] {10.1093/mnras/stv324}, \href
  {http://adsabs.harvard.edu/abs/2015MNRAS.449.1057G} {449, 1057}

\bibitem[\protect\citeauthoryear{{Gatto} et~al.,}{{Gatto}
  et~al.}{2017}]{Gatto+2017}
{Gatto} A.,  et~al., 2017, \mn@doi [\mnras] {10.1093/mnras/stw3209}, \href
  {https://ui.adsabs.harvard.edu/abs/2017MNRAS.466.1903G} {466, 1903}

\bibitem[\protect\citeauthoryear{{Genzel}, {Stacey}, {Harris}, {Townes},
  {Geis}, {Graf}, {Poglitsch}  \& {Stutzki}}{{Genzel}
  et~al.}{1990}]{Genzel+1990}
{Genzel} R.,  {Stacey} G.~J.,  {Harris} A.~I.,  {Townes} C.~H.,  {Geis} N.,
  {Graf} U.~U.,  {Poglitsch} A.,   {Stutzki} J.,  1990, \mn@doi [\apj]
  {10.1086/168827}, \href
  {https://ui.adsabs.harvard.edu/abs/1990ApJ...356..160G} {356, 160}

\bibitem[\protect\citeauthoryear{{Genzel}, {Eisenhauer}  \&
  {Gillessen}}{{Genzel} et~al.}{2010}]{Genzel+2010}
{Genzel} R.,  {Eisenhauer} F.,   {Gillessen} S.,  2010, \mn@doi [\rmp]
  {10.1103/RevModPhys.82.3121}, \href
  {http://adsabs.harvard.edu/abs/2010RvMP...82.3121G} {82, 3121}

\bibitem[\protect\citeauthoryear{{Ghez} et~al.,}{{Ghez}
  et~al.}{2008}]{Ghez+2008}
{Ghez} A.~M.,  et~al., 2008, \mn@doi [\apj] {10.1086/592738}, \href
  {http://adsabs.harvard.edu/abs/2008ApJ...689.1044G} {689, 1044}

\bibitem[\protect\citeauthoryear{{Gillessen} et~al.,}{{Gillessen}
  et~al.}{2017}]{Gillessen+2017}
{Gillessen} S.,  et~al., 2017, \mn@doi [\apj] {10.3847/1538-4357/aa5c41}, \href
  {http://adsabs.harvard.edu/abs/2017ApJ...837...30G} {837, 30}

\bibitem[\protect\citeauthoryear{{Ginsburg} et~al.,}{{Ginsburg}
  et~al.}{2016}]{Ginsburg+2016}
{Ginsburg} A.,  et~al., 2016, \mn@doi [\aap] {10.1051/0004-6361/201526100},
  \href {https://ui.adsabs.harvard.edu/abs/2016A&A...586A..50G} {586, A50}

\bibitem[\protect\citeauthoryear{{Girichidis}, {Seifried}, {Naab}, {Peters},
  {Walch}, {W{\"u}nsch}, {Glover}  \& {Klessen}}{{Girichidis}
  et~al.}{2018}]{Girichidis+2018}
{Girichidis} P.,  {Seifried} D.,  {Naab} T.,  {Peters} T.,  {Walch} S.,
  {W{\"u}nsch} R.,  {Glover} S. C.~O.,   {Klessen} R.~S.,  2018, \mn@doi
  [\mnras] {10.1093/mnras/sty2016}, \href
  {https://ui.adsabs.harvard.edu/abs/2018MNRAS.480.3511G} {480, 3511}

\bibitem[\protect\citeauthoryear{{Glover} \& {Clark}}{{Glover} \&
  {Clark}}{2012}]{gc12}
{Glover} S.~C.~O.,  {Clark} P.~C.,  2012, \mn@doi [\mnras]
  {10.1111/j.1365-2966.2011.20260.x}, \href
  {http://adsabs.harvard.edu/abs/2012MNRAS.421..116G} {421, 116}

\bibitem[\protect\citeauthoryear{{Glover} \& {Mac Low}}{{Glover} \& {Mac
  Low}}{2007a}]{gm07a}
{Glover} S.~C.~O.,  {Mac Low} M.-M.,  2007a, \mn@doi [\apjs] {10.1086/512238},
  \href {http://adsabs.harvard.edu/abs/2007ApJS..169..239G} {169, 239}

\bibitem[\protect\citeauthoryear{{Glover} \& {Mac Low}}{{Glover} \& {Mac
  Low}}{2007b}]{gm07b}
{Glover} S.~C.~O.,  {Mac Low} M.-M.,  2007b, \mn@doi [\apj] {10.1086/512227},
  \href {http://adsabs.harvard.edu/abs/2007ApJ...659.1317G} {659, 1317}

\bibitem[\protect\citeauthoryear{{Glover}, {Federrath}, {Mac Low}  \&
  {Klessen}}{{Glover} et~al.}{2010}]{glo10}
{Glover} S.~C.~O.,  {Federrath} C.,  {Mac Low} M.-M.,   {Klessen} R.~S.,  2010,
  \mn@doi [\mnras] {10.1111/j.1365-2966.2009.15718.x}, \href
  {http://adsabs.harvard.edu/abs/2010MNRAS.404....2G} {404, 2}

\bibitem[\protect\citeauthoryear{{Goldreich} \& {Lynden-Bell}}{{Goldreich} \&
  {Lynden-Bell}}{1965}]{GoldreichLyndenbell1965}
{Goldreich} P.,  {Lynden-Bell} D.,  1965, \mn@doi [\mnras]
  {10.1093/mnras/130.2.125}, \href
  {https://ui.adsabs.harvard.edu/abs/1965MNRAS.130..125G} {130, 125}

\bibitem[\protect\citeauthoryear{{Goldsmith} \& {Langer}}{{Goldsmith} \&
  {Langer}}{1978}]{gl78}
{Goldsmith} P.~F.,  {Langer} W.~D.,  1978, \mn@doi [\apj] {10.1086/156206},
  \href {http://adsabs.harvard.edu/abs/1978ApJ...222..881G} {222, 881}

\bibitem[\protect\citeauthoryear{{Gravity Collaboration} et~al.,}{{Gravity
  Collaboration} et~al.}{2019}]{Gravity2019}
{Gravity Collaboration} et~al., 2019, \mn@doi [\aap]
  {10.1051/0004-6361/201935656}, \href
  {https://ui.adsabs.harvard.edu/abs/2019A&A...625L..10G} {625, L10}

\bibitem[\protect\citeauthoryear{{Henshaw} et~al.,}{{Henshaw}
  et~al.}{2016a}]{Henshaw+16a}
{Henshaw} J.~D.,  et~al., 2016a, \mn@doi [\mnras] {10.1093/mnras/stw121}, \href
  {http://adsabs.harvard.edu/abs/2016MNRAS.457.2675H} {457, 2675}

\bibitem[\protect\citeauthoryear{{Henshaw} et~al.,}{{Henshaw}
  et~al.}{2016b}]{Henshaw+2016}
{Henshaw} J.~D.,  et~al., 2016b, \mn@doi [\mnras] {10.1093/mnras/stw121}, \href
  {https://ui.adsabs.harvard.edu/abs/2016MNRAS.457.2675H} {457, 2675}

\bibitem[\protect\citeauthoryear{{Henshaw} et~al.,}{{Henshaw}
  et~al.}{2019}]{Henshaw+2019}
{Henshaw} J.~D.,  et~al., 2019, \mn@doi [\mnras] {10.1093/mnras/stz471}, \href
  {https://ui.adsabs.harvard.edu/abs/2019MNRAS.485.2457H} {485, 2457}

\bibitem[\protect\citeauthoryear{{Herrnstein} \& {Ho}}{{Herrnstein} \&
  {Ho}}{2005}]{HerrnsteinHo2005}
{Herrnstein} R.~M.,  {Ho} P. T.~P.,  2005, \mn@doi [\apj] {10.1086/426047},
  \href {https://ui.adsabs.harvard.edu/abs/2005ApJ...620..287H} {620, 287}

\bibitem[\protect\citeauthoryear{{Heyer} \& {Dame}}{{Heyer} \&
  {Dame}}{2015}]{HeyerDame2015}
{Heyer} M.,  {Dame} T.~M.,  2015, \mn@doi [\araa]
  {10.1146/annurev-astro-082214-122324}, \href
  {http://adsabs.harvard.edu/abs/2015ARA%26A..53..583H} {53, 583}

\bibitem[\protect\citeauthoryear{{Hopkins}, {Kere{\v s}}, {O{\~n}orbe},
  {Faucher-Gigu{\`e}re}, {Quataert}, {Murray}  \& {Bullock}}{{Hopkins}
  et~al.}{2014}]{Hopkins+2014}
{Hopkins} P.~F.,  {Kere{\v s}} D.,  {O{\~n}orbe} J.,  {Faucher-Gigu{\`e}re}
  C.-A.,  {Quataert} E.,  {Murray} N.,   {Bullock} J.~S.,  2014, \mn@doi
  [\mnras] {10.1093/mnras/stu1738}, \href
  {http://adsabs.harvard.edu/abs/2014MNRAS.445..581H} {445, 581}

\bibitem[\protect\citeauthoryear{{Hsieh}, {Koch}, {Ho}, {Kim}, {Tang}, {Wang},
  {Yen}  \& {Hwang}}{{Hsieh} et~al.}{2017}]{Hsieh+2017}
{Hsieh} P.-Y.,  {Koch} P.~M.,  {Ho} P. T.~P.,  {Kim} W.-T.,  {Tang} Y.-W.,
  {Wang} H.-H.,  {Yen} H.-W.,   {Hwang} C.-Y.,  2017, \mn@doi [\apj]
  {10.3847/1538-4357/aa8329}, \href
  {https://ui.adsabs.harvard.edu/abs/2017ApJ...847....3H} {847, 3}

\bibitem[\protect\citeauthoryear{{Immer}, {Schuller}, {Omont}  \&
  {Menten}}{{Immer} et~al.}{2012}]{Immer+2012}
{Immer} K.,  {Schuller} F.,  {Omont} A.,   {Menten} K.~M.,  2012, \mn@doi
  [\aap] {10.1051/0004-6361/201117857}, \href
  {https://ui.adsabs.harvard.edu/abs/2012A&A...537A.121I} {537, A121}

\bibitem[\protect\citeauthoryear{{Immer}, {Kauffmann}, {Pillai}, {Ginsburg}  \&
  {Menten}}{{Immer} et~al.}{2016}]{Immer+2016}
{Immer} K.,  {Kauffmann} J.,  {Pillai} T.,  {Ginsburg} A.,   {Menten} K.~M.,
  2016, \mn@doi [\aap] {10.1051/0004-6361/201628777}, \href
  {https://ui.adsabs.harvard.edu/abs/2016A&A...595A..94I} {595, A94}

\bibitem[\protect\citeauthoryear{{Inutsuka}, {Inoue}, {Iwasaki}  \&
  {Hosokawa}}{{Inutsuka} et~al.}{2015}]{Inutsuka+2015}
{Inutsuka} S.-i.,  {Inoue} T.,  {Iwasaki} K.,   {Hosokawa} T.,  2015, \mn@doi
  [\aap] {10.1051/0004-6361/201425584}, \href
  {https://ui.adsabs.harvard.edu/abs/2015A&A...580A..49I} {580, A49}

\bibitem[\protect\citeauthoryear{{Izumi} et~al.,}{{Izumi}
  et~al.}{2013}]{Izumi+2013}
{Izumi} T.,  et~al., 2013, \mn@doi [\pasj] {10.1093/pasj/65.5.100}, \href
  {http://adsabs.harvard.edu/abs/2013PASJ...65..100I} {65, 100}

\bibitem[\protect\citeauthoryear{{Jones} et~al.,}{{Jones}
  et~al.}{2012}]{Jones+2012}
{Jones} P.~A.,  et~al., 2012, \mn@doi [\mnras]
  {10.1111/j.1365-2966.2011.19941.x}, \href
  {http://adsabs.harvard.edu/abs/2012MNRAS.419.2961J} {419, 2961}

\bibitem[\protect\citeauthoryear{{Julian} \& {Toomre}}{{Julian} \&
  {Toomre}}{1966}]{JulianToomre1966}
{Julian} W.~H.,  {Toomre} A.,  1966, \mn@doi [\apj] {10.1086/148957}, \href
  {https://ui.adsabs.harvard.edu/abs/1966ApJ...146..810J} {146, 810}

\bibitem[\protect\citeauthoryear{{Kalberla} \& {Dedes}}{{Kalberla} \&
  {Dedes}}{2008}]{KalberlaDedes2008}
{Kalberla} P.~M.~W.,  {Dedes} L.,  2008, \mn@doi [\aap]
  {10.1051/0004-6361:20079240}, \href
  {http://adsabs.harvard.edu/abs/2008A%26A...487..951K} {487, 951}

\bibitem[\protect\citeauthoryear{{Kataoka} et~al.,}{{Kataoka}
  et~al.}{2013}]{Kataoka+2013}
{Kataoka} J.,  et~al., 2013, \mn@doi [\apj] {10.1088/0004-637X/779/1/57}, \href
  {https://ui.adsabs.harvard.edu/abs/2013ApJ...779...57K} {779, 57}

\bibitem[\protect\citeauthoryear{{Katz}}{{Katz}}{1992}]{Katz1992}
{Katz} N.,  1992, \mn@doi [\apj] {10.1086/171366}, \href
  {http://adsabs.harvard.edu/abs/1992ApJ...391..502K} {391, 502}

\bibitem[\protect\citeauthoryear{{Katz}, {Weinberg}  \& {Hernquist}}{{Katz}
  et~al.}{1996}]{Katz+1996}
{Katz} N.,  {Weinberg} D.~H.,   {Hernquist} L.,  1996, \mn@doi [\apjs]
  {10.1086/192305}, \href {http://adsabs.harvard.edu/abs/1996ApJS..105...19K}
  {105, 19}

\bibitem[\protect\citeauthoryear{{Kauffmann}, {Pillai}, {Zhang}, {Menten},
  {Goldsmith}, {Lu}, {Guzm{\'a}n}  \& {Schmiedeke}}{{Kauffmann}
  et~al.}{2017a}]{Kauffmann+2017b}
{Kauffmann} J.,  {Pillai} T.,  {Zhang} Q.,  {Menten} K.~M.,  {Goldsmith} P.~F.,
   {Lu} X.,  {Guzm{\'a}n} A.~E.,   {Schmiedeke} A.,  2017a, \mn@doi [\aap]
  {10.1051/0004-6361/201628089}, \href
  {https://ui.adsabs.harvard.edu/abs/2017A&A...603A..90K} {603, A90}

\bibitem[\protect\citeauthoryear{{Kauffmann}, {Goldsmith}, {Melnick}, {Tolls},
  {Guzman}  \& {Menten}}{{Kauffmann} et~al.}{2017b}]{Kauffmann+2017a}
{Kauffmann} J.,  {Goldsmith} P.~F.,  {Melnick} G.,  {Tolls} V.,  {Guzman} A.,
  {Menten} K.~M.,  2017b, \mn@doi [\aap] {10.1051/0004-6361/201731123}, \href
  {https://ui.adsabs.harvard.edu/abs/2017A&A...605L...5K} {605, L5}

\bibitem[\protect\citeauthoryear{{Kennicutt}}{{Kennicutt}}{1998}]{Kennicutt1998}
{Kennicutt} Robert~C. J.,  1998, \mn@doi [\apj] {10.1086/305588}, \href
  {https://ui.adsabs.harvard.edu/abs/1998ApJ...498..541K} {498, 541}

\bibitem[\protect\citeauthoryear{{Kim} \& {Ostriker}}{{Kim} \&
  {Ostriker}}{2015}]{KimOstriker2015}
{Kim} C.-G.,  {Ostriker} E.~C.,  2015, \mn@doi [\apj]
  {10.1088/0004-637X/802/2/99}, \href
  {http://adsabs.harvard.edu/abs/2015ApJ...802...99K} {802, 99}

\bibitem[\protect\citeauthoryear{{Kim} \& {Ostriker}}{{Kim} \&
  {Ostriker}}{2017}]{KimOstriker2017}
{Kim} C.-G.,  {Ostriker} E.~C.,  2017, \mn@doi [\apj]
  {10.3847/1538-4357/aa8599}, \href
  {http://adsabs.harvard.edu/abs/2017ApJ...846..133K} {846, 133}

\bibitem[\protect\citeauthoryear{{Kimm} \& {Cen}}{{Kimm} \&
  {Cen}}{2014}]{Kimm+2014}
{Kimm} T.,  {Cen} R.,  2014, \mn@doi [\apj] {10.1088/0004-637X/788/2/121},
  \href {http://adsabs.harvard.edu/abs/2014ApJ...788..121K} {788, 121}

\bibitem[\protect\citeauthoryear{{Klessen} \& {Burkert}}{{Klessen} \&
  {Burkert}}{2000}]{KlessenBurkert2000}
{Klessen} R.~S.,  {Burkert} A.,  2000, \mn@doi [\apjs] {10.1086/313371}, \href
  {https://ui.adsabs.harvard.edu/abs/2000ApJS..128..287K} {128, 287}

\bibitem[\protect\citeauthoryear{{Krieger} et~al.,}{{Krieger}
  et~al.}{2017}]{Krieger+2017}
{Krieger} N.,  et~al., 2017, \mn@doi [\apj] {10.3847/1538-4357/aa951c}, \href
  {https://ui.adsabs.harvard.edu/abs/2017ApJ...850...77K} {850, 77}

\bibitem[\protect\citeauthoryear{{Kroupa}}{{Kroupa}}{2001}]{Kroupa2001}
{Kroupa} P.,  2001, \mn@doi [\mnras] {10.1046/j.1365-8711.2001.04022.x}, \href
  {https://ui.adsabs.harvard.edu/abs/2001MNRAS.322..231K} {322, 231}

\bibitem[\protect\citeauthoryear{{Kruijssen}, {Dale}  \&
  {Longmore}}{{Kruijssen} et~al.}{2015}]{Kruijssen+2015}
{Kruijssen} J.~M.~D.,  {Dale} J.~E.,   {Longmore} S.~N.,  2015, \mn@doi
  [\mnras] {10.1093/mnras/stu2526}, \href
  {http://adsabs.harvard.edu/abs/2015MNRAS.447.1059K} {447, 1059}

\bibitem[\protect\citeauthoryear{{Kruijssen} et~al.,}{{Kruijssen}
  et~al.}{2019a}]{Kruijssen+2019}
{Kruijssen} J.~M.~D.,  et~al., 2019a, \mn@doi [\mnras] {10.1093/mnras/stz381},
  \href {https://ui.adsabs.harvard.edu/abs/2019MNRAS.484.5734K} {484, 5734}

\bibitem[\protect\citeauthoryear{{Kruijssen} et~al.,}{{Kruijssen}
  et~al.}{2019b}]{Kruijssen+2019b}
{Kruijssen} J.~M.~D.,  et~al., 2019b, \mn@doi [\nat]
  {10.1038/s41586-019-1194-3}, \href
  {https://ui.adsabs.harvard.edu/abs/2019Natur.569..519K} {569, 519}

\bibitem[\protect\citeauthoryear{{Krumholz} \& {Federrath}}{{Krumholz} \&
  {Federrath}}{2019}]{KrumholzFederrath2019}
{Krumholz} M.~R.,  {Federrath} C.,  2019, \mn@doi [Frontiers in Astronomy and
  Space Sciences] {10.3389/fspas.2019.00007}, \href
  {https://ui.adsabs.harvard.edu/abs/2019FrASS...6....7K} {6, 7}

\bibitem[\protect\citeauthoryear{{Krumholz} \& {Kruijssen}}{{Krumholz} \&
  {Kruijssen}}{2015}]{KrumholzKruijssen2015}
{Krumholz} M.~R.,  {Kruijssen} J.~M.~D.,  2015, \mn@doi [\mnras]
  {10.1093/mnras/stv1670}, \href
  {https://ui.adsabs.harvard.edu/abs/2015MNRAS.453..739K} {453, 739}

\bibitem[\protect\citeauthoryear{{Krumholz}, {Dekel}  \& {McKee}}{{Krumholz}
  et~al.}{2012}]{Krumholz+2012}
{Krumholz} M.~R.,  {Dekel} A.,   {McKee} C.~F.,  2012, \mn@doi [\apj]
  {10.1088/0004-637X/745/1/69}, \href
  {https://ui.adsabs.harvard.edu/abs/2012ApJ...745...69K} {745, 69}

\bibitem[\protect\citeauthoryear{{Krumholz}, {Kruijssen}  \&
  {Crocker}}{{Krumholz} et~al.}{2017}]{Krumholz+2017}
{Krumholz} M.~R.,  {Kruijssen} J.~M.~D.,   {Crocker} R.~M.,  2017, \mn@doi
  [\mnras] {10.1093/mnras/stw3195}, \href
  {https://ui.adsabs.harvard.edu/abs/2017MNRAS.466.1213K} {466, 1213}

\bibitem[\protect\citeauthoryear{{Le Petit}, {Ruaud}, {Bron}, {Godard},
  {Roueff}, {Languignon}  \& {Le Bourlot}}{{Le Petit}
  et~al.}{2016}]{LePetit+2016}
{Le Petit} F.,  {Ruaud} M.,  {Bron} E.,  {Godard} B.,  {Roueff} E.,
  {Languignon} D.,   {Le Bourlot} J.,  2016, \mn@doi [\aap]
  {10.1051/0004-6361/201526658}, \href
  {https://ui.adsabs.harvard.edu/abs/2016A&A...585A.105L} {585, A105}

\bibitem[\protect\citeauthoryear{{Lee} et~al.,}{{Lee} et~al.}{2008}]{Lee+2008}
{Lee} S.,  et~al., 2008, \mn@doi [\apj] {10.1086/524128}, \href
  {https://ui.adsabs.harvard.edu/abs/2008ApJ...674..247L} {674, 247}

\bibitem[\protect\citeauthoryear{{Li}, {Shen}  \& {Schive}}{{Li}
  et~al.}{2020}]{Li+2020}
{Li} Z.,  {Shen} J.,   {Schive} H.-Y.,  2020, \mn@doi [\apj]
  {10.3847/1538-4357/ab6598}, \href
  {https://ui.adsabs.harvard.edu/abs/2020ApJ...889...88L} {889, 88}

\bibitem[\protect\citeauthoryear{{Lin} \& {Shu}}{{Lin} \&
  {Shu}}{1964}]{LinShu1964}
{Lin} C.~C.,  {Shu} F.~H.,  1964, \mn@doi [\apj] {10.1086/147955}, \href
  {http://adsabs.harvard.edu/abs/1964ApJ...140..646L} {140, 646}

\bibitem[\protect\citeauthoryear{{Liu}, {Hsieh}, {Ho}, {Su}, {Wright}, {Sun}
  \& {Minh}}{{Liu} et~al.}{2012}]{Liu+2012}
{Liu} H.~B.,  {Hsieh} P.-Y.,  {Ho} P. T.~P.,  {Su} Y.-N.,  {Wright} M.,  {Sun}
  A.-L.,   {Minh} Y.~C.,  2012, \mn@doi [\apj] {10.1088/0004-637X/756/2/195},
  \href {https://ui.adsabs.harvard.edu/abs/2012ApJ...756..195L} {756, 195}

\bibitem[\protect\citeauthoryear{{Longmore} et~al.,}{{Longmore}
  et~al.}{2012}]{Longmore+2012}
{Longmore} S.~N.,  et~al., 2012, \mn@doi [\apj] {10.1088/0004-637X/746/2/117},
  \href {https://ui.adsabs.harvard.edu/abs/2012ApJ...746..117L} {746, 117}

\bibitem[\protect\citeauthoryear{{Longmore} et~al.,}{{Longmore}
  et~al.}{2013}]{Longmore+2013a}
{Longmore} S.~N.,  et~al., 2013, \mn@doi [\mnras] {10.1093/mnras/sts376}, \href
  {http://adsabs.harvard.edu/abs/2013MNRAS.429..987L} {429, 987}

\bibitem[\protect\citeauthoryear{{Longmore} et~al.,}{{Longmore}
  et~al.}{2017}]{Longmore+2017}
{Longmore} S.~N.,  et~al., 2017, \mn@doi [\mnras] {10.1093/mnras/stx1226},
  \href {https://ui.adsabs.harvard.edu/abs/2017MNRAS.470.1462L} {470, 1462}

\bibitem[\protect\citeauthoryear{{Lu} et~al.,}{{Lu} et~al.}{2017}]{Lu+2017}
{Lu} X.,  et~al., 2017, \mn@doi [\apj] {10.3847/1538-4357/aa67f7}, \href
  {https://ui.adsabs.harvard.edu/abs/2017ApJ...839....1L} {839, 1}

\bibitem[\protect\citeauthoryear{{Mac Low} \& {Klessen}}{{Mac Low} \&
  {Klessen}}{2004}]{MacLowKlessen2004}
{Mac Low} M.-M.,  {Klessen} R.~S.,  2004, \mn@doi [Reviews of Modern Physics]
  {10.1103/RevModPhys.76.125}, \href
  {https://ui.adsabs.harvard.edu/abs/2004RvMP...76..125M} {76, 125}

\bibitem[\protect\citeauthoryear{{Maeder}}{{Maeder}}{2009}]{Maeder2009}
{Maeder} A.,  2009, {Physics, Formation and Evolution of Rotating Stars}.
Springer, Berlin Heidelberg, \mn@doi{10.1007/978-3-540-76949-1}

\bibitem[\protect\citeauthoryear{{Mangilli} et~al.,}{{Mangilli}
  et~al.}{2019}]{Mangilli+2019}
{Mangilli} A.,  et~al., 2019, \mn@doi [\aap] {10.1051/0004-6361/201935072},
  \href {https://ui.adsabs.harvard.edu/abs/2019A&A...630A..74M} {630, A74}

\bibitem[\protect\citeauthoryear{{Marsh}, {Ragan}, {Whitworth}  \&
  {Clark}}{{Marsh} et~al.}{2016}]{Marsh+2016}
{Marsh} K.~A.,  {Ragan} S.~E.,  {Whitworth} A.~P.,   {Clark} P.~C.,  2016,
  \mn@doi [\mnras] {10.1093/mnrasl/slw080}, \href
  {https://ui.adsabs.harvard.edu/abs/2016MNRAS.461L..16M} {461, L16}

\bibitem[\protect\citeauthoryear{{Marshall}, {Fux}, {Robin}  \&
  {Reyl{\'e}}}{{Marshall} et~al.}{2008}]{Marshall+2008}
{Marshall} D.~J.,  {Fux} R.,  {Robin} A.~C.,   {Reyl{\'e}} C.,  2008, \mn@doi
  [\aap] {10.1051/0004-6361:20078967}, \href
  {http://adsabs.harvard.edu/abs/2008A%26A...477L..21M} {477, L21}

\bibitem[\protect\citeauthoryear{{Martizzi}, {Faucher-Gigu{\`e}re}  \&
  {Quataert}}{{Martizzi} et~al.}{2015}]{Martizzi+2015}
{Martizzi} D.,  {Faucher-Gigu{\`e}re} C.-A.,   {Quataert} E.,  2015, \mn@doi
  [\mnras] {10.1093/mnras/stv562}, \href
  {https://ui.adsabs.harvard.edu/abs/2015MNRAS.450..504M} {450, 504}

\bibitem[\protect\citeauthoryear{{McClure-Griffiths}, {Green}, {Hill},
  {Lockman}, {Dickey}, {Gaensler}  \& {Green}}{{McClure-Griffiths}
  et~al.}{2013}]{McClureGriffiths+2013}
{McClure-Griffiths} N.~M.,  {Green} J.~A.,  {Hill} A.~S.,  {Lockman} F.~J.,
  {Dickey} J.~M.,  {Gaensler} B.~M.,   {Green} A.~J.,  2013, \mn@doi [\apjl]
  {10.1088/2041-8205/770/1/L4}, \href
  {https://ui.adsabs.harvard.edu/abs/2013ApJ...770L...4M} {770, L4}

\bibitem[\protect\citeauthoryear{{McGary}, {Coil}  \& {Ho}}{{McGary}
  et~al.}{2001}]{McGary+2001}
{McGary} R.~S.,  {Coil} A.~L.,   {Ho} P. T.~P.,  2001, \mn@doi [\apj]
  {10.1086/322390}, \href
  {https://ui.adsabs.harvard.edu/abs/2001ApJ...559..326M} {559, 326}

\bibitem[\protect\citeauthoryear{{Mills}, {G{\"u}sten}, {Requena-Torres}  \&
  {Morris}}{{Mills} et~al.}{2013}]{Mills+2013}
{Mills} E.~A.~C.,  {G{\"u}sten} R.,  {Requena-Torres} M.~A.,   {Morris} M.~R.,
  2013, \mn@doi [\apj] {10.1088/0004-637X/779/1/47}, \href
  {http://adsabs.harvard.edu/abs/2013ApJ...779...47M} {779, 47}

\bibitem[\protect\citeauthoryear{{Mills}, {Ginsburg}, {Immer}, {Barnes},
  {Wiesenfeld}, {Faure}, {Morris}  \& {Requena-Torres}}{{Mills}
  et~al.}{2018}]{Mills+2018}
{Mills} E.~A.~C.,  {Ginsburg} A.,  {Immer} K.,  {Barnes} J.~M.,  {Wiesenfeld}
  L.,  {Faure} A.,  {Morris} M.~R.,   {Requena-Torres} M.~A.,  2018, \mn@doi
  [\apj] {10.3847/1538-4357/aae581}, \href
  {https://ui.adsabs.harvard.edu/abs/2018ApJ...868....7M} {868, 7}

\bibitem[\protect\citeauthoryear{{Molinari} et~al.,}{{Molinari}
  et~al.}{2011}]{Molinari+2011}
{Molinari} S.,  et~al., 2011, \mn@doi [\apjl] {10.1088/2041-8205/735/2/L33},
  \href {http://adsabs.harvard.edu/abs/2011ApJ...735L..33M} {735, L33}

\bibitem[\protect\citeauthoryear{{Montenegro}, {Yuan}  \&
  {Elmegreen}}{{Montenegro} et~al.}{1999}]{Montenegro+99}
{Montenegro} L.~E.,  {Yuan} C.,   {Elmegreen} B.~G.,  1999, \mn@doi [\apj]
  {10.1086/307465}, \href
  {https://ui.adsabs.harvard.edu/abs/1999ApJ...520..592M} {520, 592}

\bibitem[\protect\citeauthoryear{{Morris}}{{Morris}}{2015}]{Morris2015}
{Morris} M.~R.,  2015, {Manifestations of the Galactic Center Magnetic Field}.
Springer International Publishing, p.~391,
  \mn@doi{10.1007/978-3-319-10614-4_32}

\bibitem[\protect\citeauthoryear{{Morris} \& {Serabyn}}{{Morris} \&
  {Serabyn}}{1996}]{MorrisSerabyn1996}
{Morris} M.,  {Serabyn} E.,  1996, \mn@doi [\araa]
  {10.1146/annurev.astro.34.1.645}, \href
  {http://adsabs.harvard.edu/abs/1996ARA%26A..34..645M} {34, 645}

\bibitem[\protect\citeauthoryear{{Nakashima}, {Koyama}, {Wang}  \&
  {Enokiya}}{{Nakashima} et~al.}{2019}]{Nakashima+2019}
{Nakashima} S.,  {Koyama} K.,  {Wang} Q.~D.,   {Enokiya} R.,  2019, \mn@doi
  [\apj] {10.3847/1538-4357/ab0d82}, \href
  {https://ui.adsabs.harvard.edu/abs/2019ApJ...875...32N} {875, 32}

\bibitem[\protect\citeauthoryear{{Nelson} \& {Langer}}{{Nelson} \&
  {Langer}}{1997}]{nl97}
{Nelson} R.~P.,  {Langer} W.~D.,  1997, \mn@doi [\apj] {10.1086/304167}, \href
  {http://adsabs.harvard.edu/abs/1997ApJ...482..796N} {482, 796}

\bibitem[\protect\citeauthoryear{{Offner} \& {Arce}}{{Offner} \&
  {Arce}}{2015}]{Offner+2015}
{Offner} S. S.~R.,  {Arce} H.~G.,  2015, \mn@doi [\apj]
  {10.1088/0004-637X/811/2/146}, \href
  {https://ui.adsabs.harvard.edu/abs/2015ApJ...811..146O} {811, 146}

\bibitem[\protect\citeauthoryear{{Oka}, {Geballe}, {Goto}, {Usuda}, {Benjamin},
  {McCall}  \& {Indriolo}}{{Oka} et~al.}{2019}]{Oka+2019}
{Oka} T.,  {Geballe} T.~R.,  {Goto} M.,  {Usuda} T.,  {Benjamin} {McCall} J.,
  {Indriolo} N.,  2019, \mn@doi [\apj] {10.3847/1538-4357/ab3647}, \href
  {https://ui.adsabs.harvard.edu/abs/2019ApJ...883...54O} {883, 54}

\bibitem[\protect\citeauthoryear{{Okumura}, {Ishiguro}, {Fomalont}, {Hasegawa},
  {Kasuga}, {Morita}, {Kawabe}  \& {Kobayashi}}{{Okumura}
  et~al.}{1991}]{Okumura+1991}
{Okumura} S.~K.,  {Ishiguro} M.,  {Fomalont} E.~B.,  {Hasegawa} T.,  {Kasuga}
  T.,  {Morita} K.-I.,  {Kawabe} R.,   {Kobayashi} H.,  1991, \mn@doi [\apj]
  {10.1086/170412}, \href
  {https://ui.adsabs.harvard.edu/abs/1991ApJ...378..127O} {378, 127}

\bibitem[\protect\citeauthoryear{{Ostriker} \& {Binney}}{{Ostriker} \&
  {Binney}}{1989}]{OstrikerBinney1989}
{Ostriker} E.~C.,  {Binney} J.~J.,  1989, \mn@doi [\mnras]
  {10.1093/mnras/237.3.785}, \href
  {http://adsabs.harvard.edu/abs/1989MNRAS.237..785O} {237, 785}

\bibitem[\protect\citeauthoryear{{Pakmor} \& {Springel}}{{Pakmor} \&
  {Springel}}{2013}]{PakmorSpringel2013}
{Pakmor} R.,  {Springel} V.,  2013, \mn@doi [\mnras] {10.1093/mnras/stt428},
  \href {https://ui.adsabs.harvard.edu/abs/2013MNRAS.432..176P} {432, 176}

\bibitem[\protect\citeauthoryear{{Phinney}}{{Phinney}}{1994}]{Phinney1994}
{Phinney} E.~S.,  1994, in {Shlosman} I.,  ed., Mass-Transfer Induced Activity
  in Galaxies. p.~1

\bibitem[\protect\citeauthoryear{{Pillai}, {Kauffmann}, {Tan}, {Goldsmith},
  {Carey}  \& {Menten}}{{Pillai} et~al.}{2015}]{Pillai+2015}
{Pillai} T.,  {Kauffmann} J.,  {Tan} J.~C.,  {Goldsmith} P.~F.,  {Carey} S.~J.,
    {Menten} K.~M.,  2015, \mn@doi [\apj] {10.1088/0004-637X/799/1/74}, \href
  {https://ui.adsabs.harvard.edu/abs/2015ApJ...799...74P} {799, 74}

\bibitem[\protect\citeauthoryear{{Ponti} et~al.,}{{Ponti}
  et~al.}{2019}]{Ponti+2019}
{Ponti} G.,  et~al., 2019, \mn@doi [\nat] {10.1038/s41586-019-1009-6}, \href
  {https://ui.adsabs.harvard.edu/abs/2019Natur.567..347P} {567, 347}

\bibitem[\protect\citeauthoryear{{Rahner}, {Pellegrini}, {Glover}  \&
  {Klessen}}{{Rahner} et~al.}{2018}]{Rahner+2018}
{Rahner} D.,  {Pellegrini} E.~W.,  {Glover} S. C.~O.,   {Klessen} R.~S.,  2018,
  \mn@doi [\mnras] {10.1093/mnrasl/slx149}, \href
  {https://ui.adsabs.harvard.edu/abs/2018MNRAS.473L..11R} {473, L11}

\bibitem[\protect\citeauthoryear{{Rahner}, {Pellegrini}, {Glover}  \&
  {Klessen}}{{Rahner} et~al.}{2019}]{Rahner+2019}
{Rahner} D.,  {Pellegrini} E.~W.,  {Glover} S.~C.~O.,   {Klessen} R.~S.,  2019,
  \mn@doi [\mnras] {10.1093/mnras/sty3295}, \href
  {http://adsabs.harvard.edu/abs/2019MNRAS.483.2547R} {483, 2547}

\bibitem[\protect\citeauthoryear{{Rathborne} et~al.,}{{Rathborne}
  et~al.}{2015}]{Rathborne+2015}
{Rathborne} J.~M.,  et~al., 2015, \mn@doi [\apj] {10.1088/0004-637X/802/2/125},
  \href {https://ui.adsabs.harvard.edu/abs/2015ApJ...802..125R} {802, 125}

\bibitem[\protect\citeauthoryear{{Requena-Torres} et~al.,}{{Requena-Torres}
  et~al.}{2012}]{RequenaTorres+2012}
{Requena-Torres} M.~A.,  et~al., 2012, \mn@doi [\aap]
  {10.1051/0004-6361/201219068}, \href
  {https://ui.adsabs.harvard.edu/abs/2012A&A...542L..21R} {542, L21}

\bibitem[\protect\citeauthoryear{{Ridley}, {Sormani}, {Tre{\ss}}, {Magorrian}
  \& {Klessen}}{{Ridley} et~al.}{2017}]{Ridley+2017}
{Ridley} M.,  {Sormani} M.~C.,  {Tre{\ss}} R.~G.,  {Magorrian} J.,   {Klessen}
  R.~S.,  2017, \mnras, \href
  {http://adsabs.harvard.edu/abs/2017arXiv170403665R} {469, 2251}

\bibitem[\protect\citeauthoryear{{Rosen}, {Krumholz}, {McKee}  \&
  {Klein}}{{Rosen} et~al.}{2016}]{Rosen+2016}
{Rosen} A.~L.,  {Krumholz} M.~R.,  {McKee} C.~F.,   {Klein} R.~I.,  2016,
  \mn@doi [\mnras] {10.1093/mnras/stw2153}, \href
  {https://ui.adsabs.harvard.edu/abs/2016MNRAS.463.2553R} {463, 2553}

\bibitem[\protect\citeauthoryear{{Sandage}}{{Sandage}}{1961}]{Sandage1961}
{Sandage} A.,  1961, {The Hubble Atlas of Galaxies}.
Carnegie Institution of Washington, Washington, D.C.

\bibitem[\protect\citeauthoryear{{Schmidt}}{{Schmidt}}{1959}]{Schmidt1959}
{Schmidt} M.,  1959, \mn@doi [\apj] {10.1086/146614}, \href
  {http://adsabs.harvard.edu/abs/1959ApJ...129..243S} {129, 243}

\bibitem[\protect\citeauthoryear{{Shetty}, {Beaumont}, {Burton}, {Kelly}  \&
  {Klessen}}{{Shetty} et~al.}{2012}]{Shetty+2012}
{Shetty} R.,  {Beaumont} C.~N.,  {Burton} M.~G.,  {Kelly} B.~C.,   {Klessen}
  R.~S.,  2012, \mn@doi [\mnras] {10.1111/j.1365-2966.2012.21588.x}, \href
  {https://ui.adsabs.harvard.edu/abs/2012MNRAS.425..720S} {425, 720}

\bibitem[\protect\citeauthoryear{{Shin}, {Kim}, {Baba}, {Saitoh}, {Hwang},
  {Chun}  \& {Hozumi}}{{Shin} et~al.}{2017}]{Shin+2017}
{Shin} J.,  {Kim} S.~S.,  {Baba} J.,  {Saitoh} T.~R.,  {Hwang} J.-S.,  {Chun}
  K.,   {Hozumi} S.,  2017, \mn@doi [\apj] {10.3847/1538-4357/aa7061}, \href
  {https://ui.adsabs.harvard.edu/abs/2017ApJ...841...74S} {841, 74}

\bibitem[\protect\citeauthoryear{{Simpson}, {Bryan}, {Hummels}  \&
  {Ostriker}}{{Simpson} et~al.}{2015}]{Simpson+2015}
{Simpson} C.~M.,  {Bryan} G.~L.,  {Hummels} C.,   {Ostriker} J.~P.,  2015,
  \mn@doi [\apj] {10.1088/0004-637X/809/1/69}, \href
  {https://ui.adsabs.harvard.edu/abs/2015ApJ...809...69S} {809, 69}

\bibitem[\protect\citeauthoryear{{Sjouwerman} \& {Pihlstr{\"o}m}}{{Sjouwerman}
  \& {Pihlstr{\"o}m}}{2008}]{Sjouwerman+2008}
{Sjouwerman} L.~O.,  {Pihlstr{\"o}m} Y.~M.,  2008, \mn@doi [\apj]
  {10.1086/588753}, \href
  {https://ui.adsabs.harvard.edu/abs/2008ApJ...681.1287S} {681, 1287}

\bibitem[\protect\citeauthoryear{{Sofue}}{{Sofue}}{1995}]{Sofue95}
{Sofue} Y.,  1995, \pasj, \href
  {http://adsabs.harvard.edu/abs/1995PASJ...47..527S} {47, 527}

\bibitem[\protect\citeauthoryear{{Sormani} \& {Barnes}}{{Sormani} \&
  {Barnes}}{2019}]{SormaniBarnes2019}
{Sormani} M.~C.,  {Barnes} A.~T.,  2019, \mn@doi [\mnras]
  {10.1093/mnras/stz046}, \href
  {http://adsabs.harvard.edu/abs/2019MNRAS.484.1213S} {484, 1213}

\bibitem[\protect\citeauthoryear{{Sormani} \& {Li}}{{Sormani} \&
  {Li}}{2020}]{SormaniLi2020}
{Sormani} M.~C.,  {Li} Z.,  2020, \mn@doi [\mnras] {10.1093/mnras/staa1139},
  \href {https://ui.adsabs.harvard.edu/abs/2020MNRAS.494.6030S} {494, 6030}

\bibitem[\protect\citeauthoryear{{Sormani}, {Binney}  \& {Magorrian}}{{Sormani}
  et~al.}{2015a}]{SBM2015a}
{Sormani} M.~C.,  {Binney} J.,   {Magorrian} J.,  2015a, \mn@doi [\mnras]
  {10.1093/mnras/stv441}, \href
  {http://adsabs.harvard.edu/abs/2015MNRAS.449.2421S} {449, 2421}

\bibitem[\protect\citeauthoryear{{Sormani}, {Binney}  \& {Magorrian}}{{Sormani}
  et~al.}{2015b}]{SBM2015b}
{Sormani} M.~C.,  {Binney} J.,   {Magorrian} J.,  2015b, \mn@doi [\mnras]
  {10.1093/mnras/stv1135}, \href
  {http://adsabs.harvard.edu/abs/2015MNRAS.451.3437S} {451, 3437}

\bibitem[\protect\citeauthoryear{{Sormani}, {Binney}  \& {Magorrian}}{{Sormani}
  et~al.}{2015c}]{SBM2015c}
{Sormani} M.~C.,  {Binney} J.,   {Magorrian} J.,  2015c, \mn@doi [\mnras]
  {10.1093/mnras/stv2067}, \href
  {http://adsabs.harvard.edu/abs/2015MNRAS.454.1818S} {454, 1818}

\bibitem[\protect\citeauthoryear{{Sormani}, {Tre{\ss}}, {Klessen}  \&
  {Glover}}{{Sormani} et~al.}{2017}]{Sormani+2017b}
{Sormani} M.~C.,  {Tre{\ss}} R.~G.,  {Klessen} R.~S.,   {Glover} S. C.~O.,
  2017, \mn@doi [\mnras] {10.1093/mnras/stw3205}, \href
  {https://ui.adsabs.harvard.edu/abs/2017MNRAS.466..407S} {466, 407}

\bibitem[\protect\citeauthoryear{{Sormani}, {Tre{\ss}}, {Ridley}, {Glover},
  {Klessen}, {Binney}, {Magorrian}  \& {Smith}}{{Sormani}
  et~al.}{2018}]{Sormani+2018a}
{Sormani} M.~C.,  {Tre{\ss}} R.~G.,  {Ridley} M.,  {Glover} S.~C.~O.,
  {Klessen} R.~S.,  {Binney} J.,  {Magorrian} J.,   {Smith} R.,  2018, \mn@doi
  [\mnras] {10.1093/mnras/stx3258}, \href
  {http://adsabs.harvard.edu/abs/2018MNRAS.475.2383S} {475, 2383}

\bibitem[\protect\citeauthoryear{{Sormani} et~al.,}{{Sormani}
  et~al.}{2019}]{Sormani+2019}
{Sormani} M.~C.,  et~al., 2019, \mn@doi [\mnras] {10.1093/mnras/stz2054}, \href
  {https://ui.adsabs.harvard.edu/abs/2019MNRAS.488.4663S} {488, 4663}

\bibitem[\protect\citeauthoryear{{Sormani}, {Tress}, {Glover}, {Klessen},
  {Battersby}, {Clark}, {Hatchfield}  \& {Smith}}{{Sormani}
  et~al.}{2020}]{Sormani+2020}
{Sormani} M.~C.,  {Tress} R.~G.,  {Glover} S. C.~O.,  {Klessen} R.~S.,
  {Battersby} C.~D.,  {Clark} P.~C.,  {Hatchfield} H.~P.,   {Smith} R.~J.,
  2020, \mn@doi [\mnras] {10.1093/mnras/staa1999}, \href
  {https://ui.adsabs.harvard.edu/abs/2020MNRAS.497.5024S} {497, 5024 (Paper
  II)}

\bibitem[\protect\citeauthoryear{{Springel}}{{Springel}}{2005}]{Springel2005}
{Springel} V.,  2005, \mn@doi [\mnras] {10.1111/j.1365-2966.2005.09655.x},
  \href {http://adsabs.harvard.edu/abs/2005MNRAS.364.1105S} {364, 1105}

\bibitem[\protect\citeauthoryear{{Springel}}{{Springel}}{2010}]{Springel2010}
{Springel} V.,  2010, \mn@doi [\mnras] {10.1111/j.1365-2966.2009.15715.x},
  \href {http://adsabs.harvard.edu/abs/2010MNRAS.401..791S} {401, 791}

\bibitem[\protect\citeauthoryear{{Stinson}, {Seth}, {Katz}, {Wadsley},
  {Governato}  \& {Quinn}}{{Stinson} et~al.}{2006}]{Stinson+2006}
{Stinson} G.,  {Seth} A.,  {Katz} N.,  {Wadsley} J.,  {Governato} F.,   {Quinn}
  T.,  2006, \mn@doi [\mnras] {10.1111/j.1365-2966.2006.11097.x}, \href
  {http://adsabs.harvard.edu/abs/2006MNRAS.373.1074S} {373, 1074}

\bibitem[\protect\citeauthoryear{{Su}, {Slatyer}  \& {Finkbeiner}}{{Su}
  et~al.}{2010}]{Su+2010}
{Su} M.,  {Slatyer} T.~R.,   {Finkbeiner} D.~P.,  2010, \mn@doi [\apj]
  {10.1088/0004-637X/724/2/1044}, \href
  {https://ui.adsabs.harvard.edu/abs/2010ApJ...724.1044S} {724, 1044}

\bibitem[\protect\citeauthoryear{{Takekawa}, {Oka}  \& {Tanaka}}{{Takekawa}
  et~al.}{2017}]{Takekawa+2017}
{Takekawa} S.,  {Oka} T.,   {Tanaka} K.,  2017, \mn@doi [\apj]
  {10.3847/1538-4357/834/2/121}, \href
  {https://ui.adsabs.harvard.edu/abs/2017ApJ...834..121T} {834, 121}

\bibitem[\protect\citeauthoryear{{Toomre}}{{Toomre}}{1964}]{Toomre1964}
{Toomre} A.,  1964, \mn@doi [\apj] {10.1086/147861}, \href
  {https://ui.adsabs.harvard.edu/abs/1964ApJ...139.1217T} {139, 1217}

\bibitem[\protect\citeauthoryear{{Trani}, {Mapelli}  \& {Ballone}}{{Trani}
  et~al.}{2018}]{Trani+2018}
{Trani} A.~A.,  {Mapelli} M.,   {Ballone} A.,  2018, \mn@doi [\apj]
  {10.3847/1538-4357/aad414}, \href
  {https://ui.adsabs.harvard.edu/abs/2018ApJ...864...17T} {864, 17}

\bibitem[\protect\citeauthoryear{{Tress}, {Smith}, {Sormani}, {Glover},
  {Klessen}, {Mac Low}  \& {Clark}}{{Tress} et~al.}{2020}]{Tress+2019}
{Tress} R.~G.,  {Smith} R.~J.,  {Sormani} M.~C.,  {Glover} S. C.~O.,  {Klessen}
  R.~S.,  {Mac Low} M.-M.,   {Clark} P.~C.,  2020, \mn@doi [\mnras]
  {10.1093/mnras/stz3600}, \href
  {https://ui.adsabs.harvard.edu/abs/2020MNRAS.492.2973T} {492, 2973}

\bibitem[\protect\citeauthoryear{{Truelove}, {Klein}, {McKee}, {Holliman},
  {Howell}  \& {Greenough}}{{Truelove} et~al.}{1997}]{Truelove+1997}
{Truelove} J.~K.,  {Klein} R.~I.,  {McKee} C.~F.,  {Holliman} II J.~H.,
  {Howell} L.~H.,   {Greenough} J.~A.,  1997, \mn@doi [\apjl] {10.1086/310975},
  \href {http://adsabs.harvard.edu/abs/1997ApJ...489L.179T} {489, L179}

\bibitem[\protect\citeauthoryear{{Tsuboi}, {Kitamura}, {Uehara}, {Tsutsumi},
  {Miyawaki}, {Miyoshi}  \& {Miyazaki}}{{Tsuboi} et~al.}{2018}]{Tsuboi+2018}
{Tsuboi} M.,  {Kitamura} Y.,  {Uehara} K.,  {Tsutsumi} T.,  {Miyawaki} R.,
  {Miyoshi} M.,   {Miyazaki} A.,  2018, \mn@doi [\pasj] {10.1093/pasj/psy080},
  \href {http://adsabs.harvard.edu/abs/2018PASJ...70...85T} {70, 85}

\bibitem[\protect\citeauthoryear{{Utomo} et~al.,}{{Utomo}
  et~al.}{2018}]{Utomo+2018}
{Utomo} D.,  et~al., 2018, \mn@doi [\apjl] {10.3847/2041-8213/aacf8f}, \href
  {https://ui.adsabs.harvard.edu/abs/2018ApJ...861L..18U} {861, L18}

\bibitem[\protect\citeauthoryear{{Walch} et~al.,}{{Walch}
  et~al.}{2015}]{Walch+2015}
{Walch} S.,  et~al., 2015, \mn@doi [\mnras] {10.1093/mnras/stv1975}, \href
  {http://adsabs.harvard.edu/abs/2015MNRAS.454..238W} {454, 238}

\bibitem[\protect\citeauthoryear{{Wang} \& {Wheeler}}{{Wang} \&
  {Wheeler}}{2008}]{WangWheeler2008}
{Wang} L.,  {Wheeler} J.~C.,  2008, \mn@doi [\araa]
  {10.1146/annurev.astro.46.060407.145139}, \href
  {https://ui.adsabs.harvard.edu/abs/2008ARA&A..46..433W} {46, 433}

\bibitem[\protect\citeauthoryear{{Weinberger}, {Springel}  \&
  {Pakmor}}{{Weinberger} et~al.}{2020}]{Weinberger+2020}
{Weinberger} R.,  {Springel} V.,   {Pakmor} R.,  2020, \mn@doi [\apjs]
  {10.3847/1538-4365/ab908c}, \href
  {https://ui.adsabs.harvard.edu/abs/2020ApJS..248...32W} {248, 32}

\makeatother
\end{thebibliography}

\newpage

\appendix

\section{Images of surface densities in various tracers}

In this appendix, we show the time evolution of the gas surface densities in various tracers.

\begin{figure*}
	\includegraphics[width=\textwidth]{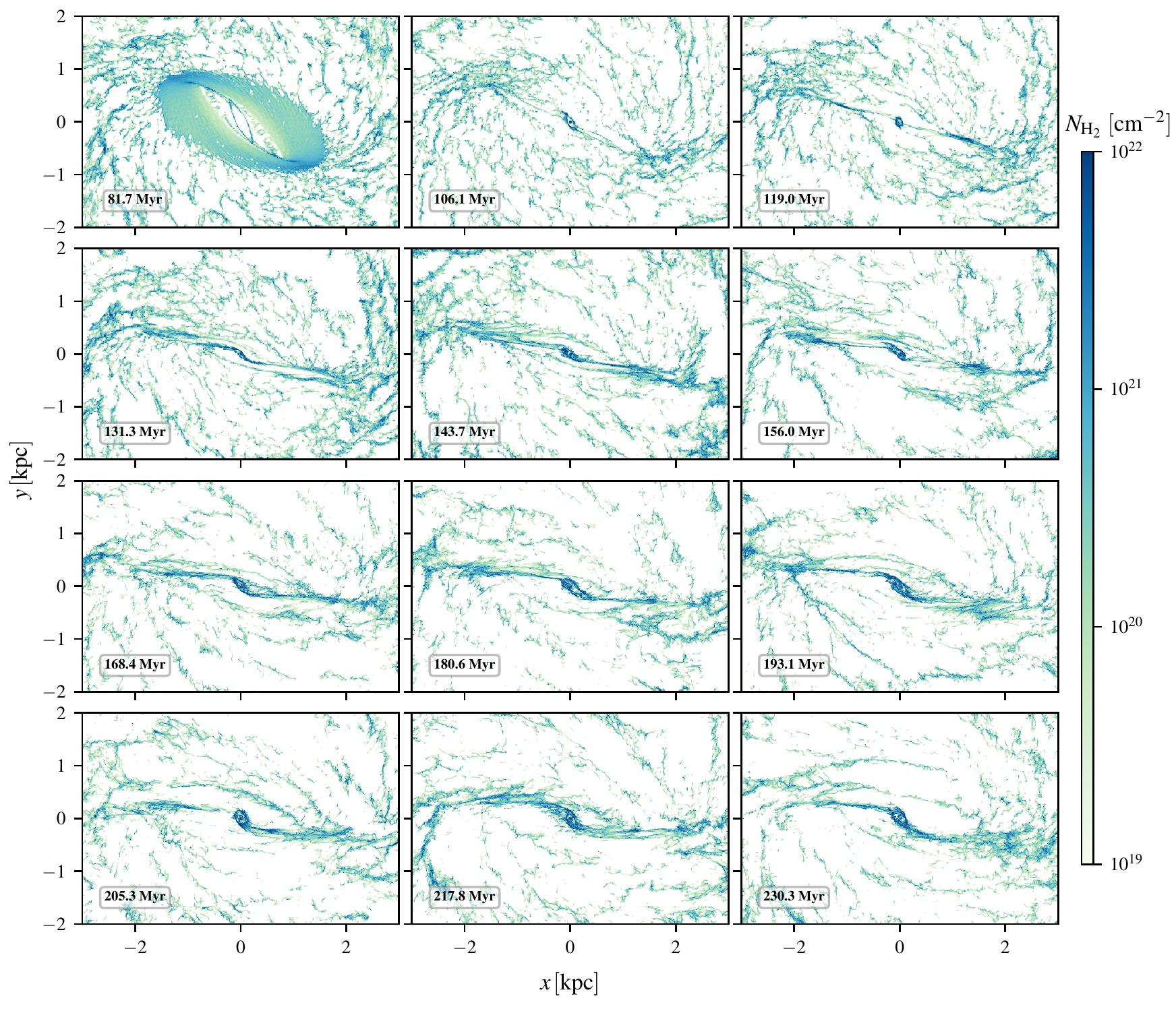}
    \caption{H$_2$ gas surface density at different times in the simulation.}
    \label{fig:NH2vstime}
\end{figure*}

\begin{figure*}
	\includegraphics[width=\textwidth]{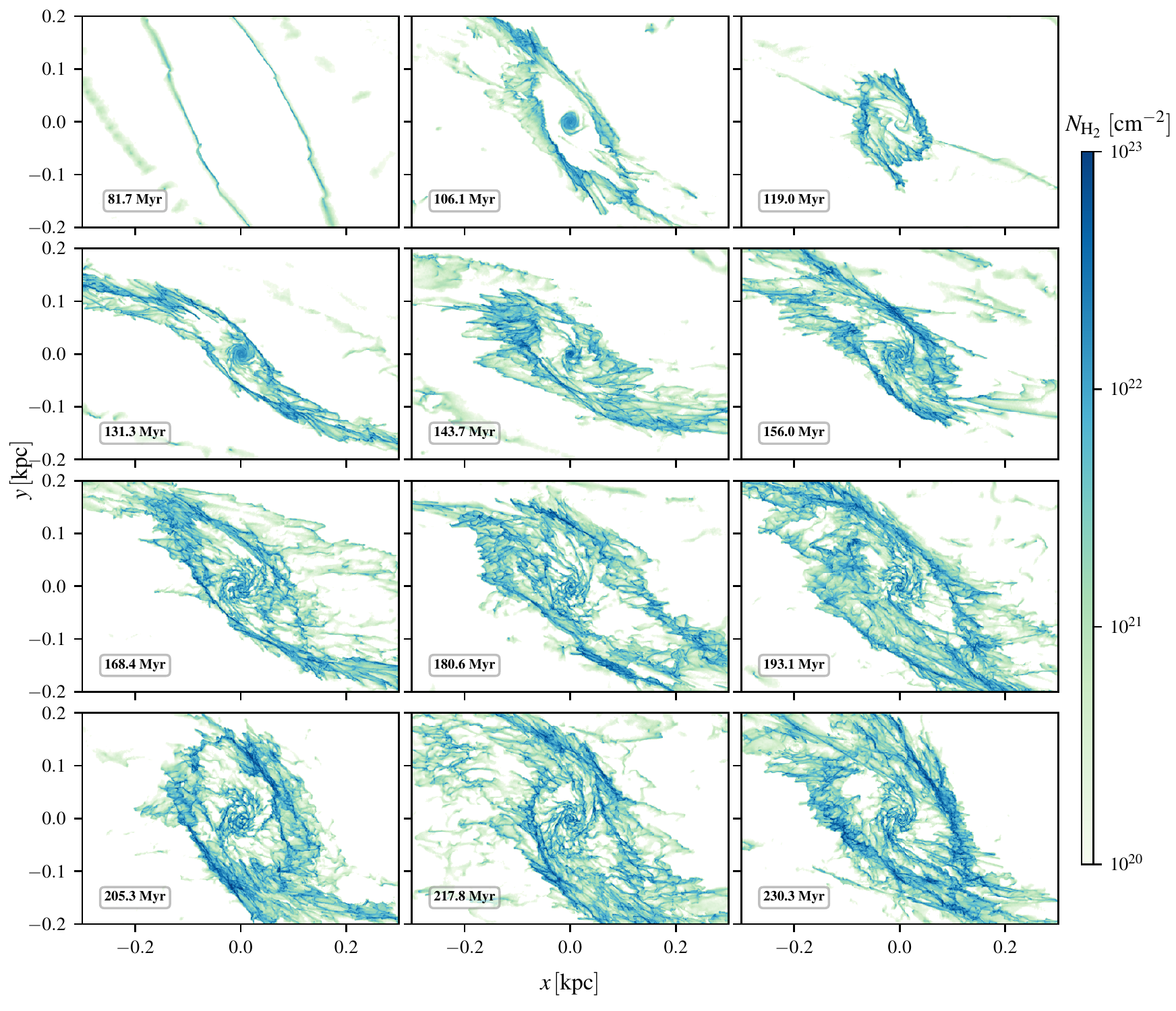}
    \caption{Zoom of Figure \ref{fig:NH2vstime} onto the CMZ.}
    \label{fig:NH2vstime_CMZ}
\end{figure*}

\begin{figure*}
	\includegraphics[width=\textwidth]{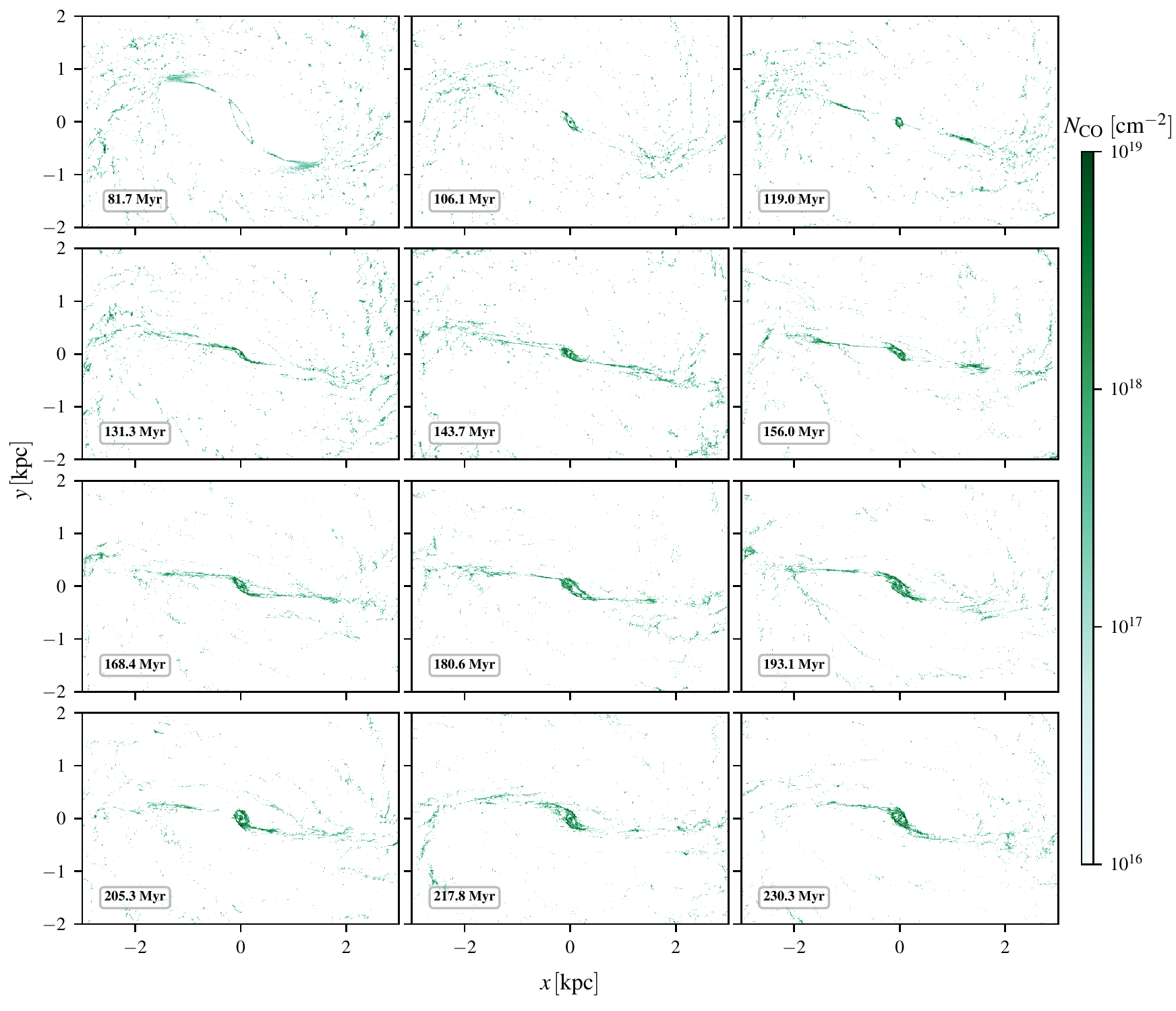}
    \caption{CO gas surface density at different times in the simulation.}
    \label{fig:NCOvstime}
\end{figure*}

\begin{figure*}
	\includegraphics[width=\textwidth]{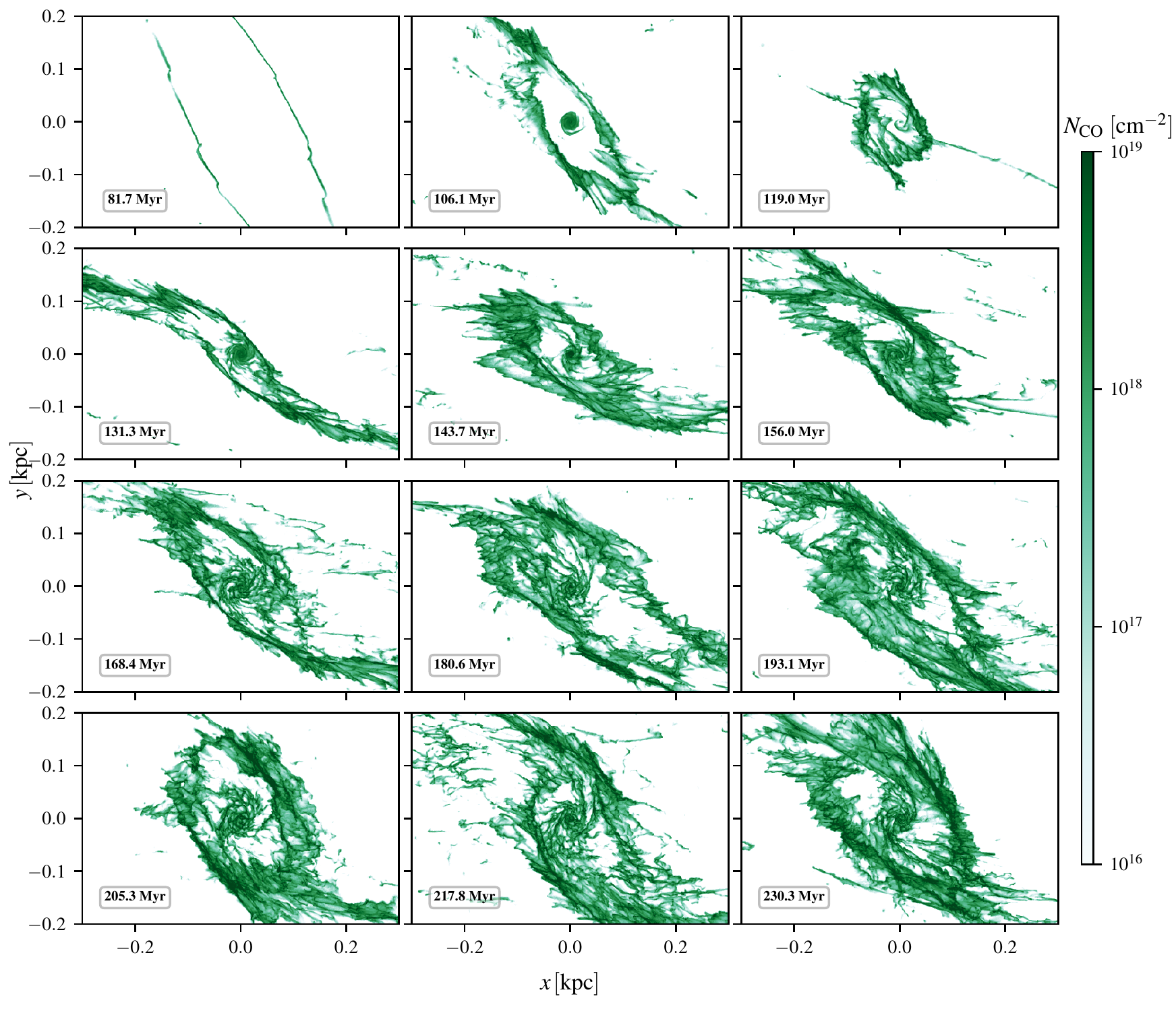}
    \caption{Zoom of Figure \ref{fig:NCOvstime} onto the CMZ.}
    \label{fig:NCOvstime_CMZ}
\end{figure*}

\begin{figure*}
	\includegraphics[width=\textwidth]{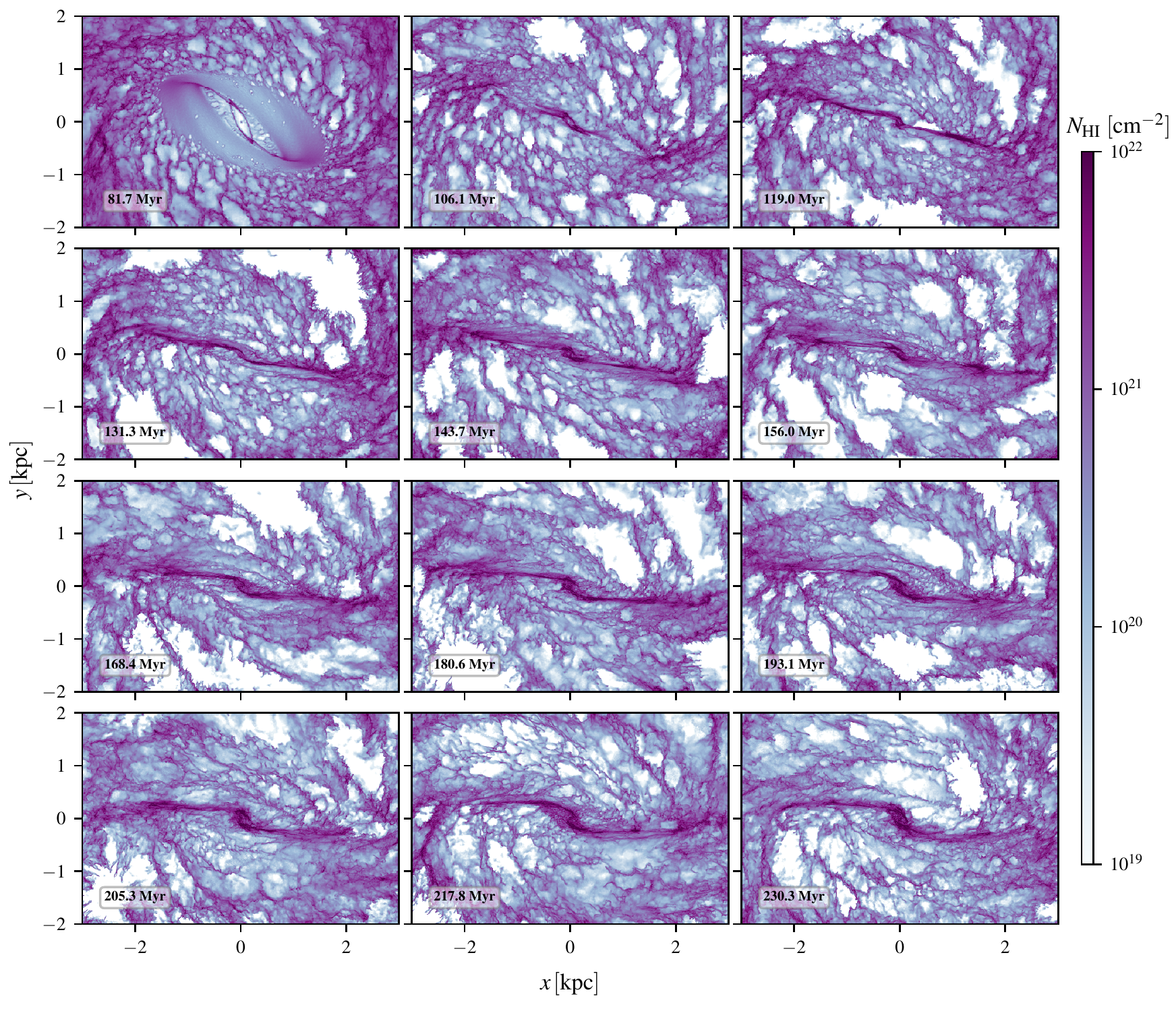}
    \caption{H{\sc I} gas surface density at different times in the simulation.}
    \label{fig:NHIvstime}
\end{figure*}

\begin{figure*}
	\includegraphics[width=\textwidth]{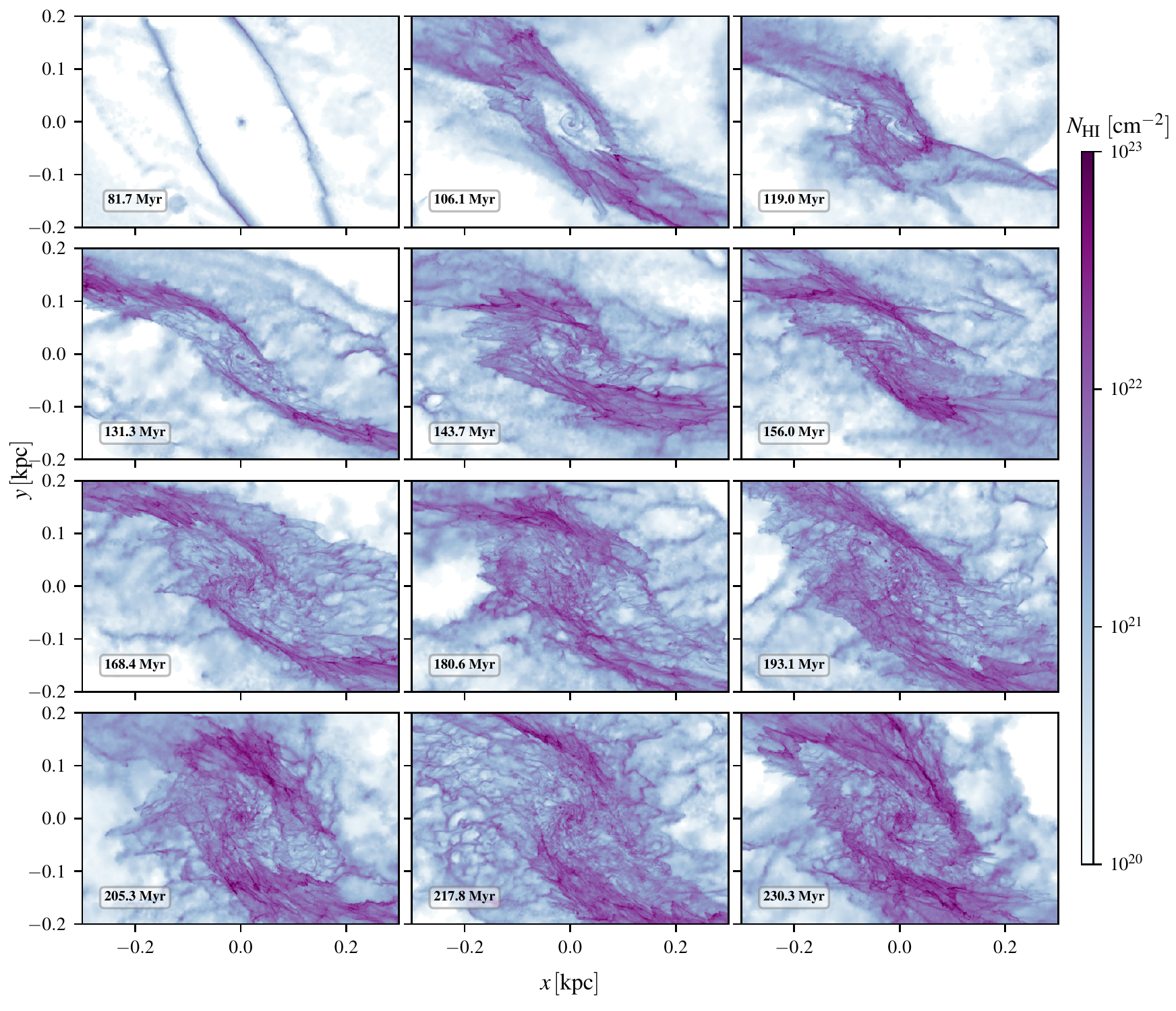}
    \caption{Zoom of Figure \ref{fig:NHIvstime} onto the CMZ.}
    \label{fig:NHIvstime_CMZ}
\end{figure*}

\begin{figure*}
	\includegraphics[width=\textwidth]{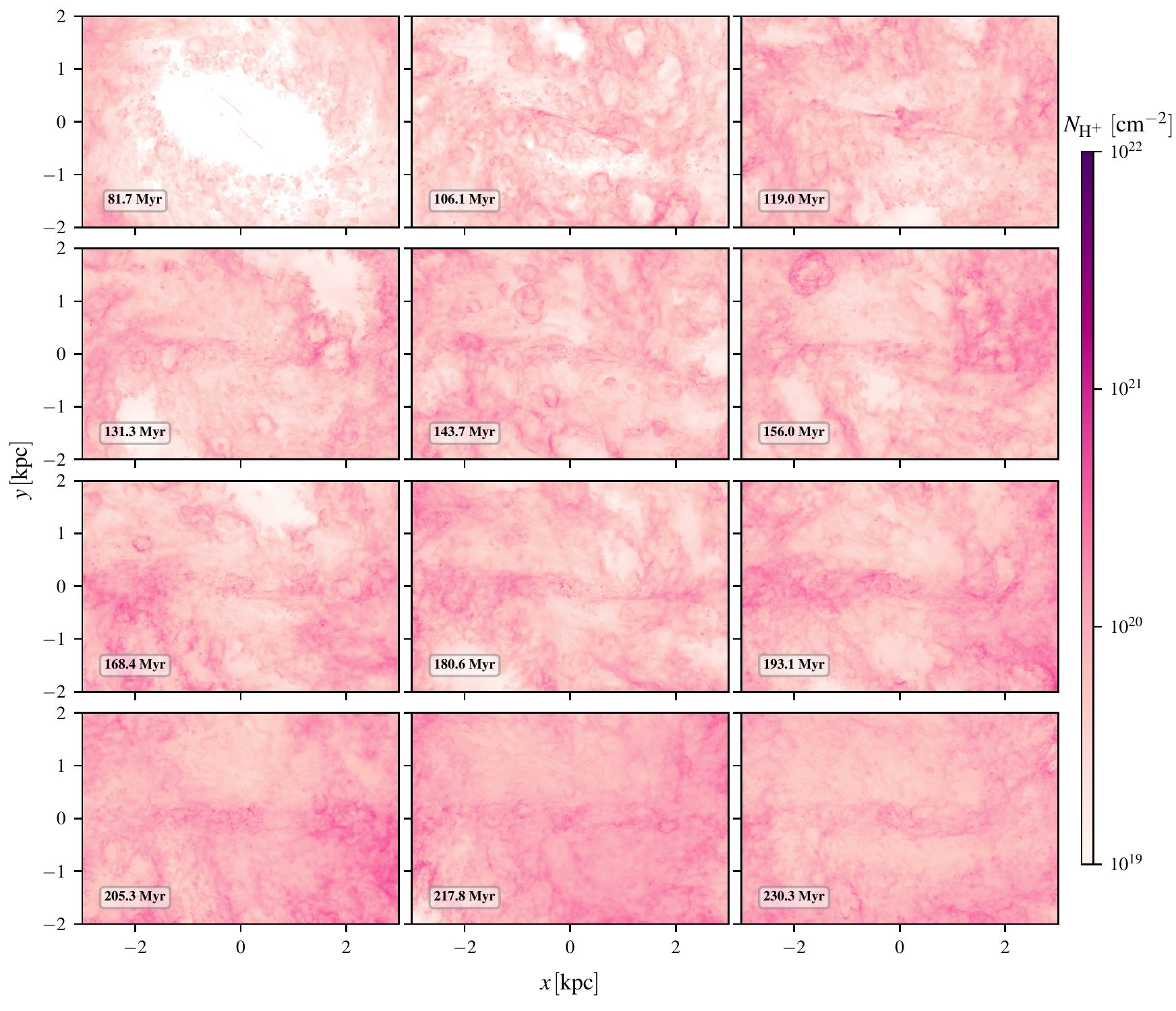}
    \caption{H$^+$ gas surface density at different times in the simulation.}
    \label{fig:NHpvstime}
\end{figure*}

\begin{figure*}
	\includegraphics[width=\textwidth]{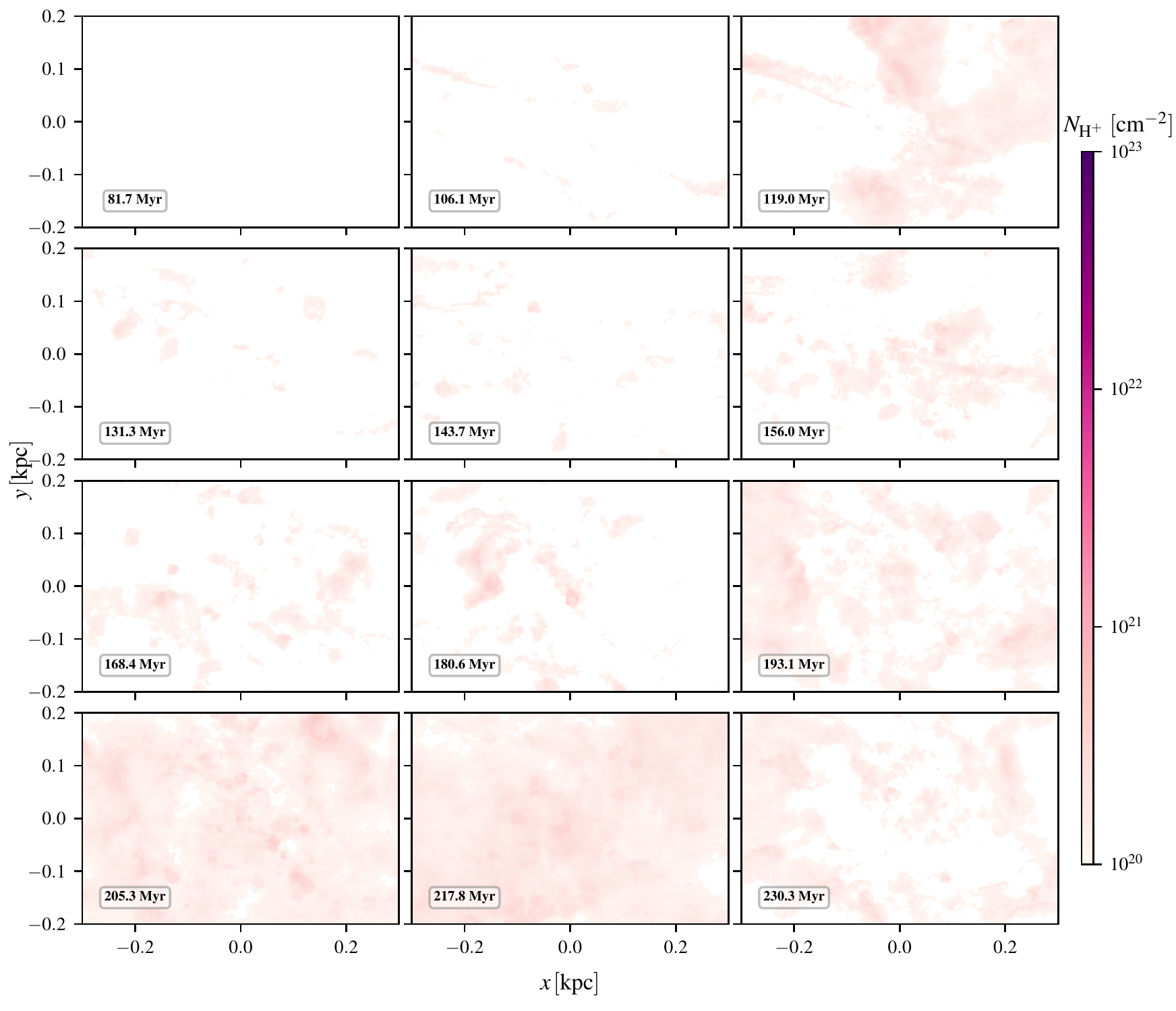}
    \caption{Zoom of Figure \ref{fig:NHpvstime} onto the CMZ.}
    \label{fig:NHpvstime_CMZ}
\end{figure*}

\end{document}